\documentclass[aps,twocolumn,english,superscriptaddress,citeautoscript,showkeys,preprintnumbers,amsmath,amssymb,floatfix,footinbib]{revtex4-2}
\usepackage{hyperref}
\hypersetup{colorlinks=true,linkcolor=red,citecolor=blue}
\usepackage{amsmath}
\usepackage{graphicx}
\usepackage{dcolumn}
\usepackage{xcolor}
\usepackage{float}
\usepackage{soul}
\usepackage{adjustbox}
\usepackage{lipsum}
\usepackage{blindtext}

\usepackage{lipsum}% example text

\mathchardef\mhyphen="2D

\begin{document}
	
	\title{Quantum critical spin-liquid-like behavior in S = 1/2 quasikagome lattice CeRh$_{1-x}$Pd$_{x}$Sn investigated using muon spin relaxation and neutron scattering}
	\author{Rajesh Tripathi}
	\email{rajeshtripathi@jncasr.ac.in}
	\affiliation{ISIS Neutron and Muon Source, STFC, Rutherford Appleton Laboratory, Chilton, Oxon OX11 0QX, United Kingdom}
	\affiliation{Jawaharlal Nehru Centre for Advanced Scientific Research, Jakkur, Bangalore 560064, India}
	
	\author{D. T. Adroja}
	\email{devashibhai.adroja@stfc.ac.uk}
	\affiliation{ISIS Neutron and Muon Source, STFC, Rutherford Appleton Laboratory, Chilton, Oxon OX11 0QX, United Kingdom}
	\affiliation{Highly Correlated Matter Research Group, Physics Department, University of Johannesburg, Auckland Park 2006, South Africa}
	\author{C. Ritter}
	\affiliation{Institut Laue-Langevin, 71 Avenue des Martyrs, CS 20156, 38042, Grenoble Cedex 9, France}

	\author{Shivani Sharma}
	\affiliation{ISIS Neutron and Muon Source, STFC, Rutherford Appleton Laboratory, Chilton, Oxon OX11 0QX, United Kingdom}
	
	\author{Chongli Yang}
	\affiliation{Beijing Academy of Quantum Information Sciences, Beijing 100083, China}
	\author{A.D. Hillier}
	\affiliation{ISIS Neutron and Muon Source, STFC, Rutherford Appleton Laboratory, Chilton, Oxon OX11 0QX, United Kingdom}
	\author{M. M. Koza}
	\affiliation{Institut Laue-Langevin, 71 Avenue des Martyrs, CS 20156, 38042, Grenoble Cedex 9, France}
	\author{F. Demmel}
	\affiliation{ISIS Neutron and Muon Source, STFC, Rutherford Appleton Laboratory, Chilton, Oxon OX11 0QX, United Kingdom}
	
	\author{ A. Sundaresan}
	\affiliation{Jawaharlal Nehru Centre for Advanced Scientific Research, Jakkur, Bangalore 560064, India}
	
	\author{S. Langridge}
	\affiliation{ISIS Neutron and Muon Source, STFC, Rutherford Appleton Laboratory, Chilton, Oxon OX11 0QX, United Kingdom}
	
	\author{Wataru Higemoto}
	\affiliation{Advanced Science Research Center, Japan Atomic Energy Agency 2-4 Shirakata, Tokai-mura, Naka-gun, Ibaraki 319-1195, Japan}
	\affiliation{Department of Physics, Tokyo Institute of Technology 2-12-1 O-Okayama, Meguro, Tokyo 152-8551, Japan}
	\author{Takashi U. Ito}
	\affiliation{Advanced Science Research Center, Japan Atomic Energy Agency 2-4 Shirakata, Tokai-mura, Naka-gun, Ibaraki 319-1195, Japan}
	
	\author{A. M. Strydom}
	\affiliation{Highly Correlated Matter Research Group, Physics Department, University of Johannesburg, Auckland Park 2006, South Africa}
	\author{G. B. G. Stenning}
	\affiliation{ISIS Neutron and Muon Source, STFC, Rutherford Appleton Laboratory, Chilton, Oxon OX11 0QX, United Kingdom}

	\author{A. Bhattacharyya}
	\affiliation{Department of Physics, Ramakrishna Mission Vivekananda Educational and Research Institute, Howrah 711202, India}
	
	\author{David Keen}
	\affiliation{ISIS Neutron and Muon Source, STFC, Rutherford Appleton Laboratory, Chilton, Oxon OX11 0QX, United Kingdom}
	\author{H. C. Walker}
	\affiliation{ISIS Neutron and Muon Source, STFC, Rutherford Appleton Laboratory, Chilton, Oxon OX11 0QX, United Kingdom}
	\author{R. S. Perry}
	
	\affiliation{London Centre for Nanotechnology and Department of Physics and Astronomy, University College London, Gower Street, London WC1E 6BT, United Kingdom}
	
	\affiliation{ISIS Neutron and Muon Source, STFC, Rutherford Appleton Laboratory, Chilton, Oxon OX11 0QX, United Kingdom}
	
	\author{Francis Pratt}
	\affiliation{ISIS Neutron and Muon Source, STFC, Rutherford Appleton Laboratory, Chilton, Oxon OX11 0QX, United Kingdom}
	
	\author{Qimiao Si}
	\affiliation{Department of Physics and Astronomy, Rice Center for Quantum Materials, Rice University, Houston, TX, USA}
	
	\author{T. Takabatake}
	\affiliation{Department of Quantum Matter, Graduate School of Advanced Science and Engineering, 
		Hiroshima University, Higashi-Hiroshima 739-8530, Japan
	}

	\date{\today}
	
	\begin{abstract}
		
		We present the results of muon spin relaxation ($\mu$SR) and neutron scattering on the Ce-based quasikagome lattice CeRh$_{1-x}$Pd$_{x}$Sn ($x=0.1$ to 0.75). Our zero-field (ZF) $\mu$SR results reveal the absence of both static long-range magnetic order and spin freezing down to 0.05~K in the single crystal sample of $x = 0.1$. The weak temperature-dependent plateaus of the dynamic spin fluctuations below 0.2~K in ZF-$\mu$SR together with its longitudinal-field (LF) dependence between 0 and 3~kG indicate the presence of dynamic spin fluctuations persisting even at $T$ = 0.05~K without static magnetic order. On the other hand, the magnetic specific heat divided by temperature $C_{\text{4f}}$/$T$ increases as --log $T$ on cooling below 0.9~K, passes through a broad maximum at 0.13~K and slightly decreases on further cooling. The ac-susceptibility also exhibits a frequency independent broad peak at 0.16~K, which is prominent with an applied field $H$ along $c$-direction. We, therefore, argue that such a behavior for $x=0.1$ (namely, a plateau in spin relaxation rate ($\lambda$) below 0.2~K and a linear $T$ dependence in $C_{\text{4f}}$ below 0.13~K) can be attributed to a metallic spin-liquid (SL) ground state near the quantum critical point in the frustrated Kondo lattice. The LF-$\mu$SR study suggests that the out of kagome plane spin fluctuations are responsible for the SL behavior. Low energy inelastic neutron scattering (INS) of $x$ = 0.1 reveals gapless magnetic excitations, which are also supported by the behavior of $C_{\text{4f}}$ proportional to $T^{1.1}$ down to 0.06~K. Our high energy INS study shows very weak and broad scattering in $x = 0$ and 0.1, which transforms into well-localized crystal field excitations with increasing $x$. The ZF-$\mu$SR results for the $x = 0.2$ polycrystalline sample exhibits similar behavior to that of $x = 0.1$. A saturation of $\lambda$ below 0.2~K suggests a spin-fluctuating SL ground state down to 0.05~K. The ZF-$\mu$SR results for the $x = 0.5$ sample are interpreted as a long-range antiferromagnetic (AFM) ground state below $T_{\text{N}}$ = 0.8~K, in which the AFM interaction of the enlarged moments probably overcomes the frustration effect. The long-range AFM ordering is also supported by the evolution of magnetic Bragg peaks in $x$ = 0.75 sample observed below 5~K in the neutron diffraction data.  
		
		\keywords {Kondo effect; Kagome lattice; Valence fluctuation; Quantum phase transition; Antiferromagnetism; Quantum spin-liquid; Muon spin rotation and relaxation}
		
	\end{abstract}
	
	\maketitle
	\section{INTRODUCTION}
	\label{Intro}
	In geometrically frustrated spin systems, the competing exchange interactions prevent a magnetically ordered ground state even at $T\rightarrow$ 0, and thus frustrated spins can form a quantum entangled ground state, so called quantum spin-liquid (QSL)~\cite{ANDERSON1973153,balents2010spin}. QSLs have been amongst the most intriguing topics in condensed matter physics since the first notion of spin-liquid (SL) was theoretically proposed by P. Anderson in 1973~\cite{ANDERSON1973153}. There has been a continual effort to explore the materials that might host QSLs, mainly in geometrically frustrated magnets~\cite{Savary_2016,chamorro2020chemistry,takagi2019concept,clark2021quantum, doi:10.1126/science.aay0668}. Among various proposed host systems for a QSL, kagome lattices are found to be the most likely candidate for the realization of a QSL ground state and topological order~\cite{balents2010spin, doi:10.1126/science.abi8794}. 
	
	Much of what we currently know about QSLs are associated with the experimental and theoretical work on insulating magnets. Little is known about their metallic counterparts, though their phase behaviors are expected to be much more diverse~\cite{JPSJ.79.011008,PhysRevLett.96.087204,PhysRevLett.110.017201,doi.org/10.1038/nmat3900,doi:10.1126/sciadv.1500001}. Ground states of $f$-electron based Kondo metals are generally classified into a nonmagnetic Fermi liquid (FL) regime and a magnetically ordered regime as the result of the competition between the Kondo effect and the Ruderman-Kittel-Kasuya-Yosida (RKKY) interactions~\cite{DONIACH1977231}. At the critical value of the coupling between $4f$ and conduction electrons ($c-f$ hybridization), magnetic ordering is suppressed to zero temperature and a quantum critical point (QCP) occurs, where Fermi-liquid theory breaks down and non-Fermi-liquid (NFL) behavior appears~\cite{DONIACH1977231,RevModPhys.79.1015,RevModPhys.92.011002}. In the case of Kondo ions arranged on a geometrically frustrated lattice, magnetic frustration suppresses both the transition temperature and the moments, and the under-screened moments may remain disordered even in the magnetic regime, forming a metallic SL state~\cite{SI200623}. In metallic systems, therefore, the frustration inherent to the Kondo lattice may lead to additional quantum fluctuations of local moments, adding to the delicate competition between the Kondo effect and RKKY interaction in the presence of magnetic frustration~\cite{Si2010QuantumCA}. As a consequence, a partial Kondo screening state~\cite{PhysRevLett.105.036403}, a valence-bond solid~\cite{doi.org/10.1007,PhysRevLett.105.036403,PhysRevB.83.214427,PhysRevLett.113.176402}, or even a QSL~\cite{SI200623,doi.org/10.1002,doi.org/10.1007,PhysRevLett.105.036403} may appear in extended phase space, competing with the magnetically ordered and FL phases. Experimentally, however, this topic is largely unexplored, mainly due to the lack of appropriate frustrated Kondo systems.

	In recent years, strongly correlated quantum matter, such as heavy-fermion (HF) metals, have been considered as prototypical systems to study metallic SL. Prominent examples are the HF compounds Pr$_2$Ir$_2$O$_7$~\cite{PhysRevLett.96.087204}, LiV$_2$O$_4$~\cite{PhysRevB.99.041113}, and Y(Sc)Mn$_2$~\cite{YScMn19931329}, which have a common feature that the transition metal ions comprise the pyrochlore lattice and are therefore subject to geometrical frustration, as inferred from the emergence of a metallic SL. Suppressing the transition temperature further results in a field-induced QSL in a finite window of the magnetic field. For example, the application of magnetic field tunes the geometrically frustrated kagome systems YbAgGe ($T_{\text{N}}$ = 0.8~K)~\cite{PhysRevLett.111.116401}, and CePdAl ($T_{\text{N}}$ = 2.7~K)~\cite{PhysRevLett.118.107204}, both of which crystallize in the hexagonal ZrNiAl-type structure with the space group of $P\bar{6}2m$, into the paramagnetic state via an intermediate QSL metal.

	\begin{figure}
		
		\includegraphics[width=10cm, keepaspectratio]{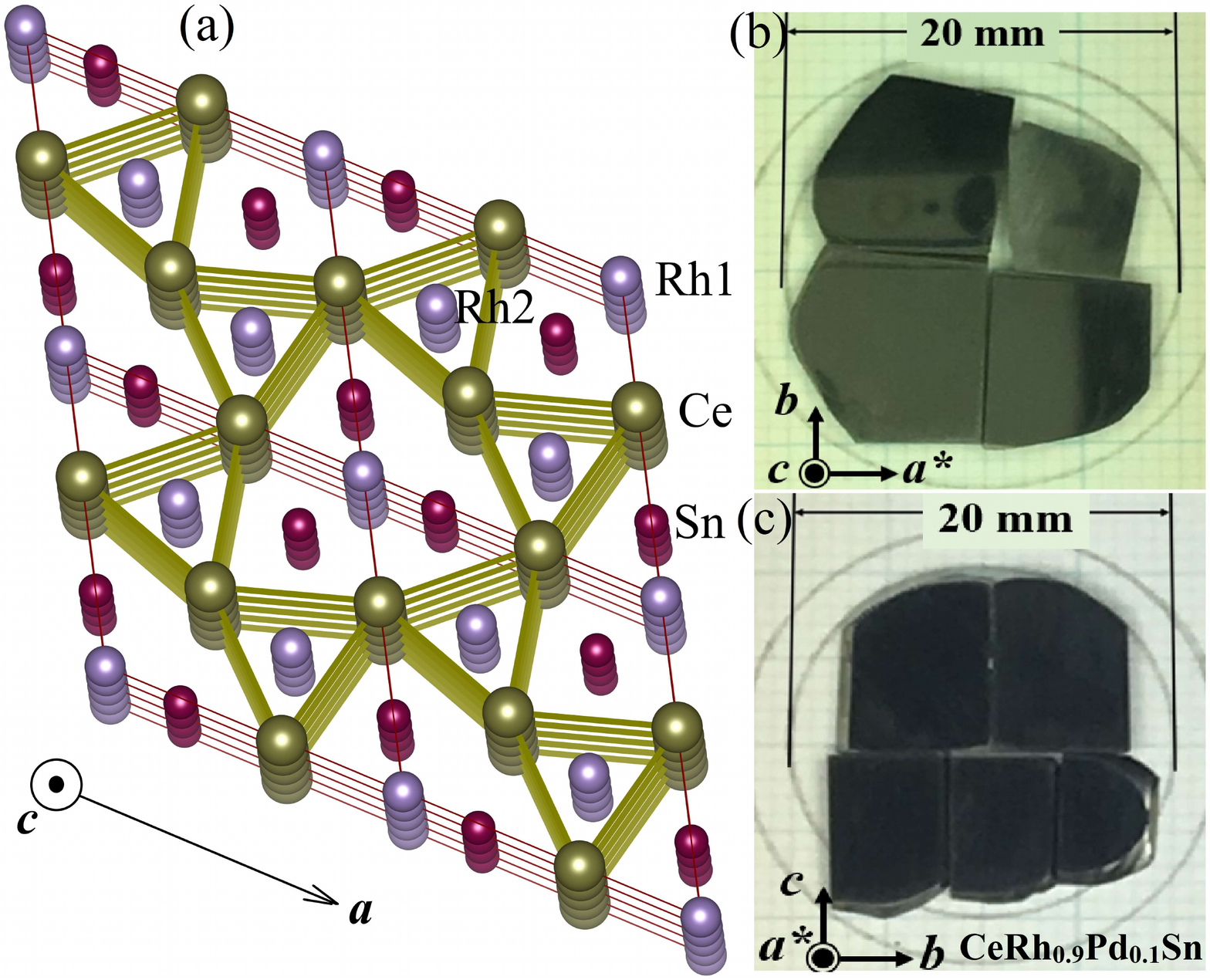}
		\caption{(a) Crystal structure of CeRhSn. Ce, Rh [with two different sites, Rh1 and Rh2, where Rh2 atoms are located inside the Ce triangular prism (Note: the $z$-parameter for Rh2 is 1/2, being deviated from the Ce-Rh1 plane)], and Sn atoms are shown as dark yellow, violet, and wine colors spheres, respectively. (b) Photograph of a CeRh$_{0.9}$Pd$_{0.1}$Sn single crystal placed such that the $c$-axis lies perpendicular to the plane of the paper, and (c) $a^*$-axis lies perpendicular to the plane of the paper.}
		\label{structure}
		
	\end{figure}

	CeRhSn is another isostructural Kondo-lattice compound as the Ce ions are arranged in a geometrically frustrated quasikagome network in the hexagonal basal plane~\cite{mishra2001stannides,PhysRevB.68.054416} [see Fig.~\ref{structure}(a)]. In contrast to the antiferromagnets CePdAl and YbAgGe, it has a large Kondo temperature $T_{\text{K}}$ = 240~K and remains paramagnetic down to at least 0.05~K, with the indication for the proximity to a magnetic QCP~\cite{doi:10.1126/sciadv.1500001}. The high-temperature susceptibility of CeRhSn shows a Curie-Weiss behavior with a Weiss temperature $\theta_{\text{P}}\sim$ --155~K ~\cite{PhysRevB.68.054416,PhysRevB.96.045139}, but no clear evidence of long-range magnetic ordering down to $T =$ 0.02~K was evident~\cite{doi:10.1143/JPSJ.73.3099}. The single crystal susceptibility of CeRhSn exhibits a strong anisotropy with an easy $c$-axis~\cite {KIM2003524} and following a power-law behavior $\chi_c\propto T^{-1.1}$ and $\chi_a\propto T^{-0.35}$ at low temperatures. Geometrical frustration has been discussed as having a profound influence on ground-state physics. Thermal expansion experiments showed that geometrical frustration is responsible for the zero-field quantum criticality~\cite{doi:10.1126/sciadv.1500001}. The application of uniaxial pressure in the hexagonal plane along the $a$-axis leads to a long-range ordered state~\cite{PhysRevB.96.241110}. This is unusual because the Kondo coupling increases with stress in Ce-based compounds, implying that the paramagnetic ground state would be stabilized. The formation of the magnetic ground state upon application of uniaxial stress in the $ab$ plane has, therefore, been interpreted in terms of a stress-induced alleviation of geometrical frustration in the hexagonal plane~\cite{PhysRevB.96.241110}. 
	
	CeRh$_{1-x}$Pd$_x$Sn has been studied extensively by C. L. Yang et al.~\cite{PhysRevB.96.045139}. A transition from paramagnetic to long-range magnetic ordering takes place when the Rh sites are substituted with Pd (i.e., $4d$-electron doping in CeRhSn). It was anticipated that the Pd substitution for Rh suppresses both frustration and the Kondo effect, leading to the development of an AFM order. Both $C/T$ and $\chi_{\text{ac}}$($T$) exhibit a maximum, whose temperature increases from 0.1~K for $x = 0.1$ to 2.5~K for $x = 0.75$. Here it is important to note that $\chi_{\text{ac}}$ of undoped CeRhSn shows a broad maximum at 0.1~K, whereas $C/T$ keeps increasing down to 0.05~K~\cite{PhysRevB.96.045139,doi:10.1126/sciadv.1500001}. The hard x-rays photoelectric spectroscopy (HAXPES) shows that both the pure CeRhSn and the sample with $x = 0.1$ belong to the valence fluctuating regime where no magnetic order should occur. With a further increase of Pd content, the $c$-$f$ hybridization is strongly suppressed, which stabilizes the trivalent state of Ce~\cite{PhysRevB.104.235150}. Therefore the exact ground state of this system is still a matter of some debate.
	
	From the bulk measurements only, it is difficult to fully understand the origin of the low temperature behavior and to separate the contributions from frustration, valence fluctuations, Kondo effect, and crystal field in CeRh$_{1-x}$Pd$_x$Sn. The investigations using microscopic measurements are, therefore, essential. For example, $\mu$SR is a powerful local probe that is able to detect tiny magnetic moments with an average ordered moment size of 0.005~$\mu_B$ (or higher) ~\cite {muonbook}. It can distinguish the random static fields associated with, for example, the dipolar coupling of the muon and quasi-static nuclear moments and dynamically fluctuating fields associated with electronic spin fluctuations~\cite{MUSR}. Therefore, with the aim to understand the static and/or dynamic behavior of CeRh$_{1-x}$Pd$_x$Sn in detail, $\mu$SR measurements were performed on the single crystal sample with $x$ = 0.1 and on the polycrystalline samples with $x$ = 0.2 and 0.5. We also performed inelastic neutron scattering (INS) measurements on all the polycrystalline samples with $x$ = 0, 0.1, 0.2, 0.5, and 0.75. The neutron diffraction (ND) measurements were performed on powder sample of $x$ = 0.75. In addition, we have performed ac magnetic susceptibility and specific heat measurements on $x$ = 0.1 single crystals under the magnetic fields along and perpendicular to the $c$-axis. 
	
	Our $\mu$SR investigation on CeRh$_{1-x}$Pd$_{x}$Sn with $x=$ 0.1 and $x$ = 0.2 reveals that the ground state is non-magnetic within the muon time scale. The temperature dependent zero-field (ZF)-$\mu$SR relaxation rate $\lambda$($T$) for both $x=$ 0.1 single crystal and $x$ = 0.2 polycrystal exhibits a typical behavior observed in QSL systems~\cite{PhysRevLett.117.097201,PhysRevLett.109.037208,PhysRevB.84.100401}. The elastic neutron data observed in the INS and ND studies support the absence of magnetic ordering in the polycrystalline samples of $x=$ 0.1 and $x=$ 0.2 down to 0.07~K and 0.12~K, respectively. The $\mu$SR study on $x=$ 0.5 and the ND study on $x =$ 0.75 polycrystalline samples provide a clear evidence of long-range magnetically ordered ground state. Low energy INS study reveals a clear presence of quasi-elastic scattering (gapless excitations) in the polycrystalline samples of $x=$ 0.1, 0.2, 0.5, and 0.75. Furthermore, the temperature-dependent quasi-elastic scattering in $x=$ 0.1 and 0.2 is very similar to that observed in NFL systems near QCP~\cite{PhysRevLett.75.725}. We have observed the energy by temperature ($E/T$) scaling of the dynamical susceptibility both for $x=0.1$ and 0.2. A high energy INS study reveals a broad inelastic excitation in $x= $0, 0.1, and 0.2 and broad crystal electric field (CEF) excitations in $x= $0.5 and 0.75.

	\begin{figure}
		\includegraphics[width=8.5cm, keepaspectratio]{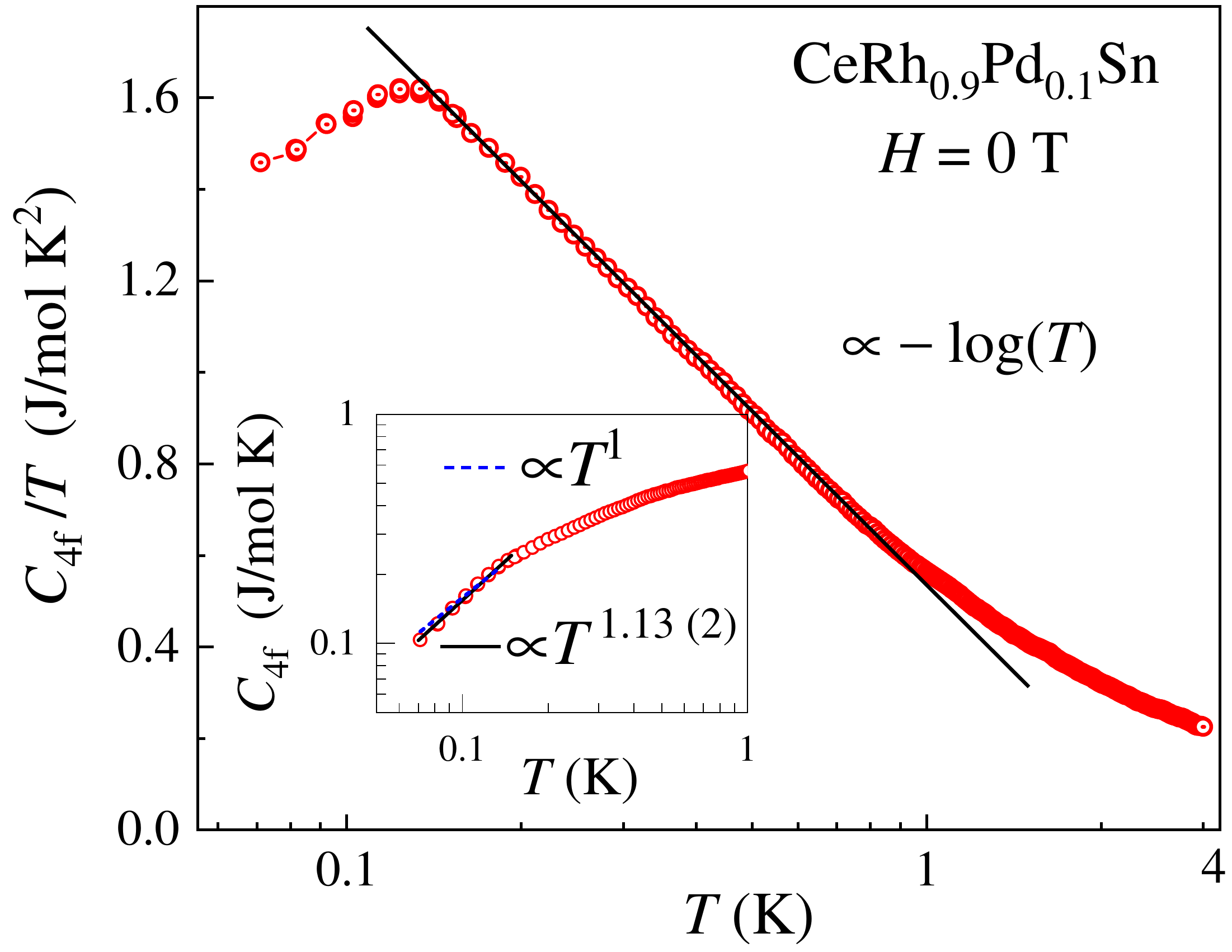}
		\caption{ The magnetic specific heat divided by temperature ($C_{\text{4f}}/T$) vs log $T$ for CeRh$_{0.9}$Pd$_{0.1}$Sn single crystal. Solid line represents the logarithmic behavior of the data. Inset shows $C_{\text{4f}}$ vs $T$ on log-log plot. The solid line is a fit to the data with $C_{\text{4f}}$ $\propto$ $T^{1.13}$. The dashed line shows $C_{\text{4f}}$ $\propto$ $T^{1}$ behavior.}
		\label{Transport2}
	\end{figure}

	\begin{figure*}
		
		\includegraphics[width=\textwidth, keepaspectratio]{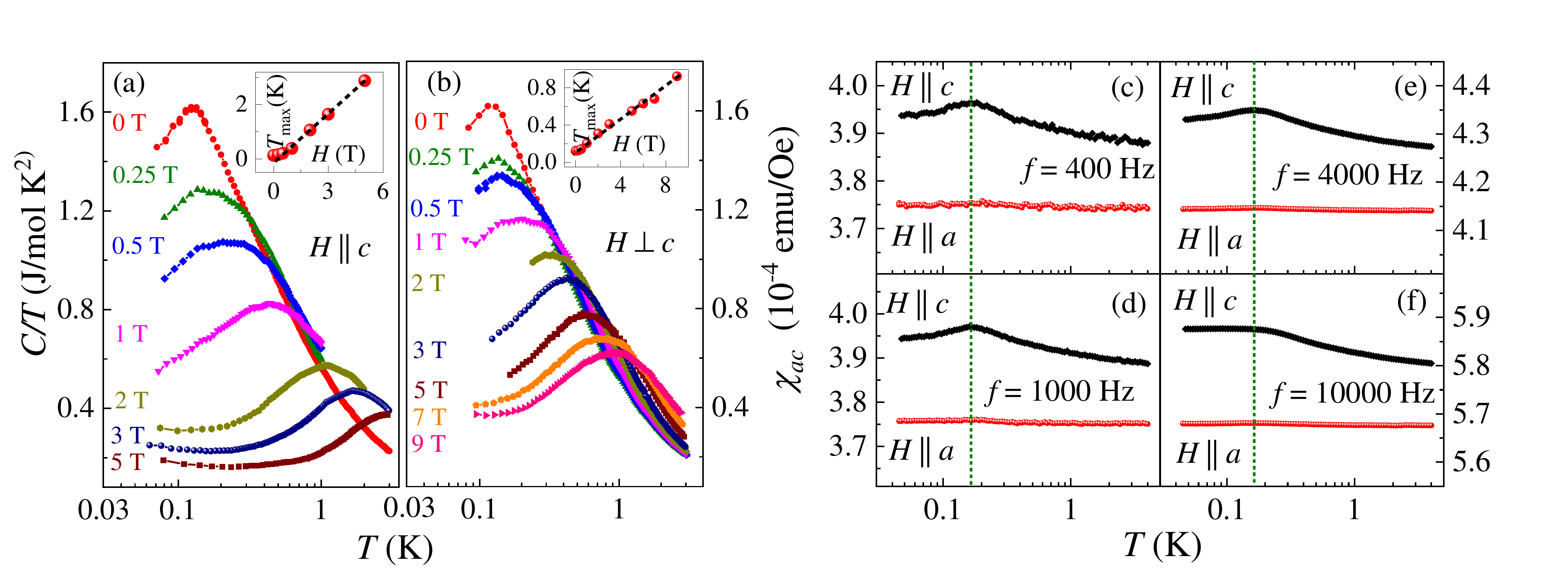}
		\caption{Specific heat and ac-susceptibility of CeRh$_{0.9}$Pd$_{0.1}$Sn single crystal: (a), (b) $C/T$ vs log $T$ for the fields applied parallel and perpendicular to the $c$-axis. Insets show the variation of $C/T$ peak ($T_{\text{max}}$) with applied magnetic fields (c)-(f) The temperature dependence of the ac-susceptibility with $H=40$~G along $a$ and $c$-axis, for a range of frequencies.}
		\label{Transport1}
		
	\end{figure*}

	\section{EXPERIMENTAL METHODS}
	\label{expt}
	
	Single crystals of CeRh$_{1-x}$Pd$_x$Sn with $x=0.1$ and the polycrystalline samples with $x$ = 0.1, 0.2, 0.5, and 0.75 were prepared according to Ref.~\cite{PhysRevB.96.045139}. Polycrystalline samples of LaRh$_{1-x}$Pd$_x$Sn with $x=0.1$ and 0.5 were prepared for phonon reference measurements in the INS study. The temperature dependent specific heat $C_{\rm p}$($T$) and ac-susceptibility $\chi_{ac}$($T$) measurements on $x=0.1$ single crystals were made using the specific heat option with mK temperature range of a physical properties measurements system (Dynacool PPMS, Quantum Design Inc). 
	
	ZF- and longitudinal-field (LF)-$\mu$SR measurements on the single crystal sample of $x$ = 0.1 [with the initial polarization of the muon beam along and perpendicular to the $c$-axis] were performed at the ISIS Neutron and Muon Source, UK, using the MuSR spectrometer. For ISIS muon measurements, the $x=0.1$ single crystals in a thin plate form with thickens of 1mm and radius of 10~mm [Figs.~\ref{structure}(b) and~\ref{structure}(c)] were mounted on a 99.999\% pure silver plate using diluted GE varnish to ensure good thermal contact and then covered with a thin silver foil. We used a dilution refrigerator to cool the samples down to 0.05~K. The $\mu$SR data collected at ISIS were analyzed with the MANTID software~\cite{ARNOLD2014156}. $\mu$SR measurements on the polycrystalline samples with $x$ = 0.2 and 0.5 were performed using the D1 muon beam line at the J-PARC, Japan and the data were analyzed with Wimda software~\cite{PRATT2000710}. 
	
	The INS experiments on polycrystalline CeRh$_{1-x}$Pd$_x$Sn ($x=0.1$, 0.2, 0.5, and 0.75) and LaRh$_{1-x}$Pd$_x$Sn ($x=0.1$, 0.5) were performed on the MERLIN time of flight (TOF) spectrometer at the UK ISIS Neutron and Muon Source~\cite{Bewley2009}. The powdered samples of these materials were wrapped in thin Al-foil and mounted inside thin-walled cylindrical Al-cans with a diameter of 30~mm and height of 40~mm. Low temperatures down to 5~K were obtained by cooling the sample mounts in a top-loading closed cycle refrigerator with He-exchange gas. The INS data were collected with repetition-rate multiplication using a neutron incident energy of $E_i$ = 60~meV and a Fermi chopper frequency of 350~Hz, which also provided data for $E_i$ = 24 and 13~meV. The elastic resolution (FWHM) was 2.58~meV for $E_i$ = 60~meV.  The INS experiments on polycrystalline CeRh$_{1-x}$Pd$_x$Sn ($x=0$) and LaRh$_{1-x}$Pd$_x$Sn ($x=0$) were performed on the MARI TOF spectrometer. The data are presented in absolute units, mb/meV/sr/f.u. using the absolute normalization obtained from the standard vanadium sample measured in identical conditions.
	
	Low energy INS data were collected using the cold IN6 TOF spectrometer with neutron incident energy of 3.1~meV at Institute Laue-Langevin (ILL), Grenoble, France. The elastic resolution (FWHM) was $80~\mu$eV. The sample with $x$ = 0.1 was mounted in an annular form in a Cu-can (16~mm diameter) and cooled down to 0.07~K using a dilution refrigerator. The samples with $x=$0, 0.2 and 0.75 were mounted in an Al-foil envelope (25 $\times$ 38~mm) and cooled down to 1.5~K using an orange He-4 cryostat. We also measured the $x = 0.5$ sample down to 5~K using low energy neutrons on the TOF inverted-geometry crystal analyzer spectrometer OSIRIS with a PG002 analyzer and selecting the final neutron energy of 1.845~meV at the ISIS Neutron and Muon Source. The elastic resolution (FWHM) was $25~\mu$eV. The powder sample was mounted in an annular form in an Al-can (20~mm diameter). The neutron powder diffraction measurements were performed on the sample with $x=0.75$ using the GEM TOF diffractometer down to 0.5~K at ISIS Neutron and Muon Source and the constant wavelength diffractometer D20 down to 1.5~K at ILL. We also performed ND measurements on the sample with $x=0.2$ using the OSIRIS TOF diffractometer down to 0.12~K at ISIS Neutron and Muon Source.
	
	\section {Experimental results and discussion}
	
	\subsection {Specific heat and ac-susceptibility}
	\label{ACSus}
	The temperature dependence of the magnetic heat capacity ($C_{\text{4f}}$) and $C_{\text{4f}}$/$T$ of $x=0.1$ single crystal at $H=0$ T are plotted in the inset and main panel of Fig.~\ref{Transport2}, respectively. Here $C_{\text{4f}}$ was estimated by subtracting the specific heat of the non-magnetic analog LaRhSn. The $C_{\text{4f}}$ data shows no lambda-type anomaly down to 0.05~K, suggesting the absence of long-range magnetic phase transition. The $C_{\text{4f}}$/$T$ increases as -- log $T$ with decreasing temperature followed by a broad peak at 0.13~K. The -- log $T$ dependence of $C_{\text{4f}}$/$T$ agrees with that expected for the system in the quantum critical regime~\cite{RevModPhys.79.1015}. 
	
	To check whether the peak might arise from magnetic ordering, we performed $\mu$SR (down to 0.05~K) and neutron scattering experiments (down to 0.08~K). The detailed discussion of these results are given below. The results show no clear evidence of magnetic ordering. Hence, it is very unlikely that the anomaly in $C_{\text{4f}}$/$T$ at 0.13~K arises from long-range AFM order. Therefore, the obvious explanation is that magnetic fluctuations dominate the low temperature heat capacity, resulting in a peak around 0.13~K, representing the onset of short-range correlations. The existence of short-range correlations is a common feature of paramagnetic HF compounds residing close to a QCP, where there is a zero-temperature transition between paramagnetic and magnetically ordered ground state~\cite{RevModPhys.79.1015}. We can therefore anticipate that the broad peak in the specific heat arises from short-range magnetic correlations, reflecting the fact that the system is close to a magnetic QCP. On the other hand, in a gapless QSL, a linear increase in magnetic specific heat is expected at low temperatures~\cite{mustonen2018spin}. As shown in the inset of Fig.~\ref{Transport2}, a fit of the low-temperature $C_{\text{4f}}$ data by the power-law $k_\text{m} T^{\alpha_\text{m}}$ yields the exponent $\alpha_\text{m}$ = 1.13 (2), confirming the linear $T$-dependence of $C_{\text{4f}}$.
	
	When a magnetic field is applied to the system in the vicinity of a QCP, the specific heat changes significantly~\cite{RevModPhys.79.1015}. Figs.~\ref{Transport1}(a) and~\ref{Transport1}(b) represent the temperature dependence of $C/T$ in various magnetic fields applied parallel and perpendicular to the $c$-axis, respectively. With the application of the magnetic field, the broad peak in $C/T$ evolves into a Schottky-type anomaly, which moves up to a higher temperature. This behavior is consistent with the field-induced splitting of the Ce ground state doublet. The temperature at the $C/T$ peak is proportional to the Zeeman energy. The insets of Figs.~\ref{Transport1}(a) and~\ref{Transport1}(b) show that the $C/T$ peak temperature $T_{\text{max}}$ increases linearly with increasing the field parallel and perpendicular to the $c$-axis. This suggests Zeeman splitting of the Ce ground state doublet, where at zero field, the splitting could arise from the fluctuating internal field due to short-range order. It is to be noted that for a ferromagnetic correlation or ordering, the specific heat peak also moves upward with increasing applied magnetic field~\cite{PhysRevB.81.235132}.
	
	To further explore the short-range correlation observed in heat capacity of $x = 0.1$ single crystal, we measured the low temperature ac-susceptibility $\chi_{\text{ac}}$($T$) at various frequencies from 400 to 10000~Hz. The $\chi_{\text{ac}}$ with applied magnetic fields along and perpendicular to the $c$-axis down to 0.05~K for a range of frequencies are shown in Figs.~\ref{Transport1}(c)--~\ref{Transport1}(f). The signal below 400~Hz is too weak to obtain reliable data, and hence the data are not presented. For the magnetic field along the $c$-direction, a broad maximum is observed at 0.16~K, close to where a weak peak is observed in $C/T$. As can be seen from Figs.~\ref{Transport1}(c) --~\ref{Transport1}(f), $\chi_{\text{ac}}$ peak does not depend on the frequency between 400 and 10000~Hz. Therefore, we conclude reliably that the single crystal sample with $x$ = 0.1 does not have a spin-glass transition.
	
	\subsection{Muon spin relaxation}
	
\begin{figure*}
		\begin{center}
			\includegraphics[width=\textwidth, keepaspectratio]{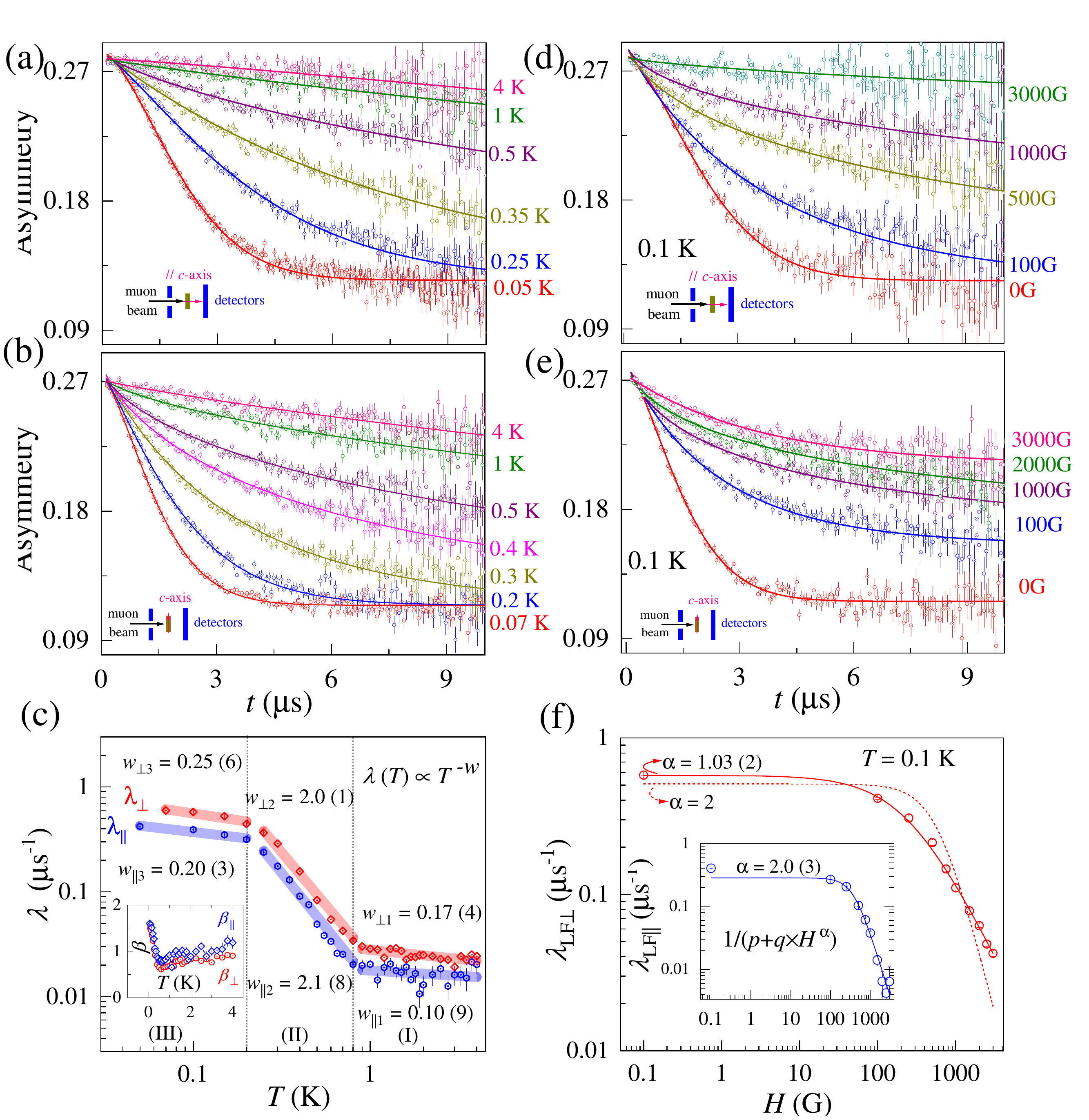}
			\caption{ $\mu$SR data of single crystal sample of CeRh$_{0.9}$Pd$_{0.1}$Sn: (a) ZF-$\mu$SR spectra with the muon beam along the $c$-axis at representative temperatures. The solid lines are the fitted curves (see the text for details). (b) ZF-$\mu$SR spectra with the muon beam $\perp c$-axis at representative temperatures. The solid lines are the fitted curves (see the text for details). (c) Temperature dependence of the muon spin relaxation rate ($\lambda$) for the muon beam along and perpendicular to the $c$-axis. The inset shows the temperature dependent stretched exponents ($\beta$) for the muon beam along and perpendicular to the $c$-axis. The blue and red solid (transparent) lines describe the power-law behavior of relaxation rate $\lambda$, i.e., $\lambda$($T$) $\propto$ $T^{- w}$, over a given temperature range for the muon beam along and perpendicular to the $c$-axis, respectively. (d) The LF-$\mu$SR spectra with the muon beam along the $c$-axis measured at 0.1~K under several longitudinal magnetic fields (e) The LF-$\mu$SR spectra with the muon beam $\perp c$-axis measured at 0.1~K under several longitudinal magnetic fields. (f) Magnetic-field dependence of the $\lambda$ for the muon beam along (inset) and perpendicular to the $c$-axis. The solid lines are the fitted curves to a power-law of the form $1/(p + qH^\alpha)$. The dotted line shows the fit with $\alpha$ = 2.}
			\label{spectra_SC}
		\end{center}
	\end{figure*}

	To gain further insights into the static and/or dynamic properties of the ground states in CeRh$_{1-x}$Pd$_{x}$Sn, we performed ZF- and LF-$\mu$SR measurements. The time-dependent ZF-$\mu$SR spectra of the single crystalline sample for $x$ = 0.1 with the initial polarisation of muon along and perpendicular to the $c$-axis are displayed in Figs.~\ref{spectra_SC}(a)  and ~\ref{spectra_SC}(b), respectively. The $\mu$SR spectra depolarizes faster as the temperature is decreased; however, neither the oscillatory signal nor a 1/3 recovery tail of the muon polarization due to a random distribution of the static field are observed with the incident muon polarization along or perpendicular to the $c$-axis down to 0.07~K and 0.05~K, respectively. This behavior suggests the absence of a well-defined or disordered static magnetic field at the muon stopping site and hence ruled out any possibilities of long-range magnetic ordering or spin freezing due to Ce moments in $x$ = 0.1 single crystal.  
	
	The time evolution of the ZF-$\mu$SR spectra at all the temperatures and for both the orientations of the crystal with respect to the muon beam can be described by a single stretched exponential model as follows:
	
	\begin{equation}
		A(t) = A_0 \exp[-(\lambda t)^{\beta}]+A_{\mathrm{bg}}   
	\end{equation}
	
	Here $A_0$ is the initial asymmetry, $A_{\text{bg}}$ is the background contribution from muons stopping on the Ag-sample holder, $\beta$ is the stretched exponent, and $\lambda$ is the muon spin relaxation rate originating from the electronic moments. The fits to the ZF-$\mu$SR spectra by Eq. (1) are shown by the solid lines in Figs.~\ref{spectra_SC}(a) and~\ref{spectra_SC}(b). We estimated $A_{\text{bg}}$ (= 0.117) from the fit to the data at the lowest temperature, and then it was kept fixed for the analysis of other data. The temperature dependence of $\lambda$ and $\beta$ with both the initial polarisation of muon along ($\lambda_\parallel$ and $\beta_\parallel$) and perpendicular ($\lambda_\perp$ and $\beta_\perp$) to the $c$-axis are shown in the main panel and in the inset of Fig.~\ref{spectra_SC}(c), respectively. The observed values of $\lambda_\parallel$ and $\lambda_\perp$ are almost similar and are very low as compared to the spin glass systems (of the order of 10--20~$\mu$s$^{-1}$) below freezing temperature~\cite{PhysRevB.31.546,PhysRevLett.73.3306,PhysRevLett.93.187201}.

	\begin{figure*}
		
		\includegraphics[width=\textwidth, keepaspectratio]{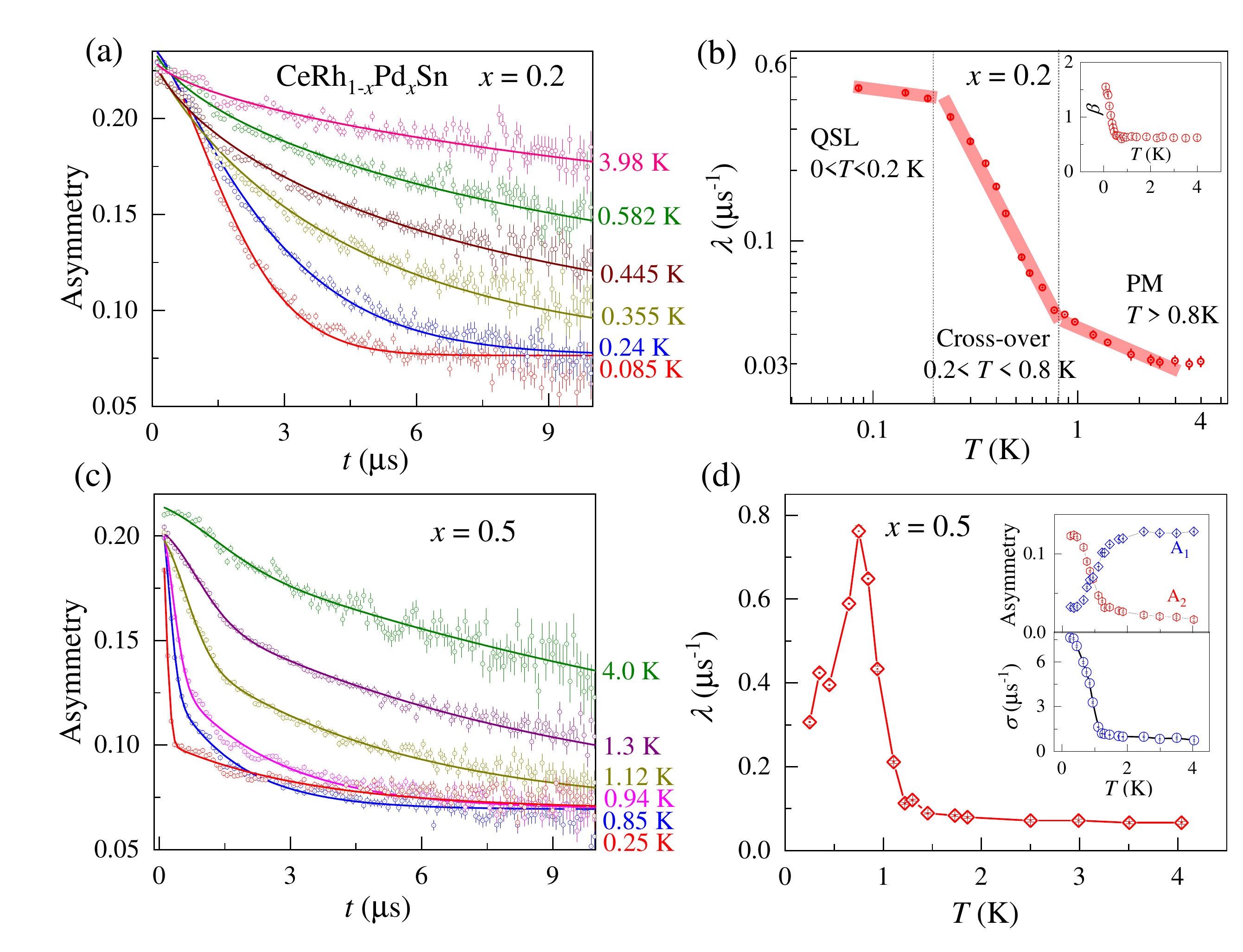}
		\caption{$\mu$SR data for the polycrystalline samples of CeRh$_{1-x}$Pd$_{x}$Sn with $x = 0.2$, 0.5: (a) ZF-$\mu$SR spectra at representative temperatures for $x = 0.2$. The solid lines are the fitted curves (see text for details). (b) Temperature dependence of the muon spin relaxation rate ($\lambda$). The grey solid (transparent) line described the power-law behavior of $\lambda$, i.e., $\lambda$($T$) $\propto$ $T^{- w}$, over a given temperature range. The inset shows the temperature dependent stretched exponent ($\beta$). (c) ZF-$\mu$SR spectra at representative temperatures for $x = 0.5$. The solid lines are the fitted curves (see text for details). (d) Temperature dependence of $\lambda$. The inset shows the temperature dependent initial asymmetries ($A_1$ and $A_2$) and Gaussian relaxation rate ($\sigma$). It is to be noted that the $\mu$SR spectra in Figs. (a and c) have slight distortion at around 2 $\mu$s, which is an artifact associated with the operation of a muon kicker at J-PARC.}
		\label{spectra_POW}
		
	\end{figure*}

	Moreover, for a 2D frustrated QSL, $\lambda$ can be described by a power-law, i.e., $\lambda$($T$) $\propto$ $T^{- w}$, over a given temperature range~\cite{Nature.471.612,Szirmai29555,PhysRevLett.102.176401}. We critically examined the behavior of $\lambda$ in the measured temperature range and found that both $\lambda_\parallel$ and $\lambda_\perp$ follow a power-law behavior in the three temperature regimes corresponding to the different values of the power exponents $w_\parallel$ and $w_\perp$, respectively. The power-law function has been fitted to the $\lambda_\parallel$ and $\lambda_\perp$ and are shown by the blue and red solid (transparent) lines, respectively. as shown in Fig.~\ref{spectra_SC}(c). The exponents $w_\parallel$ and $w_\perp$, obtained from the fit for three temperature regions are; $w_{\perp 1}=$ 0.17 (4), $w_{\parallel 1}=$ 0.10 (9) [for $T>$ 0.8~K (labeled I)], $w_{\perp 2}=$ 2.0 (1), $w_{\parallel 2}=$ 2.1 (8) [for 0.2~K $<T<$ 0.8~K (II)], and $w_{\perp 3} = $ 0.25 (6), $w_{\parallel 3} =$ 0.20 (3) [for $T<$ 0.2~K (III)]. The electronic fluctuation (1/relaxation rate) is hardly dependent on temperature in the range I, indicating that the system is close to its paramagnetic limit. Upon further cooling below 0.8~K, a first crossover occurs and in this intermediate region II, $\lambda$ increases steeply with the decrease of temperature. This increase in the relaxation rate renders evidence for a slowing down of the spin dynamics, likely resulting from the building up of short-range correlations. $\lambda$($T$) levels off with weak temperature dependence below 0.2~K and exhibits only slow changes with temperature without a critical slowing down of the electronic spin fluctuations. This constant low-temperature behavior of $\lambda$ is an universal characteristic of several SL candidates~\cite{PhysRevLett.117.097201,PhysRevB.100.241116,PhysRevLett.110.207208,Nature.471.612,PhysRevLett.109.037208,PhysRevB.84.100401,npj.6.2397.4648}. Qualitatively, this can be taken to indicate the presence of a SL ground state characterized by dynamic electronic magnetic moment fluctuations with a temperature-independent fluctuation time, which is unlike any system with a static magnetic ground state, for example, a spin glass, which is associated with a maximum in $\lambda$ near its spin freezing temperature~\cite{npj.4.2019}. 
	\\
	It is important to note that $\lambda$ presents a relatively stronger temperature dependence than organic QSL systems over the whole temperature range~\cite{Nat.Phys.15.1052,PhysRevLett.117.097201}. This reveals the presence of a substantial slowing down of spin dynamics in the metallic SL candidate systems rather than organic systems, where weak temperature dependence have been observed. The observed value of $\lambda_\text{ZF}$ is comparable to that observed for other QSL candidates, some of which are presented in the table S2 in SM.

	Furthermore, the stretch exponent $\beta$ increases abruptly from $\beta = 1$ to $\beta = 1.6$ below 0.5~K. The high-$T$ $\beta$ $\sim$ 1 (a simple exponential function) corroborates that the relaxation rate is in the fast fluctuation limit due to the dynamics of the unpaired electrons. However, the change in the value of $\beta$ around 0.5~K indicates a distribution of the muon spin relaxation rates. This increase is a very common phenomenon in lots of frustrated quantum magnets and may be caused by the magnetic spins strongly entangled in space and time, which is the basic requirements of a gapless quantum SL. Noteworthy is that such a change in the shape of the relaxation function from exponential to Gaussian has been observed in the kagome systems SrCr$_8$Ga$_4$O$_{19}$~\cite{PhysRevLett.73.3306} and Ca$_{10}$Cr$_7$O$_{28}$~\cite{Nat.Phys.12.942}, which feature a SL-like ground state.
	
	To identify the dynamics of the electronic magnetic moment fluctuations in our system, we have performed LF-dependent measurements of the relaxation. When small LF of 50--80~G is applied, the weak contribution from the nuclear magnetic moments observed in the ZF signal (static magnetism, estimate as $\lambda_{\text{ZF}}$/($\gamma_{\mu}/2\pi$), where $\gamma_{\mu} = 2\pi \times 135.53$~MHz/T is the muon gyromagnetic ratio) to the relaxation is typically eliminated. On the other hand, a large LF is needed to decouple the muon depolarization from the internal field arising from the fluctuating electronic spins. The representative LF-spectra with the initial polarisation of muon along and perpendicular to the $c$-axis at 0.1~K are displayed in Figs.~\ref{spectra_SC}(d) and~\ref{spectra_SC}(e), respectively. It is seen that even 1000~G LF is not sufficient to suppress the muon relaxation at 0.1~K completely. This means the magnetic ground state is entirely dynamic at the base temperature. Similar effects were also found at 0.25~K, i.e., in the crossover regime. However, field dependent spectra at $T =~$0.75~K and 4~K behave as expected for the high-temperature paramagnetic state (Fig.~S4~\cite{SM}). The LF spectra measured at 0.1~K under several magnetic fields can also be modeled by Eq. (1). The obtained $\lambda_{\text{LF}}$ as a function of the field for the muon beam along ($\lambda_{\text{LF}\parallel}$) and perpendicular ($\lambda_{\text{LF}\perp}$) to the $c$-axis are shown in the inset and main panel of Fig.~\ref{spectra_SC}(f), respectively. The corresponding $\beta_{\text{LF}\parallel}$ and $\beta_{\text{LF}\perp}$ are shown in Fig.~S5~\cite{SM}. The variation of $\lambda_{\text{LF}\parallel}$ and $\lambda_{\text{LF}\perp}$ with a longitudinal magnetic field can be represented by the power-law behavior as given below:
	
	\begin{equation}
		\lambda_{\mathrm{LF}}(H) = \lambda_{0}/(p + qH^\alpha)   
	\end{equation}
	
	where $p$ and $q$ depend on the fluctuation rate and fluctuating field. The $\lambda_{\text{LF}\parallel}$ can be described by the above equation with $\alpha$ = 2.0 (3) [solid line; inset of Fig.~\ref{spectra_SC}(f)]. It is worth noting that the Eq. (2) with $\alpha$ = 2, is a standard Redfield equation given below~\cite{PhysRevB.31.546, RevModPhys.69.1119}: 
	
	\begin{equation}
		\lambda_{\mathrm{LF}}(H) = \lambda_0+\frac{2\gamma^2_\mu<H^2_{\mathrm{loc}}>\tau}{1+\gamma^2_\mu H^2_{\mathrm{LF}}\tau^2}
	\end{equation}
	
	Here $\lambda_0$ is the $H$-independent depolarization rate, $\tau$ is the spin autocorrelation time of spin fluctuation, and $H_{\text{loc}}$ is the time average of the second moment of the time-varying local field $H_{\text{loc}}$($t$) at muon sites due to the fluctuations of neighboring Ce $4f$ moments. 
	\\
	From Eq. (3) it is clear that, in the absence of applied field, the relaxation rate is inversely proportional to the spin-fluctuation rate i.e., $\lambda \sim 1/\nu$, where $\nu = 1/\tau$. This further corroborates that an increase in $\lambda$ (with decreasing temperature) indicates a decrease in spin-fluctuation rate and hence a slowing down of dynamic spins. In ordinary disordered spin systems, $\lambda_{\text{LF}}$ exhibits a field-inverse square dependence. Such a spectral-weight function is commonly used to describe classical fluctuations in the paramagnetic regime. We found that Eq. (3) provides the best fit to the $\lambda_{\text{LF}\parallel}$ with $H_{\text{loc}}$ = 25 (7)~G and $\tau$ = 31 (4)~ns. This value of $\tau$ is comparable with 37 and 42~ns found in CeCoGe$_{1.8}$Si$_{1.2}$ and stoichiometric, \mbox{CeRhBi} respectively~\cite{PhysRevLett.88.046402,doi:10.7566/JPSJ.87.064708}, but is much larger compared to 0.10 -- 0.01~ns observed in metallic spin-glasses such as CuMn~\cite{PhysRevB.31.546}. Long correlation times (slow magnetic fluctuations) are generally expected in the critical region just above a magnetic transition.
	
	Moreover, our value of $\tau$ = 30~ns at 0.1~K gives $\nu$ = 33.3~MHz at 0.1~K. The $\nu$ = 470~MHz at 20~mK has been observed for $S$ = 1/2 V-based Kagome SL~\cite{Orain2014} and 9.4~MHz at 0.07~K for YbMgGaO$_4$~\cite{PhysRevLett.117.097201}, compared to high temperature paramagnetic value $\nu$ = 40~GHz. These low temperature values are much smaller than the paramagnetic limit $\nu$ = 523~GHz and 13.3~THz~\cite{Orain2014}. These results clearly indicate the slowing down of the spin fluctuations in the SL state. However, the ratio $\gamma_{\mu}H_{\text{loc}}/\nu \sim 0.1$ indicates that the Ce spin fluctuations still seem to be in fast fluctuating regime at the base temperature. Similar range of dynamic fluctuations are also seen for many systems near quantum critical regime and typical values are resented in Table-S2 of SM~\cite{SM}
	
	On the other hand, the $\lambda_{\text{LF}\perp}$ appears to be described by the Eq. (2) with $\alpha$ = 1.05 [solid line in Fig.~\ref{spectra_SC}(f)]. The dotted line in Fig.~\ref{spectra_SC}(f) with $\alpha$ = 2 (kept fixed) shows a significant deviation from the data. The observed values, $\alpha$ = 1.05, are inconsistent with the existence of a single timescale and instead suggest a more exotic spectral density, such as the one at play in a QSL~\cite{Nature.com.11.1.7}. We can, therefore, conclude from our LF-$\mu$SR study that the spin fluctuations perpendicular to the kagome plane are responsible for the SL behavior.
	
	We also studied the polycrystalline specimens of CeRh$_{1-x}$Pd$_x$Sn with $x$ = 0.2 and 0.5 to trace the development of AFM order. The ZF-$\mu$SR spectra for $x$ = 0.2 in Fig.~\ref{spectra_POW}(a), exhibits similar behavior to that of single crystalline sample of $x$ = 0.1. The fit by Eq. (1) gives $\lambda$($T$) and the stretched exponent $\beta$, which are shown in the main panel and inset of Fig.~\ref{spectra_POW}(b). Similar to $x$ = 0.1, $\lambda$($T$) can be described by a power-law, i.e., $\lambda$($T$) $\propto$ $T^{- w}$, over a given temperature range in the sample with $x = 0.2$. As shown in Fig.~\ref{spectra_POW}(b), three different regimes with identical crossover temperatures are seen in the $\lambda$($T$). The observed feature for $x = 0.2$ can be related to dynamic slowing-down of spin fluctuations at $T < 0.8$~K and a saturation of $\lambda$ at $T < 0.2$~K suggest a spin-fluctuating SL ground state down to 0.085~K. 
	
	In order to further confirm the non-magnetic ground state in $x=0.2$, we also carried out a ND study at 0.1~K and 2~K using the OSIRIS spectrometer at ISIS Neutron and Muon Source. We did not see any clear sign of the presence of magnetic Bragg peaks at 0.1~K (Fig.~S7~\cite{SM}), indicating that the ground state is either non-magnetic (as shown by ZF-$\mu$SR) or ordered state Ce moment is very small (below 0.1~$\mu_B$) to be detected by ND. Moreover, the $\lambda$ continuously increases down to the lowest temperature for a $T=0$ quantum phase transition, such as in CeRhBi~\cite{doi:10.7566/JPSJ.87.064708}, which is not the case for $x=0.1$ and 0.2 systems, indicating quantum critical SL ground state dominated by the long-range spin entanglement.
	
	The evolution of $A$($t$) for $x=$ 0.5 is markedly different from those for $x=$ 0.1 and 0.2 as shown in Fig.~\ref{spectra_POW}(c). At all the temperatures, $A$($t$) displays exponential plus Gaussian shape, i.e., $A_1 \exp[-(\sigma t)^2]$ + $A_2 \exp(-\lambda t) + A_{\text{bg}}$. The $\lambda$($T$), thus obtained, exhibits a clear peak at 0.8~K, as shown in Fig.~\ref{spectra_POW}(d), providing evidence for a static AFM ordering below $T_{\text{N}}$ = 0.8~K. The $T_{\text{N}}$ is close to the temperature of the peak in the specific heat in this system ~\cite{PhysRevB.96.045139}.
	
	\begin{figure*}
		
		\includegraphics[width=\textwidth]{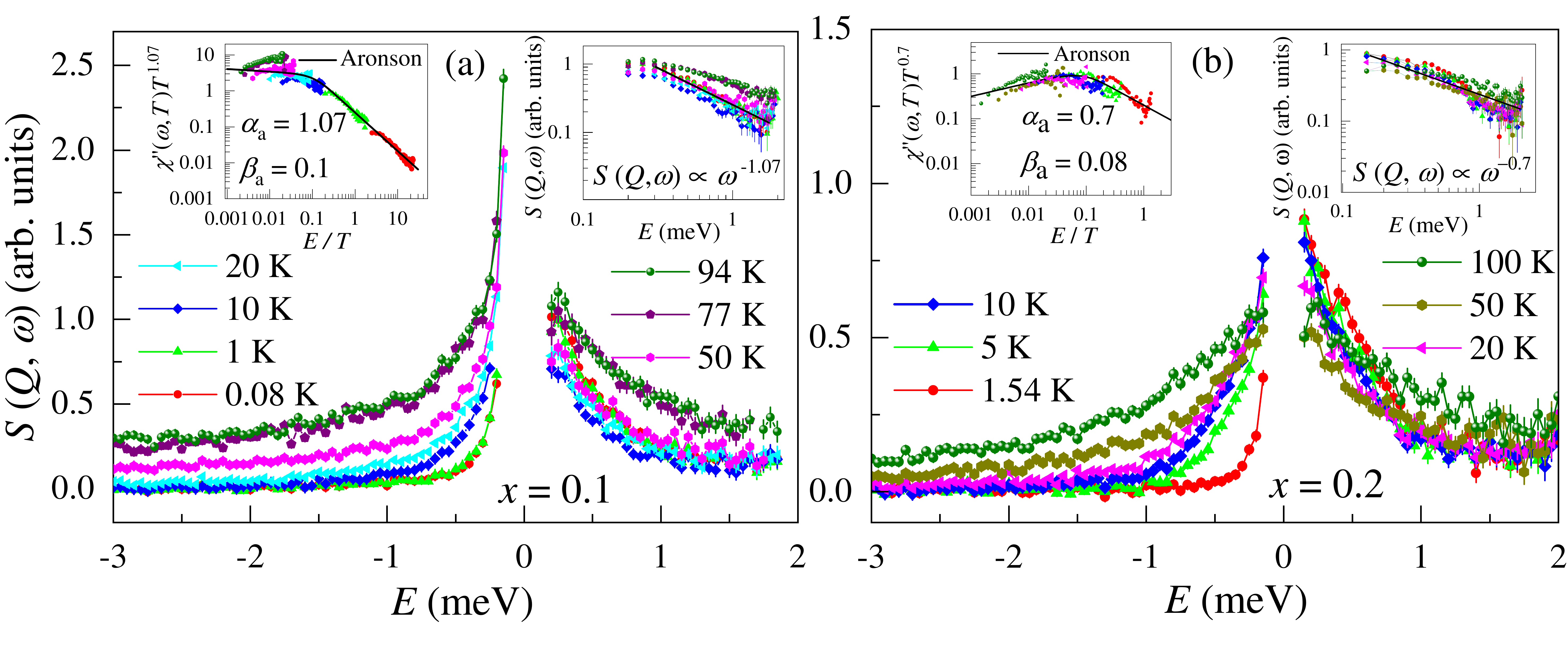}
		\caption{Low energy INS of CeRh$_{1-x}$Pd$_{x}$Sn for (a) $x$ = 0.1 and (b) $x$ = 0.2: $Q$-integrated (0 $< Q <2\AA^{-1}$) scattering intensities $S$($Q$, $\omega$) versus $E$ at different temperatures. Right insets(a,b): A log-log plot of $S$($Q$, $\omega$) data of the energy-loss part. The solid line shows $S$($Q$, $\omega$) $\propto$ $\omega^{-\alpha}$ power-law behavior. Left insets(a,b): Dynamic susceptibility $\chi^{\prime\prime}$($\omega$, $T$) plotted as $\chi^{\prime\prime}$($\omega$, $T$)$T^{\alpha}$ versus $E/T$. The solid curve is the simulation according to the Aronson scaling function~\cite{PhysRevLett.75.725}, in Eq. (5).}
		\label{LOWINS}
		
	\end{figure*}
	
	\begin{figure}
		\includegraphics[width=8.5cm, keepaspectratio]{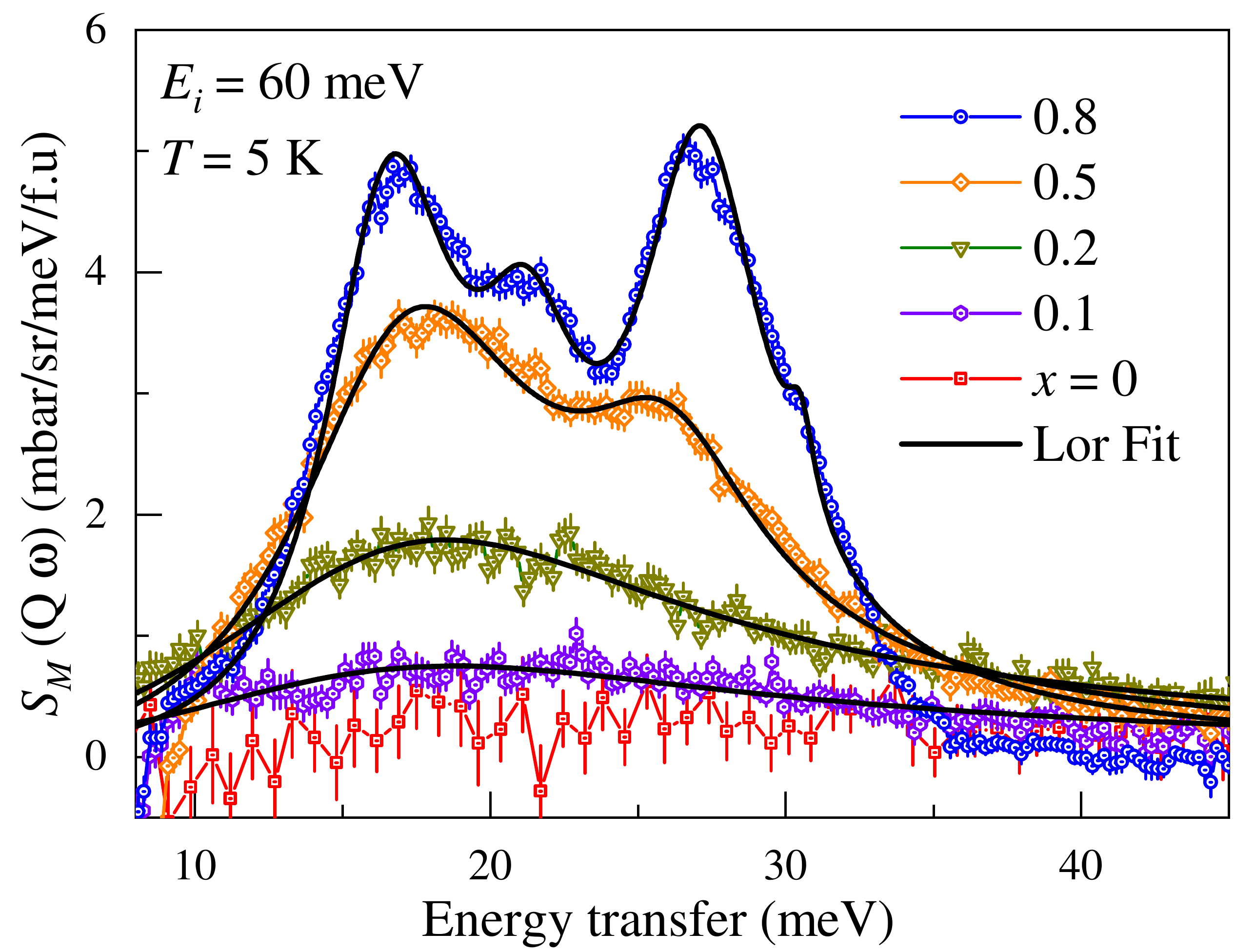}
		\caption{Magnetic INS response at 5~K after subtracting the phonon contribution for CeRh$_{1-x}$Pd$_x$Sn. The solid black lines show the fit to the Lorentzian functions multiplied by the Bose factor.}
		\label{HEINS}
	\end{figure}
	
	\section{Inelastic Neutron Scattering}
	\subsection{Low energy INS study}
	In order to examine the critical scaling of the low energy spin dynamics in CeRh$_{1-x}$Pd$_{x}$Sn, the low energy INS measurements have been performed on polycrystalline samples with $x =0$, 0.1, 0.2, and 0.75, using the IN6 TOF chopper spectrometer at ILL, Grenoble, with an incident energy $E_i$ = 3.1~meV. We also measured $x =0$ and 0.5 samples on the OSIRIS spectrometer. For $x =0$, both IN6 and OSIRIS results did not reveal a clear sign of low energy magnetic scattering (Fig.~S9a~\cite{SM}). This is also in agreement with CeRhSn single crystal study on the IRIS spectrometer by T. Sato et al.~\cite{Sato} that reveals very weak excitation intensity at 0.13~meV at $Q$ $\sim$ (1, 0, 0) below 0.05~K. They mentioned that the observed inelastic intensity is too weak to conclude the existence of CEF, and further measurements are needed. 
	
	We observed quasi-elastic scattering in $x =$ 0.1, 0.2, 0.5, and 0.75 (Fig.~S9~\cite{SM}), whose intensity decreases with increasing $Q$ (see color coded $E$ vs $Q$ plot for $x =$ 0.1 at 0.08~K in the inset of Fig.~S7\cite{SM}). Hence, in our analysis, we used $Q$-integrated intensity. Figure~\ref{LOWINS}(a) shows the $Q$-integrated (0 $< Q <$ 2 $\AA^{-1}$) scattering intensity $S$($Q$, $\omega$, $T$) versus $E$ for $x =$ 0.1 for 0.08~K $\leq T \leq$ 94~K summed over all scattering angles between 10$^{\circ}$ and 115$^{\circ}$. The clear sign of quasi-elastic scattering can be seen at all temperatures. The scattering on the neutron energy loss side (i.e., + $E$ side) between 0.08 and 50~K does not vary substantially with temperature, while that on the neutron energy gain side (i.e., -- $E$ side) increases with increasing temperature, which is due to the thermal population factor. On the other hand, the data at 77~K and 94~K do not follow the same behavior as those between 0.08 -- 50~K. These temperature and energy dependencies indicate a typical NFL response near the QCP below 50~K and support the idea that temperature is the only energy scale in the system, as observed in the dynamic susceptibility of many NFL systems~\cite{PhysRevLett.75.725}.

	According to the fluctuation-dissipation theorem, the measured scattering intensity $S$($Q, \omega, T$) is related to the imaginary part of the dynamic susceptibility $\chi^{\prime \prime}(Q, \omega)$~\cite{doi:10.1063/1.4818323} as,
	
	\begin{equation}
		S(Q, \omega, T)=\frac{\hbar}{\pi g^{2} \mu_{\mathrm{B}}^{2}} \frac{\chi^{\prime \prime}(Q, \omega, T)}{1-\exp \left(-\hbar \omega / k_{\mathrm{B}} T\right)}
	\end{equation}
	
	with $\chi^{\prime \prime}(Q, \omega, T)=\omega F^{2}(Q) \chi^{\prime}(0,0, T) P(Q, \omega, T)$ according to the Kramers-Kronig relation, where $F(Q)$ is the magnetic form factor, $\chi^{\prime}(0,0, T)=\chi_{0}$($T$) is the static bulk susceptibility and $P(Q, \omega, T)$ is the normalized spectral function.
	
	A double logarithmic plot of the magnetic scattering $S(Q, \omega, T)$ as a function of $E$ for $x=0.1$ at all the measured temperatures is shown in the right inset of Fig.~\ref{LOWINS}(a). The $S(Q, \omega, T)$ between $0.2$ and $2~\mathrm{meV}$ at all temperatures are linear in $E$ on a log-log scale suggesting a power-law behavior. The magnetic scattering between 0.08~K and 50~K thus follows a power-law behavior, $S(Q, \omega, T) \propto \omega^{-\alpha}$ with $\alpha=1.07$ ($\omega$ is related to $E$ as $E=\hbar \omega$). A very similar behavior with $\alpha$ in the range 0.33 - 0.77, has been observed in other NFL systems such as UCu$_{5-x}$ Pd$_{x}$~\cite{PhysRevLett.75.725}, CeRh$_{0.8}$Pd$_{0.2}$Sb~\cite{Park_2002}, Ce$_{0.7}$Th$_{0.3}$RhSb~\cite{SO2002472} and CePd$_{0.15}$Rh$_{0.85}$~\cite{ADROJA2007858}
	
	Further, we find a clear evidence of $E/T$ scaling in the imaginary part of the dynamic susceptibility. The $\chi^{\prime \prime}(\omega, T)$ for $x=0.1$ obtained from Eq. (4) is shown in the left inset of Fig.~\ref{LOWINS}(a) as $\chi^{\prime \prime}(\omega, T) T^{\alpha}$ versus $E/T$ plot. The $\chi^{\prime \prime}(\omega, T)$ data between 0.2 to 2~$\mathrm{meV}$ at $0.08 \leq T \leq 50 \mathrm{~K}$ collapse onto a single curve. This confirms the universal $E/T$ scaling behavior of $\chi^{\prime \prime}(\omega, T)$. The $\chi^{\prime \prime}(\omega, T)$ data are well described by the scaling relation $\chi^{\prime \prime}(\omega, T) T^{\alpha} \sim f(\omega / T)$ with $\alpha=1.07$. The solid curve in the left inset of Fig.~\ref{LOWINS}(a) represents the scaling function proposed by Aronson et al.~\cite{PhysRevLett.75.725}:
	
	\begin{equation}
		\chi^{\prime \prime}(\omega, T) T^{\alpha_a}=(T / \omega)^{\alpha_a} \tanh (\omega / \beta_a T)
	\end{equation}
	
	with $\alpha_a=1.07$ and $\beta_a=0.1$. The $E/T$ scaling behavior has been observed in several NFL systems, however with different values of $\alpha$, and sometimes with different choice of scaling function $f(\omega / T)$~\cite{PhysRevLett.75.725,PhysRevLett.104.147201,Schroder2000}. For example, the values of $\alpha$ are 1/3 and 0.2 for the quantum spin glasses UCu$_{5-x}$Pd$_{x}$~\cite{PhysRevLett.75.725} and Sc$_{1-x}$ U$_{x}$Pd$_{3}$~\cite{PhysRevLett.94.056402}, respectively. In AFM quantum critical systems CeCu$_{6-x}$Au$_{x}$~\cite{PhysRevLett.80.5623}, CeRh$_{0.8}$Pd$_{0.2}$Sb~\cite{Park_2002}, Ce$_{0.7}$Th$_{0.3}$RhSb~\cite{SO2002472}, and Ce$_{2}$PdIn$_{8}$~\cite{PhysRevB.86.094525} the values of $\alpha$ are 0.75, 0.77, 0.33, and 1.5, respectively. In the ferromagnetic quantum critical system CeRh$_{0.85}$Pd$_{0.15}$~\cite{ADROJA2007858}, $\alpha$ is 0.6. The reason for the different values of the exponent in different compounds is not well understood. The wide variation in the $\alpha$ value might indicate criticality at the distance from the QCP in the phase space and the dimensionality including the amount of chemical disorder. 
	
	Furthermore, hard x-ray photoelectron spectra (HXPES) reveal the presence of valence fluctuations in $x=0.1$~\cite{PhysRevB.104.235150}. Hence the presence of valence fluctuations at the QCP indicates the possible role of the valence fluctuations in addition to the magnetic fluctuations, requiring the novel quantum critical scheme beyond the Doniach picture~\cite{PhysRevLett.105.186403,PhysRevLett.109.086403}. It has been shown theoretically that charge responses at the Kondo destruction QCP are singular and obey $\omega$/$T$ scaling, very similar to spin responses~\cite{PhysRevLett.124.027205}. In order to see the effect of valence fluctuation on the $\omega$/$T$ scaling, we also carried out $\omega$/$T$ scaling of $x = 0.2$, in which the valence fluctuations are reduced and magnetic fluctuations have increased. The $\omega$/$T$ scaling plot is shown in the left inset of Fig.~\ref{LOWINS}(b), clearly reveals the scaling with $\alpha$ = 0.7, which is in agreement with other systems mentioned above, and is consistent with the theoretical value 0.72 calculated from a local QCP in an anisotropic Kondo lattice~\cite{PhysRevLett.91.156404,PhysRevLett.91.026401}.
	
	\begin{figure}
		
		\includegraphics[width=8.5cm, keepaspectratio]{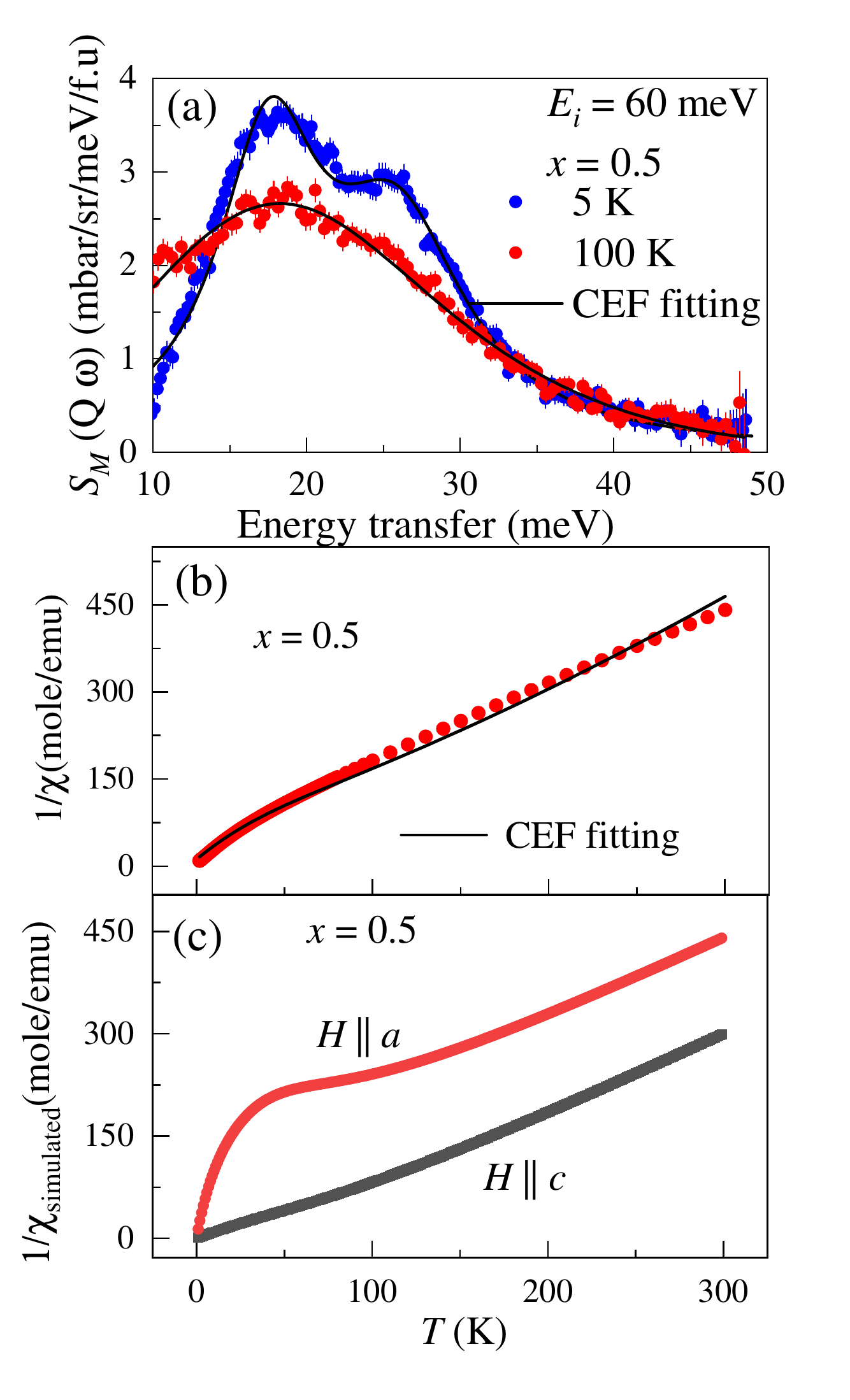}
		\caption{(a) Magnetic INS response for CeRh$_{1-x}$Pd$_x$Sn, $x$ = 0.5 at $T$ = 0.1 and 100~K. The solid black lines show the fit based on the CEF model, see text. (b) Polycrystalline inverse susceptibility for $x$ = 0.5. The solid black line shows the fit based on the CEF model, see text. (c) Simulated inverse susceptibility with the field along $a$ and $c$-direction using the $B_{n}^{m}$ parameters obtained from the INS and polycrystalline susceptibility fit.}
		\label{CEF0p5}
		
	\end{figure}
	
	\subsection{High energy INS study}
	
	The nature of $4f$-electrons and the cross-over from a valence fluctuating regime with itinerant $4f$-electrons to localized $4f$-electrons with increasing Pd doping in CeRh$_{1-x}$Pd$_{x}$Sn has been studied by high energy INS measurements. In order to estimate the phonon scattering, INS measurements were performed on the nonmagnetic LaRh$_{1-x}$Pd$_{x}$Sn sample with $x=0$, 0.1 and 0.5. The data of $x=0$ from reference~\cite{AdrojaMARI} are presented for comparison. The $Q$-integrated energy vs intensity 1D cuts from low-$Q$ = 0 to 3~$\AA^{-1}$ and high-$Q$ = 7 to 11~$\AA^{-1}$ regions from the raw data for an incident neutron energy of $E_i$ = 60~meV are given in Fig.~S10~\cite{SM}. The magnetic scattering of the 60~meV data for $x=$ 0.1 to 0.75 was estimated using a direct subtraction of the scattering of the non-magnetic reference LaRh$_{1-x}$Pd$_{x}$Sn of $x=$ 0.1 and 0.5 compounds from the Ce-data as $S$($Q$,$\omega$)$_{\text{M}}$ = $S$($Q$,$\omega$)$_{\text{Ce}}$ -- $\alpha_k$ $\times$ $S$($Q$,$\omega$)$_{\text{La}}$ (see, Fig.~S11~\cite{SM}). Here $\alpha_k$ is the scaling factor obtained from the ratio of the total scattering cross-section of the Ce by La compounds. In order to estimate the magnetic scattering for $x=0.2$ Ce sample, we used $x=0.1$ La data and for $x=0.75$ Ce sample, we used $x=0.5$ La data. In this procedure, we found that the phonon modes were still present in the magnetic scattering at high-$Q$. Hence we used $\alpha_k$ = 1 (i.e., direct subtraction of the La data from the Ce data) to estimate the magnetic scattering, which resulted in a better estimation of the magnetic scattering. As the phonon signal at high-$Q$ in LaRhSn is higher than in CeRhSn, we estimated $\alpha_k$ = 0.8 by scaling the high-$Q$ phonon spectra of LaRhSn to agree with that of CeRhSn. The estimated magnetic scattering is presented in Fig.~\ref{HEINS}. We also estimated the magnetic scattering using another method, i.e., $S$($Q$,$\omega$)$_{\text{M}}$ = $S$($Q$,$\omega$,low-$Q$)$_{\text{Ce}}$ -- $S$($Q$,$\omega$, high-$Q$)$_{\text{Ce}}$/[$S$($Q$,$\omega$, high-$Q$)$_{\text{La}}$/S($Q$,$\omega$, low-$Q$)$_{\text{La}}$]. We found very similar magnetic scattering (data not shown) as in Fig.~\ref{HEINS}. 
	
	For $x = 0$, a very broad and weak scattering is present, but no clear sign of CEF excitations. Furthermore, the high energy measurements with $E_i$ = 500~meV revealed the weak magnetic scattering in CeRhSn extended up to 300~meV~\cite{AdrojaMARI}. This is a typical response observed in valence fluctuating systems~\cite{PhysRevB.68.094425} and confirms the presence of mixed-valence nature of Ce ion in $x=0$, which is in agreement with HAXPES study~\cite{PhysRevB.104.235150}. The difference between this systems and other mixed-valence systems, e.g., CeRu$_4$Sb$_{12}$~\cite{PhysRevB.68.094425}, is that the magnetic scattering is very weak in $x=0$. Another important difference is that the $\chi$($T$) of CeRhSn exhibits strong anisotropy, $\chi$($H \parallel c$) $>$ $\chi$($H \parallel a$), with divergent behavior at low temperatures~\cite{KIM2003524}. Furthermore, the hard-axis magnetization $M$($H \parallel a$) exhibits a metamagnetic behavior at around $H \parallel a$ = 3 T. In conventional mixed-valence systems $\chi$($T$) exhibits a broad maximum at a certain temperature (or at a characteristic temperature, which is related to $T_{\text{K}}$) and decreases on cooling. On the other hand, for $x = 0.1$ and 0.2 the magnetic response is still broad centered near 20~meV, but its intensity increases and line-width decreases with $x$. 
	
	For $x=0.5$, two well-defined magnetic excitations appear near 18 and 25~meV, which are attributed to the excitations from the ground state multiplet $J=5/2$ of Ce$^{3+}$ splitting into three CEF doublets. For $x=0.75$, in addition to two strong excitations near 17 and 27~meV, there are two weak CEF excitations present near 21 and 30~meV. For the orthorhombic point symmetry ($m 2 m$, C$_{\text{2v}}$) of Ce ions in CeRh$_{1-x}$Pd$_x$Sn with the hexagonal crystal structure, we expect two CEF excitations in the paramagnetic state. Therefore, the presence of four CEF excitations in $x=0.75$ suggests the coexisting of two hexagonal phases with very similar lattice parameters due to the inhomogeneous distribution of Rh/Pd concentration. The support of two phases comes from the magnetic ND analysis presented below. We have fitted the INS spectra of  CeRh$_{1-x}$Pd$_x$Sn using a Lorentzian line shape, and the fits are shown by the solid black lines in Fig.~\ref{HEINS}. The fit parameters are given in Table-I of the SM~\cite{SM}.
	
	Now we present the INS data analysis of $x=0.5$ based on the CEF model. The CEF Hamiltonian for the orthorhombic point symmetry ($C_{\text{2v}}$) of the Ce$^{3+}$ ions is given by
	\begin{equation}\label{H-CEF}
		H_{\rm CEF} = B_{2}^{0}O_{2}^{0} + B_{2}^{2}O_{2}^{2}+B_{4}^{0}O_{4}^{0} +B_{4}^{2}O_{4}^{2}  + B_{4}^{4}O_{4}^{4}
	\end{equation}
	where $B_{n}^{m}$ are CEF parameters and $O_{n}^{m}$ are the Stevens operators~\cite{Stevens_1952}. The parameters $B_{n}^{m}$ need to be estimated by fitting the experimental data, such as single crystal susceptibility and/or INS data. For the analysis of INS data, we use a Lorentzian line shape for both quasi-elastic (QE) and inelastic excitations.
	
	In order to obtain a set of CEF parameters that consistently fit the INS data and polycrystal susceptibility (as the single crystal susceptibility data are not available for $x=$0.5), we performed a simultaneous fit of INS (at 5 and 100~K) and the polycrystal susceptibility data using the program in the Mantid software~\cite{MantidCEF}. Fits to the INS data at 5 and 100~K, and $\chi$($T$) from 5--300~K are shown by the solid black curve in Figs.~\ref{CEF0p5}(a) and~\ref{CEF0p5}(b), respectively. The obtained CEF parameters are (in meV) $B_2^0$ = -- 1.01, $B_2^2$ = 0.244, $B_4^0$ = -- 0.017, $B_4^2$ = -- 0.222 and $B_4^4$ = 0.159. The analysis gives the first excited doublet ($\Delta_1$) at 17.5~meV and the second one ($\Delta_2$) at 25.9~meV. The ground state wave functions are $\Psi_{0}$ = 0.90$|\mp \frac{5}{2}\rangle$ -- 0.292$|\pm \frac{3}{2}\rangle$ + 0.324$|\mp \frac{1}{2}\rangle$. The values of magnetic moments for the ordered state are $\mu_{x}$ = 0.60~$\mu_{\text{B}}$ and $\mu_{z}$ = 1.67~$\mu_{\text{B}}$ indicating the easy axis of the magnetization along the $c$-axis.
	We next used the Mantid software to simulate the $\chi^{x,y,z}$($T$) using the $B_{n}^{m}$ parameters. We have then calculated $\chi_{\text{CEF}}^a$ from $\chi^x$ and $\chi^y$, and $\chi_{\text{CEF}}^c$ from $\chi^z$ using the equations given below~\cite{Isikawa}. 
	
	\begin{equation}
		\frac{1}{\chi^a_{\mathrm{CFE}}} =  \frac{2}{\chi^x+\chi^y}+ \lambda^a \quad {\mathrm{and}} \quad \frac{1}{\chi^c_{\mathrm{CFE}}} = \frac{1}{\chi^z}+ \lambda^z,
	\end{equation}
	
	where $\chi^x$, $\chi^y$, and $\chi^z$ are the single ion crystal field susceptibility along $x$, $y$, and $z$-axes simulated using the CEF parameters obtained from the INS data analysis. The simulation was carried out without the molecular field parameters, $\lambda^a = \lambda^z$ = 0, using Mantid plot~\cite{MantidCEF}. The results of the simulation are presented in Fig~\ref{CEF0p5}(c). The simulation shows that the $c$-axis is an easy axis of the magnetization in $x = 0.5$.
	
	\begin{figure*}
		
		\includegraphics[width=16cm]{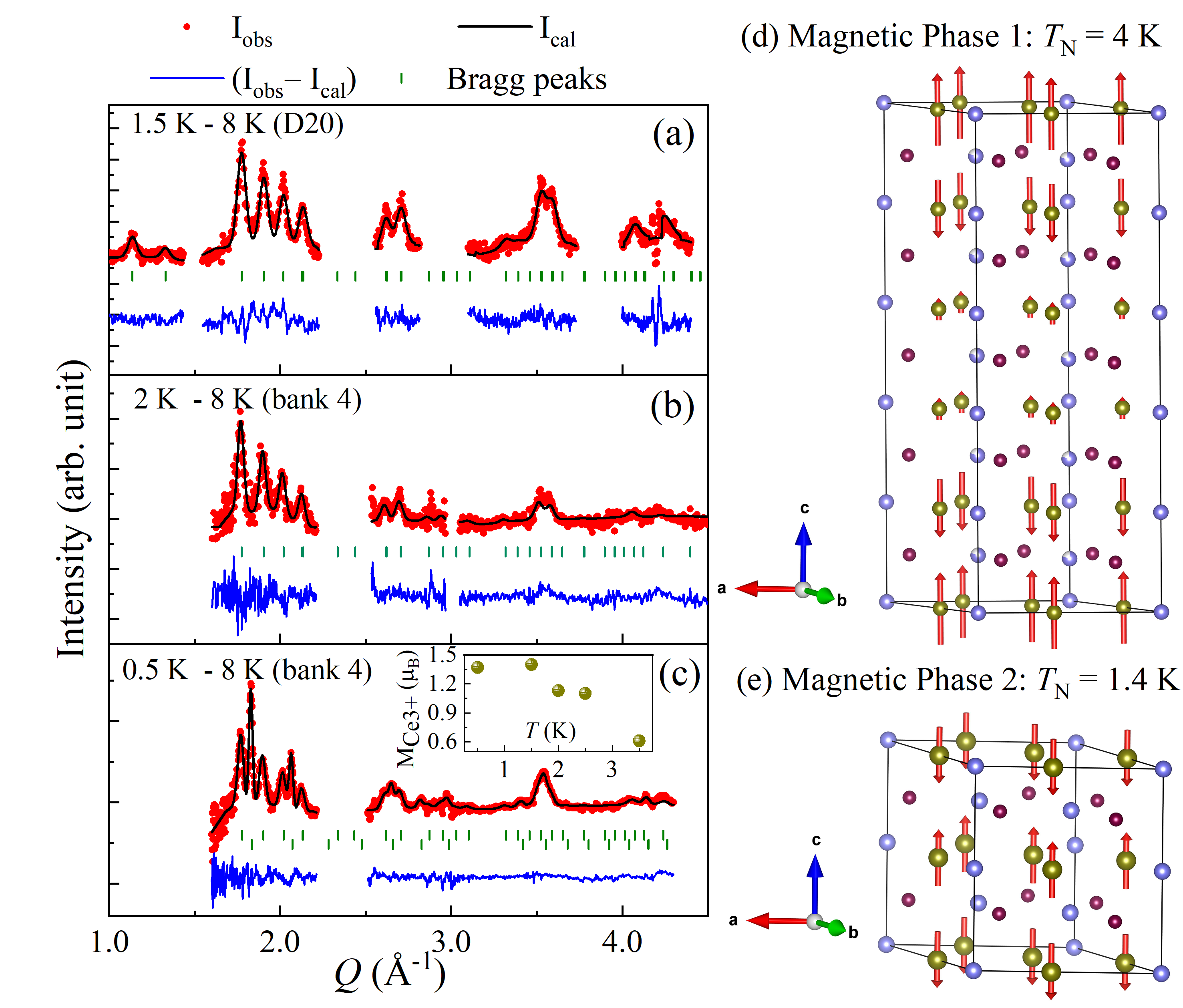}
		\caption{Temperature difference ND patterns along with the magnetic structural refinement profile of CeRh$_{1-x}$Pd$_x$Sn with $x=$ 0.75 (a) from D20 1.5~K -- 8~K, (b) from GEM 2~K -- 8~K and (c) from GEM 0.5~K -- 8~K. The solid black lines show the fit. The difference between the experimental and calculated intensities are shown by the blue curves at the bottom. The light green vertical ticks show the position of the magnetic Bragg peaks. For 0.5~K -- 8~K data in (c), two sets of vertical ticks are due to two magnetic phases. The inset shows the temperature dependence of the estimated ordered state magnetic moment of Ce ions for the incommensurate magnetic structure. (d) Incommensurate magnetic structure (longitudinal spin density wave) of CeRh$_{1-x}$Pd$_x$Sn with $x=$ 0.75 obtained from the refinement of ND pattern at 1.5~K -- 8~K (D20) and 2~K -- 8~K (GEM). The red arrows show the Ce moments. (e) The second magnetic phase having the commensurate magnetic structure below 1.4~K obtained from the refinement of 0.5~K -- 8~K data of GEM.}
		\label{ND}
		
	\end{figure*}

	\section{Neutron diffraction study}
	
	We have performed the powder ND study on the polycrystalline sample of CeRh$_{1-x}$Pd$_x$Sn with $x=$ 0.75 using the TOF neutron diffractometer GEM at ISIS and constant wavelength neutron diffractometer D20 at ILL. The room temperature ND results show that the sample crystallizes in the ZrNiAl-type hexagonal structure with space group $P\bar{6}2m$ as the parent system CeRhSn (Fig.~S8)~\cite{SM}. To investigate the magnetic structure, we collected ND data from 8~K to 0.5~K on GEM and down to 1.5~K on D20. At 1.5~K / 2~K, many new magnetic Bragg peaks have appeared compared to the 8~K data, which contain only nuclear peaks. In Figs.~\ref{ND}(a) and~\ref{ND}(b), we have plotted temperature differences [1.5~K -- 8~K (D20) and 2~K -- 8~K (GEM)] ND patterns, which show the presence of more than ten magnetic Bragg peaks. From the positions of seven strong magnetic peaks, the magnetic propagation vector was determined using the program $K$-Search, that is part of the Fullprof suite of programs~\cite{RODRIGUEZCARVAJAL199355}. All the magnetic peaks appearing below about $T_{\text{N1}}$ = 5~K (and above $T_{\text{N2}}<$ 1.3~K) can be indexed with an incommensurate propagation vector $k_1$ = [0, 0, 0.4]. In order to refine the magnetic structure, we performed magnetic symmetry analysis using the Basireps program~\cite{RodriguezFullProf,ritter2011} for Ce on the Wyckoff site $3f$ ($x$ 0 0) of space group $P\bar{6}2m$. There exist three allowed irreducible representations (IREPS) ($\Gamma$$_1$$^1$, $\Gamma$$_2$$^1$, $\Gamma$$_3$$^2$) having respectively one, two and six basis vectors (BV).
	
	Testing these three IREPS against the data, it was found that only $\Gamma$$_2$$^1$ which has two BVs can fit the magnetic Bragg peaks very well, see Figs.~\ref{ND}(a) and \ref{ND}(b). It should be noted here that the refinement of the difference data sets, which contain only the magnetic diffraction intensity, necessitates the use of a fixed scale factor which is previously determined from the refinement of the nuclear structure using the 8~K paramagnetic data. Only the BV, which describes the coupling between the magnetic components pointing along the $c$-direction, is needed for the refinement, while the second BV, which creates and couples components of the Ce moments in the $ab$-plane has zero intensity. The magnetic structure obtained from the refinement corresponds to a longitudinal spin density wave as $k$-vector and spin direction are parallel (Fig.~\ref{ND}(d)). The Ce moments are ferromagnetically coupled in the $ab$-plane. The value of the magnetic moment of Ce at 1.5~K amounts to about 1.22 (1) $\mu_{\text{B}}$/Ce-atom. The magnetic structure of $x=$ 0.75 is different from the one found in the related compound \mbox{CePdAl}~\cite{Donni1996GeometricallyFM} as will be discussed below.
	As the temperature is lowered below 1.5~K additional magnetic Bragg peaks, different from those indexed with $k_1$ = [0 0 0.4], appear in $x=$ 0.75. The intensity of the magnetic peaks with $k_1$ = [0 0 0.4] increases below 5~K and saturates at about 1.5~K, while the new magnetic Bragg peaks appear at 1.1~K and increases on cooling to 0.5~K. At 0.5~K, two sets of magnetic Bragg peaks coexist, which cannot be indexed using a single propagation vector. Using again the $K$-Search program the new magnetic peaks were indexed with the second magnetic propagation vector $k_2$ = [0, 0, 1/2]. The presence of two $k$-vectors at 0.5~K points strongly to a phase segregation scenario where one part of the volume of the sample adopts one magnetic structure while the other part follows a different one. Correspondingly, the temperature difference 0.5~K -- 8~K data were refined using two magnetic phases. Magnetic symmetry analysis for $k_2$ = [0, 0, $\frac{1}{2}$] returns five allowed IREPS ($\Gamma$$_1$$^1$, $\Gamma$$_2$$^1$, $\Gamma$$_3$$^2$, $\Gamma$$_4$$^2$, $\Gamma$$_5$$^2$) with respectively one, one, one, four, and two BVs. Only $\Gamma$$_2$$^1$ creates the correct intensity on the new magnetic peaks.
	The refinement was proceeded by partitioning the scale factor between the two magnetic phases using the assumption that the size of the Ce magnetic moment is the same in both. There is no absolute physical reason for having the same magnetic moment values in the two magnetic phase fractions, but it represents the only way to do this kind of refinement as scale factor (phase fraction) and magnetic moment values are totally correlated. The refinement was proceeded very carefully by varying manually the moment value and scale factor partitioning to avoid divergence of the fit. Only when close to the final solution, both the scale factor partitioning and the magnetic moment were given free to vary at the same time. The final refinement of the 0.5~K data shown in Fig.~\ref{ND}(c) has $R_{\text{mag}}$ = 3.4 for the majority phase and $R_{\text{mag}}$ = 3.2 for the minority phase. 
	The $k_1$ = [0, 0, 0.4] phase is the majority phase as it occupies about 76\% of the sample volume. The magnetic structure of the 24\% minority phase having $k_2$ = [0, 0, $\frac{1}{2}$] is shown in Fig.~\ref{ND}(e). It resembles strongly to the majority phase with magnetic moments pointing as well solely along the $c$-axis and again a ferromagnetic coupling within the $ab$-planes. Assuming the same magnetic moment in both volume fractions, the refinement of the 0.5~K data returns a value of $\mu_{\text{Ce}}$ = 1.37 (1)~$\mu_{\text{B}}$. This value is slightly smaller than the ordered  state moment ($\mu_{\text{z}}$ = 1.67~$\mu_{\text{B}}$) estimated from the ground state CEF wave function for $x = 0.5$, which could be due to the presence of the Kondo effect or a change in the CEF ground state wave function for $x = 0.75$. The inset of Fig.~\ref{ND}(c) shows the temperature dependence of the Ce moment of the majority phase obtained from the refinement of the ND at various temperatures assuming that the volume fraction of this phase stays constant through and above $T_{\text{N2}}$ at 76\%. About equal values at 0.5~K [1.37 (1)$\mu_{\text{B}}$] and 1.5~K [1.40 (1)$\mu_{\text{B}}$] confirm that the $k_1$ phase is already saturated at 1.5~K. Above 3.5~K [0.61 (2)$\mu_{\text{B}}$], it is no longer possible to do a consistent refinement as the magnetic peaks decrease strongly in intensity and increase in width. The two magnetic structures observed in $x = 0.75$ at 0.5~K are both not frustrated. It is furthermore perhaps indicate to argue that the phase segregation could be due to small local differences in the cation distribution Rh/Pd, which, however, is not leading to any macroscopically visible phase separation as the room temperature ND data do not show the presence of two phases (Fig.~S8) ~\cite{SM}.
	
	It is interesting to compare the magnetic structures of isostructural ZrNiA1 type hexagonal compounds \mbox{CePdAl}, PrPdA1, NdPdAl and NdRhSn, which provide prominent case subjects to the competition between magnetic frustration, Kondo effect, and the RKKY interactions. For example, \mbox{CePdAl} orders antiferromagnetically below $T_{\text{N}}$ = 2.7~K with an incommensurate propagation vector $k$ = [1, 0, 0.35] and a longitudinal sine-wave modulated spin arrangement oriented along the hexagonal $c$-direction~\cite{D_nni_1996}. Due to the geometrical frustration present in the $ab$-plane, two-third of the magnetic moments [Ce(1) and Ce(3)] order and the remaining one-third [Ce(2)] is screened by the Kondo effect or magnetic frustration, which leads to heavy-fermion behaviors~\cite{D_nni_1996,PhysRevB.89.054416}. The experimentally determined magnetic structure of \mbox{CePdAl} is in agreement with group theoretical symmetry analysis of the propagation vector, which confirms that for the Ce(2) site an ordered magnetic moment parallel to the magnetically easy $c$-axis is forbidden by symmetry. This is not the case for $x = 0.75$ system and hence all Ce moments are ordered. A powder neutron diffraction study on PrPdA1 revealed two magnetic phase transitions at $T_{\text{N1}} \sim$ 4.2~K and at $T_{\text{N2}} \sim $ 1.5~K~\cite{KELLER1997660, KELLER2000672} with an average incommensurate propagation vector $k$ = [1, 0, 0.398] at 1.5~K. Two third of the magnetic moments are close to the full magnetic moment of free Pr$^{3+}$ ion i.e., 3.1~$\mu_{\text{B}}$. The remaining one-third of the moments are significantly reduced to 1.9~$\mu_{\text{B}}$ due to quantum fluctuations arising from frustration and the Kondo effect, unlike \mbox{CePdAl} where Ce(2) are fully Kondo screened. The heat capacity of NdPdA1 reveals two magnetic phase transitions at $T_{\text{N1}} \sim$ 5.0~K and at $T_{\text{N2}} \sim $ 4.0~K and neutron diffraction study shows a pronounced temperature dependence magnetic propagation vector above $T_{\text{N2}}$ and locks in to $k$ = [$\frac{1}{4}$, 0, 0.444] below $T_{\text{N2}}$ ~\cite{KELLER1997660}. On the other hand, NdRhSn system first orders AFM at $T_{\text{N}}$ = 9.8 (1)~K and then orders FM at $T_{\text{C}}$ = 7.6~K with an antiferromagnetic phase with a propagation vector (0, 0, $\frac{1}{11}$) between $T_{\text{N}}$ and $T_{\text{C}}$~\cite{PhysRevB.83.104403}. In both magnetic phases the magnetic moments are aligned along the $c$-axis, which is the easy-magnetization axis of the system.

	\section{Comparison of SL properties between C$\text{e}$-based metallic and other insulating systems.}
	
	We have shown that the CeRh$_{1-x}$Pd$_x$Sn series offers a possible route to a metallic SL phase in an effective $S =1/2$ quasikagome system. The various experimental results~\cite{doi:10.1126/sciadv.1500001,doi:10.1143/JPSJ.73.3099} on the paramagnetic quasi-kagome Kondo lattice CeRhSn, where the Kondo ions are arranged on distorted kagome planes stacked along the $c$-axis, show that this system represents an example of a quantum criticality induced by geometrical frustration. The existence of local magnetic moments is suggested by spin flip metamagnetism at low temperatures, despite the large Kondo temperature of 200~K~\cite{doi:10.1126/sciadv.1500001}. This excludes an itinerant scenario and suggests that quantum criticality is related to local moments in a SL-like state. However, $\mu$SR results of CeRhSn are inconsistent with the  behavior observed for a typical SL phase~\cite{doi:10.1143/JPSJ.73.3099}. The ZF-$\mu$SR results do not show any sign of long-range ordering down to 50~mK. However, the relaxation rate remains almost constant between 2~K and 1~K and exhibits weak temperature dependence below 1~K~\cite{doi:10.1143/JPSJ.73.3099}. On the other hand, the TF-$\mu$SR relaxation rates exhibit a power-law type behaviour between 50~mK and 30~K for both $H_{ext}$ $\parallel$ and $H_{ext}$ $\perp c$.

	We find that substituting 10\% of Pd for Rh into the crystal structure of quasi-kagome CeRhSn leads to an evolution towards a SL regime possibly arising from the competing exchange interaction between the Ce spins on the kagome lattice. As a result, the electronic magnetic moments remain dynamically fluctuating and saturate below 0.12~K, which occurs in the same $T$ range as the change in the low-$T$ specific heat. This behavior is similar to a gapless spin-liquid state. The low energy inelastic neutron measurements on the powder sample of CeRh$_{0.9}$Pd$_{0.1}$Sn do reveal the presence of gapless magnetic excitations, as found in U(1) SL. Powder averaging avoided to observe a clear sign of the $Q$-dependence. The low energy INS measurement on the large single crystals are highly essential to further shed light on the nature of the SL ground state for $x = 0.1$. In addition, the $T$-linear contribution is often reported as evidence for the presence of lowenergy gapless spinon excitations in the SL candidate materials. For example, $C_\text{mag} \sim T^\alpha$ with $\alpha \sim 1$ was seen in the spin-1/2 kagome lattice antiferromagnet ZnCu$_3$(OH)$_6$Cl$_2$~\cite{PhysRevLett.98.107204}, Kitaev honeycomb iridate Cu$_2$IrO$_3$~\cite{PhysRevLett.122.167202}, triangle-based iridate Ba$_3$IrTi$_2$O$_9$~\cite{PhysRevB.96.014432}, as well as $\alpha \sim 1.25$ in the hexagonal TbInO$_3$ with quasitwo-dimensional triangular spin lattice~\cite{PhysRevX.9.031005}. However, little deviation from $T$-linear behavior ($\alpha =1.13$) in our case can be attributed to: (i) CeRh$_{0.9}$Pd$_{0.1}$Sn do not realize a perfect kagome lattice, (ii) chemical disorder, that may alters extremely low-energy excitations, or (iii) the uncertainty in evaluating the lattice contribution, as LaRhSn shows superconductor below 2~K.

	Furthermore, few other compounds of Ce$TX$ ($T$ = Rh, Ir, Pd, Pt ; and $X$ = Sn, Al) family have also been found to exhibit SL behavior when subjected to pressure, doping, and/or magnetic field, where the Kondo coupling and magnetic frustration result in this exotic quantum behavior. For example, the AFM corelation at very low temperature was observed in the intermediate valence compound CeIrSn with a high Kondo temperature of $T_\text{K} \sim 480$~K and proposed to be caused by the competition between Kondo singlet formation and geometrical frustration~\cite{PhysRevLett.126.217202}. CePdAl is another isostructural frustrated Kondo lattice with partial long-range order at $T_{\text{N}} = 2.7$~K. The AFM order in CePdAl has been tuned by both chemical or hydrostatic pressure. Very recently $\mu$SR investigation under pressure proposed a SL like state at $p_\text{c} \leq p \leq 1.75p_\text{c}$ ($p_\text{c} = 0.92$ GPa) i.e., close to quantum critical regime, where AFM order disappears~\cite{PhysRevB.105.L180402}. The SL behavior was interpreted as a divergent relaxation rate and quantum critical time over field scaling. On the other hand, the AFM order can also be suppressed by chemical substitution in the alloying series CePd$_{1-x}$Ni$_x$Al~\cite{PhysRevB.94.220405,zhao2019quantum}, yielding a QCP at $x_\text{c} \sim 0.14$. The frustration effect induces substantial short-range magnetic correlations even above $x_\text{c}$, and a frustration-dominated SL state may intrinsically emerge above $x=0.15$~\cite{PhysRevB.94.220405,zhao2019quantum}, which requires further investigation.
	
	The spin-glass state can also mimic a QSL in many aspects. For instance, it lacks the hallmark of a long-range magnetic order in the magnetic susceptibility, specific heat, and neutron diffraction and maintains short-range spin-spin correlations~\cite{RevModPhys.58.801,Mydosh,Mydosh1,Wu_2011,Ma_2018}. A spin-glass phase can also produce continuous INS spectra as seen in YbZnGaO$_4$~\cite{PhysRevLett.120.087201}, which is arguably the strongest evidence for a QSL so far, but from ac-susceptibility results (frequency-dependent peak position) the ground state was conformed to the spin-glass ground state. Moreover, spin-1/2 triangular lattice magnet YbMgGaO$_4$ is a promising candidate for a gapless QSL~\cite{shen2016evidence,paddison2017continuous}. However, magnetic excitation is absent in the thermal conductivity measurement, which contradicts the gapless QSL picture~\cite{PhysRevLett.117.267202}. The spin-glass phase is also suggested for YbMgGaO$_4$~\cite{RevModPhys.58.801,Mydosh,Mydosh1}, which is supported from the ac-susceptibility~\cite{PhysRevLett.120.087201}, while inelastic neutron scattering reveals continuous spin excitations~\cite{shen2016evidence,paddison2017continuous}. Also, the disorder was suggested to play an important role for the most heavily studied kagome compound ZnCu$_3$(OH)$_6$Cl$_2$~\cite{han2012fractionalized,RevModPhys.88.041002}. For CeRh$_{0.9}$Pd$_{0.1}$Sn, in fact, the spin-glass phase with frozen short-range correlations is completely ruled out by the a.c. susceptibility and $\mu$SR measurements.
	
	\section{Conclusions}
	
	We have investigated quasikagome system CeRh$_{1-x}$Pd$_{x}$Sn using heat capacity, ac-susceptibility, $\mu$SR, and neutron scattering measurements. For $x = 0.1$ single crystal sample, the -- log $T$ dependence of the $C_{\text{4f}}$/$T$ below 0.9~K is followed by a broad anomaly at 0.13~K, representing the onset of short-range correlations. Upon further cooling to 0.05~K, $C_{\text{4f}}$ shows a nearly linear $T$-dependence. The ac-susceptibility also exhibits a frequency independent broad peak at 0.16~K which is prominent with $H$ along $c$-direction. The ZF-$\mu$SR relaxation rate suggests the presence of dynamic spin fluctuations persisting even at $T$ = 0.05~K without static magnetic order in the single crystal sample with $x = 0.1$. We, therefore, argue that such behavior for the ground state of $x=0.1$ (namely, a constant $\lambda$ and a linear $T$-dependence in $C_{\text{4f}}$) can be attributed to a metallic spin-liquid near the QCP in the frustrated Kondo lattice. The LF-$\mu$SR results of the single crystal samples of $x = 0.1$ with the muon beam along and perpendicular to the $c$-axis suggest that the out of kagome plane spin fluctuations are responsible for the SL behavior. Our low energy INS study in $x$ = 0.1 and 0.2 indicates gapless excitations, a typical behavior observed for the NFL systems near QCP where the neutron dynamical susceptibility exhibits $E/T$ scaling. Our high energy INS shows a very weak and broad scattering in $x$ = 0 and 0.1, which transforms into well-localized CEF excitations with increasing $x$. The ZF-$\mu$SR spectra for $x$ = 0.2, exhibits similar behavior to that of $x$ = 0.1 and a saturation of $\lambda$ at $T < 0.2$~K was attributed to a spin-fluctuating SL ground state down to 0.085~K. The ZF-$\mu$SR results for the $x = 0.5$ sample are interpreted as a long-range AFM ground state below $T_{\text{N}}$ = 0.8~K, in which the AFM interaction probably overcomes the frustration effect. This has also been supported by the ND observation of incommensurate longitudinal spin density wave ground state with Ce moment along the $c$-axis below 4~K in $x$ = 0.75 sample. The present work will generate considerable interest in kagome-based HF materials and pave the way to understand the QSL behavior near the QCP, where both the spin and charge excitations play an important role.
	
	Data for the ISIS Neutron and muon measurements are available here~\cite{MERLIN1,MARI1,MUSR1} and for the ILL are available here ~\cite{IN6}. 
	
	\begin{acknowledgments}
		We gratefully acknowledge the ISIS Neutron and Muon Source for the beam time on MERLIN (RB1620476), MARI (RB14460), MUSR (RB1710457), ILL for the beam time on IN6 Experiment No.4-01-1552, and J-PARC for muon beam time on D1, experiment no 2018A0149. DTA would like to thank the Royal Society of London for International Exchange funding between the UK and Japan, Newton Advanced Fellowship funding between UK and China, and EPSRC UK for the funding (Grant No. EP/W00562X/1). RT thanks the Indian Nanomission for a post-doctoral fellowship. TT acknowledges the financial support from JSPS, grant numbers JP26400363, JP15K05180, JP16H01076, and JP21K03473. The work at Rice University was supported by the AFOSR Grant No. FA9550-21-1-0356 and the Robert A. Welch Foundation grant No. C-1411, with travel support provided by the NSF Grant No. DMR-1920740. We would like to thank Dr Matthias  Gutmann for the interesting discussion on GEM data analysis, Prof. Brian Rainford on CEF analysis, and Dr Andrea Severing on HAXPES. We also thank Profs. Keith McEwen and Je-Geun Park for participating in MARI experiment on CeRhSn and Prof. T. Sato for interesting discussion on IRIS results of single crystal of CeRhSn. AMS thanks the SA-NRF (93549) and the URC/FRC of UJ for generous financial assistance. For the purpose of open access, the authors have applied a CC BY public copyright licence to any Author Accepted Manuscript version arising.
	\end{acknowledgments}
	
	\bibliography{bibliography}

\begin{thebibliography}{121}%
\makeatletter
\providecommand \@ifxundefined [1]{%
 \@ifx{#1\undefined}
}%
\providecommand \@ifnum [1]{%
 \ifnum #1\expandafter \@firstoftwo
 \else \expandafter \@secondoftwo
 \fi
}%
\providecommand \@ifx [1]{%
 \ifx #1\expandafter \@firstoftwo
 \else \expandafter \@secondoftwo
 \fi
}%
\providecommand \natexlab [1]{#1}%
\providecommand \enquote  [1]{``#1''}%
\providecommand \bibnamefont  [1]{#1}%
\providecommand \bibfnamefont [1]{#1}%
\providecommand \citenamefont [1]{#1}%
\providecommand \href@noop [0]{\@secondoftwo}%
\providecommand \href [0]{\begingroup \@sanitize@url \@href}%
\providecommand \@href[1]{\@@startlink{#1}\@@href}%
\providecommand \@@href[1]{\endgroup#1\@@endlink}%
\providecommand \@sanitize@url [0]{\catcode `\\12\catcode `\$12\catcode
  `\&12\catcode `\#12\catcode `\^12\catcode `\_12\catcode `\%12\relax}%
\providecommand \@@startlink[1]{}%
\providecommand \@@endlink[0]{}%
\providecommand \url  [0]{\begingroup\@sanitize@url \@url }%
\providecommand \@url [1]{\endgroup\@href {#1}{\urlprefix }}%
\providecommand \urlprefix  [0]{URL }%
\providecommand \Eprint [0]{\href }%
\providecommand \doibase [0]{https://doi.org/}%
\providecommand \selectlanguage [0]{\@gobble}%
\providecommand \bibinfo  [0]{\@secondoftwo}%
\providecommand \bibfield  [0]{\@secondoftwo}%
\providecommand \translation [1]{[#1]}%
\providecommand \BibitemOpen [0]{}%
\providecommand \bibitemStop [0]{}%
\providecommand \bibitemNoStop [0]{.\EOS\space}%
\providecommand \EOS [0]{\spacefactor3000\relax}%
\providecommand \BibitemShut  [1]{\csname bibitem#1\endcsname}%
\let\auto@bib@innerbib\@empty
%</preamble>
\bibitem [{\citenamefont {Anderson}(1973)}]{ANDERSON1973153}%
  \BibitemOpen
  \bibfield  {author} {\bibinfo {author} {\bibfnamefont {P.}~\bibnamefont
  {Anderson}},\ }\href
  {https://doi.org/https://doi.org/10.1016/0025-5408(73)90167-0} {\bibfield
  {journal} {\bibinfo  {journal} {Materials Research Bulletin}\ }\textbf
  {\bibinfo {volume} {8}},\ \bibinfo {pages} {153} (\bibinfo {year}
  {1973})}\BibitemShut {NoStop}%
\bibitem [{\citenamefont {Balents}(2010)}]{balents2010spin}%
  \BibitemOpen
  \bibfield  {author} {\bibinfo {author} {\bibfnamefont {L.}~\bibnamefont
  {Balents}},\ }\href {https://doi.org/10.1038/nature08917} {\bibfield
  {journal} {\bibinfo  {journal} {Nature}\ }\textbf {\bibinfo {volume} {464}},\
  \bibinfo {pages} {199} (\bibinfo {year} {2010})}\BibitemShut {NoStop}%
\bibitem [{\citenamefont {Savary}\ and\ \citenamefont
  {Balents}(2016)}]{Savary_2016}%
  \BibitemOpen
  \bibfield  {author} {\bibinfo {author} {\bibfnamefont {L.}~\bibnamefont
  {Savary}}\ and\ \bibinfo {author} {\bibfnamefont {L.}~\bibnamefont
  {Balents}},\ }\href {https://doi.org/10.1088/0034-4885/80/1/016502}
  {\bibfield  {journal} {\bibinfo  {journal} {Reports on Progress in Physics}\
  }\textbf {\bibinfo {volume} {80}},\ \bibinfo {pages} {016502} (\bibinfo
  {year} {2016})}\BibitemShut {NoStop}%
\bibitem [{\citenamefont {Chamorro}\ \emph {et~al.}(2020)\citenamefont
  {Chamorro}, \citenamefont {McQueen},\ and\ \citenamefont
  {Tran}}]{chamorro2020chemistry}%
  \BibitemOpen
  \bibfield  {author} {\bibinfo {author} {\bibfnamefont {J.~R.}\ \bibnamefont
  {Chamorro}}, \bibinfo {author} {\bibfnamefont {T.~M.}\ \bibnamefont
  {McQueen}},\ and\ \bibinfo {author} {\bibfnamefont {T.~T.}\ \bibnamefont
  {Tran}},\ }\href {https://doi.org/10.1021/acs.chemrev.0c00641} {\bibfield
  {journal} {\bibinfo  {journal} {Chem. Rev.}\ }\textbf {\bibinfo {volume}
  {121}},\ \bibinfo {pages} {2898} (\bibinfo {year} {2020})}\BibitemShut
  {NoStop}%
\bibitem [{\citenamefont {Takagi}\ \emph {et~al.}(2019)\citenamefont {Takagi},
  \citenamefont {Takayama}, \citenamefont {Jackeli}, \citenamefont
  {Khaliullin},\ and\ \citenamefont {Nagler}}]{takagi2019concept}%
  \BibitemOpen
  \bibfield  {author} {\bibinfo {author} {\bibfnamefont {H.}~\bibnamefont
  {Takagi}}, \bibinfo {author} {\bibfnamefont {T.}~\bibnamefont {Takayama}},
  \bibinfo {author} {\bibfnamefont {G.}~\bibnamefont {Jackeli}}, \bibinfo
  {author} {\bibfnamefont {G.}~\bibnamefont {Khaliullin}},\ and\ \bibinfo
  {author} {\bibfnamefont {S.~E.}\ \bibnamefont {Nagler}},\ }\href
  {https://doi.org/10.1038/s42254-019-0038-2} {\bibfield  {journal} {\bibinfo
  {journal} {Nature Reviews Physics}\ }\textbf {\bibinfo {volume} {1}},\
  \bibinfo {pages} {264} (\bibinfo {year} {2019})}\BibitemShut {NoStop}%
\bibitem [{\citenamefont {Clark}\ and\ \citenamefont
  {Abdeldaim}(2021)}]{clark2021quantum}%
  \BibitemOpen
  \bibfield  {author} {\bibinfo {author} {\bibfnamefont {L.}~\bibnamefont
  {Clark}}\ and\ \bibinfo {author} {\bibfnamefont {A.~H.}\ \bibnamefont
  {Abdeldaim}},\ }\href {https://doi.org/10.1146/annurev-matsci-080819-011453}
  {\bibfield  {journal} {\bibinfo  {journal} {Annual Review of Materials
  Research}\ }\textbf {\bibinfo {volume} {51}},\ \bibinfo {pages} {495}
  (\bibinfo {year} {2021})}\BibitemShut {NoStop}%
\bibitem [{\citenamefont {Broholm}\ \emph {et~al.}(2020)\citenamefont
  {Broholm}, \citenamefont {Cava}, \citenamefont {Kivelson}, \citenamefont
  {Nocera}, \citenamefont {Norman},\ and\ \citenamefont
  {Senthil}}]{doi:10.1126/science.aay0668}%
  \BibitemOpen
  \bibfield  {author} {\bibinfo {author} {\bibfnamefont {C.}~\bibnamefont
  {Broholm}}, \bibinfo {author} {\bibfnamefont {R.~J.}\ \bibnamefont {Cava}},
  \bibinfo {author} {\bibfnamefont {S.~A.}\ \bibnamefont {Kivelson}}, \bibinfo
  {author} {\bibfnamefont {D.~G.}\ \bibnamefont {Nocera}}, \bibinfo {author}
  {\bibfnamefont {M.~R.}\ \bibnamefont {Norman}},\ and\ \bibinfo {author}
  {\bibfnamefont {T.}~\bibnamefont {Senthil}},\ }\href
  {https://doi.org/10.1126/science.aay0668} {\bibfield  {journal} {\bibinfo
  {journal} {Science}\ }\textbf {\bibinfo {volume} {367}},\ \bibinfo {pages}
  {eaay0668} (\bibinfo {year} {2020})}\BibitemShut {NoStop}%
\bibitem [{\citenamefont {Semeghini}\ \emph {et~al.}(2021)\citenamefont
  {Semeghini}, \citenamefont {Levine}, \citenamefont {Keesling}, \citenamefont
  {Ebadi}, \citenamefont {Wang}, \citenamefont {Bluvstein}, \citenamefont
  {Verresen}, \citenamefont {Pichler}, \citenamefont {Kalinowski},
  \citenamefont {Samajdar}, \citenamefont {Omran}, \citenamefont {Sachdev},
  \citenamefont {Vishwanath}, \citenamefont {Greiner}, \citenamefont
  {Vuletić},\ and\ \citenamefont {Lukin}}]{doi:10.1126/science.abi8794}%
  \BibitemOpen
  \bibfield  {author} {\bibinfo {author} {\bibfnamefont {G.}~\bibnamefont
  {Semeghini}}, \bibinfo {author} {\bibfnamefont {H.}~\bibnamefont {Levine}},
  \bibinfo {author} {\bibfnamefont {A.}~\bibnamefont {Keesling}}, \bibinfo
  {author} {\bibfnamefont {S.}~\bibnamefont {Ebadi}}, \bibinfo {author}
  {\bibfnamefont {T.~T.}\ \bibnamefont {Wang}}, \bibinfo {author}
  {\bibfnamefont {D.}~\bibnamefont {Bluvstein}}, \bibinfo {author}
  {\bibfnamefont {R.}~\bibnamefont {Verresen}}, \bibinfo {author}
  {\bibfnamefont {H.}~\bibnamefont {Pichler}}, \bibinfo {author} {\bibfnamefont
  {M.}~\bibnamefont {Kalinowski}}, \bibinfo {author} {\bibfnamefont
  {R.}~\bibnamefont {Samajdar}}, \bibinfo {author} {\bibfnamefont
  {A.}~\bibnamefont {Omran}}, \bibinfo {author} {\bibfnamefont
  {S.}~\bibnamefont {Sachdev}}, \bibinfo {author} {\bibfnamefont
  {A.}~\bibnamefont {Vishwanath}}, \bibinfo {author} {\bibfnamefont
  {M.}~\bibnamefont {Greiner}}, \bibinfo {author} {\bibfnamefont
  {V.}~\bibnamefont {Vuletić}},\ and\ \bibinfo {author} {\bibfnamefont
  {M.~D.}\ \bibnamefont {Lukin}},\ }\href
  {https://doi.org/10.1126/science.abi8794} {\bibfield  {journal} {\bibinfo
  {journal} {Science}\ }\textbf {\bibinfo {volume} {374}},\ \bibinfo {pages}
  {1242} (\bibinfo {year} {2021})}\BibitemShut {NoStop}%
\bibitem [{\citenamefont {Lacroix}(2010)}]{JPSJ.79.011008}%
  \BibitemOpen
  \bibfield  {author} {\bibinfo {author} {\bibfnamefont {C.}~\bibnamefont
  {Lacroix}},\ }\href {https://doi.org/10.1143/JPSJ.79.011008} {\bibfield
  {journal} {\bibinfo  {journal} {J. Phys. Soc. Jpn.}\ }\textbf {\bibinfo
  {volume} {79}},\ \bibinfo {pages} {011008} (\bibinfo {year}
  {2010})}\BibitemShut {NoStop}%
\bibitem [{\citenamefont {Nakatsuji}\ \emph {et~al.}(2006)\citenamefont
  {Nakatsuji}, \citenamefont {Machida}, \citenamefont {Maeno}, \citenamefont
  {Tayama}, \citenamefont {Sakakibara}, \citenamefont {Duijn}, \citenamefont
  {Balicas}, \citenamefont {Millican}, \citenamefont {Macaluso},\ and\
  \citenamefont {Chan}}]{PhysRevLett.96.087204}%
  \BibitemOpen
  \bibfield  {author} {\bibinfo {author} {\bibfnamefont {S.}~\bibnamefont
  {Nakatsuji}}, \bibinfo {author} {\bibfnamefont {Y.}~\bibnamefont {Machida}},
  \bibinfo {author} {\bibfnamefont {Y.}~\bibnamefont {Maeno}}, \bibinfo
  {author} {\bibfnamefont {T.}~\bibnamefont {Tayama}}, \bibinfo {author}
  {\bibfnamefont {T.}~\bibnamefont {Sakakibara}}, \bibinfo {author}
  {\bibfnamefont {J.~v.}\ \bibnamefont {Duijn}}, \bibinfo {author}
  {\bibfnamefont {L.}~\bibnamefont {Balicas}}, \bibinfo {author} {\bibfnamefont
  {J.~N.}\ \bibnamefont {Millican}}, \bibinfo {author} {\bibfnamefont {R.~T.}\
  \bibnamefont {Macaluso}},\ and\ \bibinfo {author} {\bibfnamefont {J.~Y.}\
  \bibnamefont {Chan}},\ }\href {https://doi.org/10.1103/PhysRevLett.96.087204}
  {\bibfield  {journal} {\bibinfo  {journal} {Phys. Rev. Lett.}\ }\textbf
  {\bibinfo {volume} {96}},\ \bibinfo {pages} {087204} (\bibinfo {year}
  {2006})}\BibitemShut {NoStop}%
\bibitem [{\citenamefont {Kim}\ and\ \citenamefont
  {Aronson}(2013)}]{PhysRevLett.110.017201}%
  \BibitemOpen
  \bibfield  {author} {\bibinfo {author} {\bibfnamefont {M.~S.}\ \bibnamefont
  {Kim}}\ and\ \bibinfo {author} {\bibfnamefont {M.~C.}\ \bibnamefont
  {Aronson}},\ }\href {https://doi.org/10.1103/PhysRevLett.110.017201}
  {\bibfield  {journal} {\bibinfo  {journal} {Phys. Rev. Lett.}\ }\textbf
  {\bibinfo {volume} {110}},\ \bibinfo {pages} {017201} (\bibinfo {year}
  {2013})}\BibitemShut {NoStop}%
\bibitem [{\citenamefont {Tokiwa}\ \emph {et~al.}(2014)\citenamefont {Tokiwa},
  \citenamefont {Ishikawa}, \citenamefont {Nakatsuji},\ and\ \citenamefont
  {Gegenwart}}]{doi.org/10.1038/nmat3900}%
  \BibitemOpen
  \bibfield  {author} {\bibinfo {author} {\bibfnamefont {Y.}~\bibnamefont
  {Tokiwa}}, \bibinfo {author} {\bibfnamefont {J.~J.}\ \bibnamefont
  {Ishikawa}}, \bibinfo {author} {\bibfnamefont {S.}~\bibnamefont
  {Nakatsuji}},\ and\ \bibinfo {author} {\bibfnamefont {P.}~\bibnamefont
  {Gegenwart}},\ }\href {https://doi.org/10.1038/nmat3900} {\bibfield
  {journal} {\bibinfo  {journal} {Nature Materials}\ }\textbf {\bibinfo
  {volume} {13}},\ \bibinfo {pages} {356} (\bibinfo {year} {2014})}\BibitemShut
  {NoStop}%
\bibitem [{\citenamefont {Tokiwa}\ \emph {et~al.}(2015)\citenamefont {Tokiwa},
  \citenamefont {Stingl}, \citenamefont {Kim}, \citenamefont {Takabatake},\
  and\ \citenamefont {Gegenwart}}]{doi:10.1126/sciadv.1500001}%
  \BibitemOpen
  \bibfield  {author} {\bibinfo {author} {\bibfnamefont {Y.}~\bibnamefont
  {Tokiwa}}, \bibinfo {author} {\bibfnamefont {C.}~\bibnamefont {Stingl}},
  \bibinfo {author} {\bibfnamefont {M.-S.}\ \bibnamefont {Kim}}, \bibinfo
  {author} {\bibfnamefont {T.}~\bibnamefont {Takabatake}},\ and\ \bibinfo
  {author} {\bibfnamefont {P.}~\bibnamefont {Gegenwart}},\ }\href
  {https://doi.org/10.1126/sciadv.1500001} {\bibfield  {journal} {\bibinfo
  {journal} {Science Advances}\ }\textbf {\bibinfo {volume} {1}},\ \bibinfo
  {pages} {e1500001} (\bibinfo {year} {2015})}\BibitemShut {NoStop}%
\bibitem [{\citenamefont {Doniach}(1977)}]{DONIACH1977231}%
  \BibitemOpen
  \bibfield  {author} {\bibinfo {author} {\bibfnamefont {S.}~\bibnamefont
  {Doniach}},\ }\href
  {https://doi.org/https://doi.org/10.1016/0378-4363(77)90190-5} {\bibfield
  {journal} {\bibinfo  {journal} {Physica B+C}\ }\textbf {\bibinfo {volume}
  {91}},\ \bibinfo {pages} {231} (\bibinfo {year} {1977})}\BibitemShut
  {NoStop}%
\bibitem [{\citenamefont {L\"ohneysen}\ \emph {et~al.}(2007)\citenamefont
  {L\"ohneysen}, \citenamefont {Rosch}, \citenamefont {Vojta},\ and\
  \citenamefont {W\"olfle}}]{RevModPhys.79.1015}%
  \BibitemOpen
  \bibfield  {author} {\bibinfo {author} {\bibfnamefont {H.~v.}\ \bibnamefont
  {L\"ohneysen}}, \bibinfo {author} {\bibfnamefont {A.}~\bibnamefont {Rosch}},
  \bibinfo {author} {\bibfnamefont {M.}~\bibnamefont {Vojta}},\ and\ \bibinfo
  {author} {\bibfnamefont {P.}~\bibnamefont {W\"olfle}},\ }\href
  {https://doi.org/10.1103/RevModPhys.79.1015} {\bibfield  {journal} {\bibinfo
  {journal} {Rev. Mod. Phys.}\ }\textbf {\bibinfo {volume} {79}},\ \bibinfo
  {pages} {1015} (\bibinfo {year} {2007})}\BibitemShut {NoStop}%
\bibitem [{\citenamefont {Kirchner}\ \emph {et~al.}(2020)\citenamefont
  {Kirchner}, \citenamefont {Paschen}, \citenamefont {Chen}, \citenamefont
  {Wirth}, \citenamefont {Feng}, \citenamefont {Thompson},\ and\ \citenamefont
  {Si}}]{RevModPhys.92.011002}%
  \BibitemOpen
  \bibfield  {author} {\bibinfo {author} {\bibfnamefont {S.}~\bibnamefont
  {Kirchner}}, \bibinfo {author} {\bibfnamefont {S.}~\bibnamefont {Paschen}},
  \bibinfo {author} {\bibfnamefont {Q.}~\bibnamefont {Chen}}, \bibinfo {author}
  {\bibfnamefont {S.}~\bibnamefont {Wirth}}, \bibinfo {author} {\bibfnamefont
  {D.}~\bibnamefont {Feng}}, \bibinfo {author} {\bibfnamefont {J.~D.}\
  \bibnamefont {Thompson}},\ and\ \bibinfo {author} {\bibfnamefont
  {Q.}~\bibnamefont {Si}},\ }\href
  {https://doi.org/10.1103/RevModPhys.92.011002} {\bibfield  {journal}
  {\bibinfo  {journal} {Rev. Mod. Phys.}\ }\textbf {\bibinfo {volume} {92}},\
  \bibinfo {pages} {011002} (\bibinfo {year} {2020})}\BibitemShut {NoStop}%
\bibitem [{\citenamefont {Si}(2006)}]{SI200623}%
  \BibitemOpen
  \bibfield  {author} {\bibinfo {author} {\bibfnamefont {Q.}~\bibnamefont
  {Si}},\ }\href {https://doi.org/https://doi.org/10.1016/j.physb.2006.01.156}
  {\bibfield  {journal} {\bibinfo  {journal} {Physica B: Condensed Matter}\
  }\textbf {\bibinfo {volume} {378-380}},\ \bibinfo {pages} {23} (\bibinfo
  {year} {2006})}\BibitemShut {NoStop}%
\bibitem [{\citenamefont {Si}(2010)}]{Si2010QuantumCA}%
  \BibitemOpen
  \bibfield  {author} {\bibinfo {author} {\bibfnamefont {Q.}~\bibnamefont
  {Si}},\ }\href {https://doi.org/https://doi.org/10.1002/pssb.200983082}
  {\bibfield  {journal} {\bibinfo  {journal} {Physica Status Solidi (b)}\
  }\textbf {\bibinfo {volume} {247}},\ \bibinfo {pages} {476} (\bibinfo {year}
  {2010})}\BibitemShut {NoStop}%
\bibitem [{\citenamefont {Motome}\ \emph {et~al.}(2010)\citenamefont {Motome},
  \citenamefont {Nakamikawa}, \citenamefont {Yamaji},\ and\ \citenamefont
  {Udagawa}}]{PhysRevLett.105.036403}%
  \BibitemOpen
  \bibfield  {author} {\bibinfo {author} {\bibfnamefont {Y.}~\bibnamefont
  {Motome}}, \bibinfo {author} {\bibfnamefont {K.}~\bibnamefont {Nakamikawa}},
  \bibinfo {author} {\bibfnamefont {Y.}~\bibnamefont {Yamaji}},\ and\ \bibinfo
  {author} {\bibfnamefont {M.}~\bibnamefont {Udagawa}},\ }\href
  {https://doi.org/10.1103/PhysRevLett.105.036403} {\bibfield  {journal}
  {\bibinfo  {journal} {Phys. Rev. Lett.}\ }\textbf {\bibinfo {volume} {105}},\
  \bibinfo {pages} {036403} (\bibinfo {year} {2010})}\BibitemShut {NoStop}%
\bibitem [{\citenamefont {Coleman}\ and\ \citenamefont
  {Nevidomskyy}(2010)}]{doi.org/10.1007}%
  \BibitemOpen
  \bibfield  {author} {\bibinfo {author} {\bibfnamefont {P.}~\bibnamefont
  {Coleman}}\ and\ \bibinfo {author} {\bibfnamefont {A.}~\bibnamefont
  {Nevidomskyy}},\ }\href {https://doi.org/10.1007/s10909-010-0213-4}
  {\bibfield  {journal} {\bibinfo  {journal} {J. Low. Temp. Phys.}\ }\textbf
  {\bibinfo {volume} {161}},\ \bibinfo {pages} {182–202} (\bibinfo {year}
  {2010})}\BibitemShut {NoStop}%
\bibitem [{\citenamefont {Bernhard}\ \emph {et~al.}(2011)\citenamefont
  {Bernhard}, \citenamefont {Coqblin},\ and\ \citenamefont
  {Lacroix}}]{PhysRevB.83.214427}%
  \BibitemOpen
  \bibfield  {author} {\bibinfo {author} {\bibfnamefont {B.~H.}\ \bibnamefont
  {Bernhard}}, \bibinfo {author} {\bibfnamefont {B.}~\bibnamefont {Coqblin}},\
  and\ \bibinfo {author} {\bibfnamefont {C.}~\bibnamefont {Lacroix}},\ }\href
  {https://doi.org/10.1103/PhysRevB.83.214427} {\bibfield  {journal} {\bibinfo
  {journal} {Phys. Rev. B}\ }\textbf {\bibinfo {volume} {83}},\ \bibinfo
  {pages} {214427} (\bibinfo {year} {2011})}\BibitemShut {NoStop}%
\bibitem [{\citenamefont {Pixley}\ \emph {et~al.}(2014)\citenamefont {Pixley},
  \citenamefont {Yu},\ and\ \citenamefont {Si}}]{PhysRevLett.113.176402}%
  \BibitemOpen
  \bibfield  {author} {\bibinfo {author} {\bibfnamefont {J.~H.}\ \bibnamefont
  {Pixley}}, \bibinfo {author} {\bibfnamefont {R.}~\bibnamefont {Yu}},\ and\
  \bibinfo {author} {\bibfnamefont {Q.}~\bibnamefont {Si}},\ }\bibfield
  {title} {\bibinfo {title} {Quantum phases of the shastry-sutherland kondo
  lattice: Implications for the global phase diagram of heavy-fermion metals},\
  }\href {https://doi.org/10.1103/PhysRevLett.113.176402} {\bibfield  {journal}
  {\bibinfo  {journal} {Phys. Rev. Lett.}\ }\textbf {\bibinfo {volume} {113}},\
  \bibinfo {pages} {176402} (\bibinfo {year} {2014})}\BibitemShut {NoStop}%
\bibitem [{\citenamefont {Si}\ and\ \citenamefont
  {Paschen}(2013)}]{doi.org/10.1002}%
  \BibitemOpen
  \bibfield  {author} {\bibinfo {author} {\bibfnamefont {Q.}~\bibnamefont
  {Si}}\ and\ \bibinfo {author} {\bibfnamefont {S.}~\bibnamefont {Paschen}},\
  }\href {https://doi.org/https://doi.org/10.1002/pssb.201300005} {\bibfield
  {journal} {\bibinfo  {journal} {Physica Status Solidi (b)}\ }\textbf
  {\bibinfo {volume} {250}},\ \bibinfo {pages} {425} (\bibinfo {year}
  {2013})}\BibitemShut {NoStop}%
\bibitem [{\citenamefont {Okabe}\ \emph {et~al.}(2019)\citenamefont {Okabe},
  \citenamefont {Hiraishi}, \citenamefont {Koda}, \citenamefont {Kojima},
  \citenamefont {Takeshita}, \citenamefont {Yamauchi}, \citenamefont
  {Matsushita}, \citenamefont {Kuramoto},\ and\ \citenamefont
  {Kadono}}]{PhysRevB.99.041113}%
  \BibitemOpen
  \bibfield  {author} {\bibinfo {author} {\bibfnamefont {H.}~\bibnamefont
  {Okabe}}, \bibinfo {author} {\bibfnamefont {M.}~\bibnamefont {Hiraishi}},
  \bibinfo {author} {\bibfnamefont {A.}~\bibnamefont {Koda}}, \bibinfo {author}
  {\bibfnamefont {K.~M.}\ \bibnamefont {Kojima}}, \bibinfo {author}
  {\bibfnamefont {S.}~\bibnamefont {Takeshita}}, \bibinfo {author}
  {\bibfnamefont {I.}~\bibnamefont {Yamauchi}}, \bibinfo {author}
  {\bibfnamefont {Y.}~\bibnamefont {Matsushita}}, \bibinfo {author}
  {\bibfnamefont {Y.}~\bibnamefont {Kuramoto}},\ and\ \bibinfo {author}
  {\bibfnamefont {R.}~\bibnamefont {Kadono}},\ }\href
  {https://doi.org/10.1103/PhysRevB.99.041113} {\bibfield  {journal} {\bibinfo
  {journal} {Phys. Rev. B}\ }\textbf {\bibinfo {volume} {99}},\ \bibinfo
  {pages} {041113} (\bibinfo {year} {2019})}\BibitemShut {NoStop}%
\bibitem [{\citenamefont {Shiga}\ \emph {et~al.}(1993)\citenamefont {Shiga},
  \citenamefont {Fujisawa},\ and\ \citenamefont {Wada}}]{YScMn19931329}%
  \BibitemOpen
  \bibfield  {author} {\bibinfo {author} {\bibfnamefont {M.}~\bibnamefont
  {Shiga}}, \bibinfo {author} {\bibfnamefont {K.}~\bibnamefont {Fujisawa}},\
  and\ \bibinfo {author} {\bibfnamefont {H.}~\bibnamefont {Wada}},\ }\href
  {https://doi.org/10.1143/JPSJ.62.1329} {\bibfield  {journal} {\bibinfo
  {journal} {J. Phys. Soc. Jpn.}\ }\textbf {\bibinfo {volume} {62}},\ \bibinfo
  {pages} {1329} (\bibinfo {year} {1993})}\BibitemShut {NoStop}%
\bibitem [{\citenamefont {Tokiwa}\ \emph {et~al.}(2013)\citenamefont {Tokiwa},
  \citenamefont {Garst}, \citenamefont {Gegenwart}, \citenamefont {Bud'ko},\
  and\ \citenamefont {Canfield}}]{PhysRevLett.111.116401}%
  \BibitemOpen
  \bibfield  {author} {\bibinfo {author} {\bibfnamefont {Y.}~\bibnamefont
  {Tokiwa}}, \bibinfo {author} {\bibfnamefont {M.}~\bibnamefont {Garst}},
  \bibinfo {author} {\bibfnamefont {P.}~\bibnamefont {Gegenwart}}, \bibinfo
  {author} {\bibfnamefont {S.~L.}\ \bibnamefont {Bud'ko}},\ and\ \bibinfo
  {author} {\bibfnamefont {P.~C.}\ \bibnamefont {Canfield}},\ }\href
  {https://doi.org/10.1103/PhysRevLett.111.116401} {\bibfield  {journal}
  {\bibinfo  {journal} {Phys. Rev. Lett.}\ }\textbf {\bibinfo {volume} {111}},\
  \bibinfo {pages} {116401} (\bibinfo {year} {2013})}\BibitemShut {NoStop}%
\bibitem [{\citenamefont {Lucas}\ \emph {et~al.}(2017)\citenamefont {Lucas},
  \citenamefont {Grube}, \citenamefont {Huang}, \citenamefont {Sakai},
  \citenamefont {Wunderlich}, \citenamefont {Green}, \citenamefont {Wosnitza},
  \citenamefont {Fritsch}, \citenamefont {Gegenwart}, \citenamefont
  {Stockert},\ and\ \citenamefont {v.~L\"ohneysen}}]{PhysRevLett.118.107204}%
  \BibitemOpen
  \bibfield  {author} {\bibinfo {author} {\bibfnamefont {S.}~\bibnamefont
  {Lucas}}, \bibinfo {author} {\bibfnamefont {K.}~\bibnamefont {Grube}},
  \bibinfo {author} {\bibfnamefont {C.-L.}\ \bibnamefont {Huang}}, \bibinfo
  {author} {\bibfnamefont {A.}~\bibnamefont {Sakai}}, \bibinfo {author}
  {\bibfnamefont {S.}~\bibnamefont {Wunderlich}}, \bibinfo {author}
  {\bibfnamefont {E.~L.}\ \bibnamefont {Green}}, \bibinfo {author}
  {\bibfnamefont {J.}~\bibnamefont {Wosnitza}}, \bibinfo {author}
  {\bibfnamefont {V.}~\bibnamefont {Fritsch}}, \bibinfo {author} {\bibfnamefont
  {P.}~\bibnamefont {Gegenwart}}, \bibinfo {author} {\bibfnamefont
  {O.}~\bibnamefont {Stockert}},\ and\ \bibinfo {author} {\bibfnamefont
  {H.}~\bibnamefont {v.~L\"ohneysen}},\ }\href
  {https://doi.org/10.1103/PhysRevLett.118.107204} {\bibfield  {journal}
  {\bibinfo  {journal} {Phys. Rev. Lett.}\ }\textbf {\bibinfo {volume} {118}},\
  \bibinfo {pages} {107204} (\bibinfo {year} {2017})}\BibitemShut {NoStop}%
\bibitem [{\citenamefont {Mishra}\ \emph {et~al.}(2001)\citenamefont {Mishra},
  \citenamefont {P{\"o}ttgen}, \citenamefont {Hoffmann}, \citenamefont {Trill},
  \citenamefont {Mosel}, \citenamefont {Piotrowski},\ and\ \citenamefont
  {Zumdick}}]{mishra2001stannides}%
  \BibitemOpen
  \bibfield  {author} {\bibinfo {author} {\bibfnamefont {R.}~\bibnamefont
  {Mishra}}, \bibinfo {author} {\bibfnamefont {R.}~\bibnamefont {P{\"o}ttgen}},
  \bibinfo {author} {\bibfnamefont {R.-D.}\ \bibnamefont {Hoffmann}}, \bibinfo
  {author} {\bibfnamefont {H.}~\bibnamefont {Trill}}, \bibinfo {author}
  {\bibfnamefont {B.~D.}\ \bibnamefont {Mosel}}, \bibinfo {author}
  {\bibfnamefont {H.}~\bibnamefont {Piotrowski}},\ and\ \bibinfo {author}
  {\bibfnamefont {M.~F.}\ \bibnamefont {Zumdick}},\ }\href
  {https://doi.org/doi:10.1515/znb-2001-0705} {\bibfield  {journal} {\bibinfo
  {journal} {Zeitschrift f{\"u}r Naturforschung B}\ }\textbf {\bibinfo {volume}
  {56}},\ \bibinfo {pages} {589} (\bibinfo {year} {2001})}\BibitemShut
  {NoStop}%
\bibitem [{\citenamefont {Kim}\ \emph {et~al.}(2003{\natexlab{a}})\citenamefont
  {Kim}, \citenamefont {Echizen}, \citenamefont {Umeo}, \citenamefont
  {Kobayashi}, \citenamefont {Sera}, \citenamefont {Salamakha}, \citenamefont
  {Sologub}, \citenamefont {Takabatake}, \citenamefont {Chen}, \citenamefont
  {Tayama}, \citenamefont {Sakakibara}, \citenamefont {Jung},\ and\
  \citenamefont {Maple}}]{PhysRevB.68.054416}%
  \BibitemOpen
  \bibfield  {author} {\bibinfo {author} {\bibfnamefont {M.~S.}\ \bibnamefont
  {Kim}}, \bibinfo {author} {\bibfnamefont {Y.}~\bibnamefont {Echizen}},
  \bibinfo {author} {\bibfnamefont {K.}~\bibnamefont {Umeo}}, \bibinfo {author}
  {\bibfnamefont {S.}~\bibnamefont {Kobayashi}}, \bibinfo {author}
  {\bibfnamefont {M.}~\bibnamefont {Sera}}, \bibinfo {author} {\bibfnamefont
  {P.~S.}\ \bibnamefont {Salamakha}}, \bibinfo {author} {\bibfnamefont {O.~L.}\
  \bibnamefont {Sologub}}, \bibinfo {author} {\bibfnamefont {T.}~\bibnamefont
  {Takabatake}}, \bibinfo {author} {\bibfnamefont {X.}~\bibnamefont {Chen}},
  \bibinfo {author} {\bibfnamefont {T.}~\bibnamefont {Tayama}}, \bibinfo
  {author} {\bibfnamefont {T.}~\bibnamefont {Sakakibara}}, \bibinfo {author}
  {\bibfnamefont {M.~H.}\ \bibnamefont {Jung}},\ and\ \bibinfo {author}
  {\bibfnamefont {M.~B.}\ \bibnamefont {Maple}},\ }\href
  {https://doi.org/10.1103/PhysRevB.68.054416} {\bibfield  {journal} {\bibinfo
  {journal} {Phys. Rev. B}\ }\textbf {\bibinfo {volume} {68}},\ \bibinfo
  {pages} {054416} (\bibinfo {year} {2003}{\natexlab{a}})}\BibitemShut
  {NoStop}%
\bibitem [{\citenamefont {Yang}\ \emph {et~al.}(2017)\citenamefont {Yang},
  \citenamefont {Tsuda}, \citenamefont {Umeo}, \citenamefont {Yamane},
  \citenamefont {Onimaru}, \citenamefont {Takabatake}, \citenamefont
  {Kikugawa}, \citenamefont {Terashima},\ and\ \citenamefont
  {Uji}}]{PhysRevB.96.045139}%
  \BibitemOpen
  \bibfield  {author} {\bibinfo {author} {\bibfnamefont {C.~L.}\ \bibnamefont
  {Yang}}, \bibinfo {author} {\bibfnamefont {S.}~\bibnamefont {Tsuda}},
  \bibinfo {author} {\bibfnamefont {K.}~\bibnamefont {Umeo}}, \bibinfo {author}
  {\bibfnamefont {Y.}~\bibnamefont {Yamane}}, \bibinfo {author} {\bibfnamefont
  {T.}~\bibnamefont {Onimaru}}, \bibinfo {author} {\bibfnamefont
  {T.}~\bibnamefont {Takabatake}}, \bibinfo {author} {\bibfnamefont
  {N.}~\bibnamefont {Kikugawa}}, \bibinfo {author} {\bibfnamefont
  {T.}~\bibnamefont {Terashima}},\ and\ \bibinfo {author} {\bibfnamefont
  {S.}~\bibnamefont {Uji}},\ }\href
  {https://doi.org/10.1103/PhysRevB.96.045139} {\bibfield  {journal} {\bibinfo
  {journal} {Phys. Rev. B}\ }\textbf {\bibinfo {volume} {96}},\ \bibinfo
  {pages} {045139} (\bibinfo {year} {2017})}\BibitemShut {NoStop}%
\bibitem [{\citenamefont {Schenck}\ \emph {et~al.}(2004)\citenamefont
  {Schenck}, \citenamefont {N.~Gygax}, \citenamefont {S.~Kim},\ and\
  \citenamefont {Takabatake}}]{doi:10.1143/JPSJ.73.3099}%
  \BibitemOpen
  \bibfield  {author} {\bibinfo {author} {\bibfnamefont {A.}~\bibnamefont
  {Schenck}}, \bibinfo {author} {\bibfnamefont {F.}~\bibnamefont {N.~Gygax}},
  \bibinfo {author} {\bibfnamefont {M.}~\bibnamefont {S.~Kim}},\ and\ \bibinfo
  {author} {\bibfnamefont {T.}~\bibnamefont {Takabatake}},\ }\href
  {https://doi.org/10.1143/JPSJ.73.3099} {\bibfield  {journal} {\bibinfo
  {journal} {J. Phys. Soc. Jpn.}\ }\textbf {\bibinfo {volume} {73}},\ \bibinfo
  {pages} {3099} (\bibinfo {year} {2004})}\BibitemShut {NoStop}%
\bibitem [{\citenamefont {Kim}\ \emph {et~al.}(2003{\natexlab{b}})\citenamefont
  {Kim}, \citenamefont {Echizen}, \citenamefont {Umeo}, \citenamefont {Tayama},
  \citenamefont {Sakakibara},\ and\ \citenamefont {Takabatake}}]{KIM2003524}%
  \BibitemOpen
  \bibfield  {author} {\bibinfo {author} {\bibfnamefont {M.}~\bibnamefont
  {Kim}}, \bibinfo {author} {\bibfnamefont {Y.}~\bibnamefont {Echizen}},
  \bibinfo {author} {\bibfnamefont {K.}~\bibnamefont {Umeo}}, \bibinfo {author}
  {\bibfnamefont {T.}~\bibnamefont {Tayama}}, \bibinfo {author} {\bibfnamefont
  {T.}~\bibnamefont {Sakakibara}},\ and\ \bibinfo {author} {\bibfnamefont
  {T.}~\bibnamefont {Takabatake}},\ }\href
  {https://doi.org/https://doi.org/10.1016/S0921-4526(02)02326-8} {\bibfield
  {journal} {\bibinfo  {journal} {Physica B: Condensed Matter}\ }\textbf
  {\bibinfo {volume} {329-333}},\ \bibinfo {pages} {524} (\bibinfo {year}
  {2003}{\natexlab{b}})}\BibitemShut {NoStop}%
\bibitem [{\citenamefont {K\"uchler}\ \emph {et~al.}(2017)\citenamefont
  {K\"uchler}, \citenamefont {Stingl}, \citenamefont {Tokiwa}, \citenamefont
  {Kim}, \citenamefont {Takabatake},\ and\ \citenamefont
  {Gegenwart}}]{PhysRevB.96.241110}%
  \BibitemOpen
  \bibfield  {author} {\bibinfo {author} {\bibfnamefont {R.}~\bibnamefont
  {K\"uchler}}, \bibinfo {author} {\bibfnamefont {C.}~\bibnamefont {Stingl}},
  \bibinfo {author} {\bibfnamefont {Y.}~\bibnamefont {Tokiwa}}, \bibinfo
  {author} {\bibfnamefont {M.~S.}\ \bibnamefont {Kim}}, \bibinfo {author}
  {\bibfnamefont {T.}~\bibnamefont {Takabatake}},\ and\ \bibinfo {author}
  {\bibfnamefont {P.}~\bibnamefont {Gegenwart}},\ }\href
  {https://doi.org/10.1103/PhysRevB.96.241110} {\bibfield  {journal} {\bibinfo
  {journal} {Phys. Rev. B}\ }\textbf {\bibinfo {volume} {96}},\ \bibinfo
  {pages} {241110} (\bibinfo {year} {2017})}\BibitemShut {NoStop}%
\bibitem [{\citenamefont {Sundermann}\ \emph {et~al.}(2021)\citenamefont
  {Sundermann}, \citenamefont {Marino}, \citenamefont {Gloskovskii},
  \citenamefont {Yang}, \citenamefont {Shimura}, \citenamefont {Takabatake},\
  and\ \citenamefont {Severing}}]{PhysRevB.104.235150}%
  \BibitemOpen
  \bibfield  {author} {\bibinfo {author} {\bibfnamefont {M.}~\bibnamefont
  {Sundermann}}, \bibinfo {author} {\bibfnamefont {A.}~\bibnamefont {Marino}},
  \bibinfo {author} {\bibfnamefont {A.}~\bibnamefont {Gloskovskii}}, \bibinfo
  {author} {\bibfnamefont {C.}~\bibnamefont {Yang}}, \bibinfo {author}
  {\bibfnamefont {Y.}~\bibnamefont {Shimura}}, \bibinfo {author} {\bibfnamefont
  {T.}~\bibnamefont {Takabatake}},\ and\ \bibinfo {author} {\bibfnamefont
  {A.}~\bibnamefont {Severing}},\ }\href
  {https://doi.org/10.1103/PhysRevB.104.235150} {\bibfield  {journal} {\bibinfo
   {journal} {Phys. Rev. B}\ }\textbf {\bibinfo {volume} {104}},\ \bibinfo
  {pages} {235150} (\bibinfo {year} {2021})}\BibitemShut {NoStop}%
\bibitem [{\citenamefont {Blundell}\ \emph {et~al.}(2022)\citenamefont
  {Blundell}, \citenamefont {De~Renzi}, \citenamefont {Lancaster},\ and\
  \citenamefont {Pratt}}]{muonbook}%
  \BibitemOpen
  \bibfield  {author} {\bibinfo {author} {\bibfnamefont {S.}~\bibnamefont
  {Blundell}}, \bibinfo {author} {\bibfnamefont {R.}~\bibnamefont {De~Renzi}},
  \bibinfo {author} {\bibfnamefont {T.}~\bibnamefont {Lancaster}},\ and\
  \bibinfo {author} {\bibfnamefont {F.}~\bibnamefont {Pratt}},\ }\href
  {https://books.google.co.uk/books?id=mEF8zgEACAAJ} {\emph {\bibinfo {title}
  {Muon Spectroscopy: An Introduction}}}\ (\bibinfo  {publisher} {Oxford
  University Press},\ \bibinfo {year} {2022})\BibitemShut {NoStop}%
\bibitem [{\citenamefont {Lee}\ \emph {et~al.}(1998)\citenamefont {Lee},
  \citenamefont {Kilcoyne},\ and\ \citenamefont {Cywinski}}]{MUSR}%
  \BibitemOpen
  \bibfield  {author} {\bibinfo {author} {\bibfnamefont {S.~L.}\ \bibnamefont
  {Lee}}, \bibinfo {author} {\bibfnamefont {S.~H.}\ \bibnamefont {Kilcoyne}},\
  and\ \bibinfo {author} {\bibfnamefont {R.}~\bibnamefont {Cywinski}},\ }\href
  {https://inspirehep.net/literature/508883} {\bibfield  {journal} {\bibinfo
  {journal} {Proc. Fifty First Scottish Universities Summer School in Physics,
  NATO Advanced Study Institute, August 17-28 St. Andrews, UK}\ } (\bibinfo
  {year} {1998})}\BibitemShut {NoStop}%
\bibitem [{\citenamefont {Li}\ \emph {et~al.}(2016)\citenamefont {Li},
  \citenamefont {Adroja}, \citenamefont {Biswas}, \citenamefont {Baker},
  \citenamefont {Zhang}, \citenamefont {Liu}, \citenamefont {Tsirlin},
  \citenamefont {Gegenwart},\ and\ \citenamefont
  {Zhang}}]{PhysRevLett.117.097201}%
  \BibitemOpen
  \bibfield  {author} {\bibinfo {author} {\bibfnamefont {Y.}~\bibnamefont
  {Li}}, \bibinfo {author} {\bibfnamefont {D.}~\bibnamefont {Adroja}}, \bibinfo
  {author} {\bibfnamefont {P.~K.}\ \bibnamefont {Biswas}}, \bibinfo {author}
  {\bibfnamefont {P.~J.}\ \bibnamefont {Baker}}, \bibinfo {author}
  {\bibfnamefont {Q.}~\bibnamefont {Zhang}}, \bibinfo {author} {\bibfnamefont
  {J.}~\bibnamefont {Liu}}, \bibinfo {author} {\bibfnamefont {A.~A.}\
  \bibnamefont {Tsirlin}}, \bibinfo {author} {\bibfnamefont {P.}~\bibnamefont
  {Gegenwart}},\ and\ \bibinfo {author} {\bibfnamefont {Q.}~\bibnamefont
  {Zhang}},\ }\href {https://doi.org/10.1103/PhysRevLett.117.097201} {\bibfield
   {journal} {\bibinfo  {journal} {Phys. Rev. Lett.}\ }\textbf {\bibinfo
  {volume} {117}},\ \bibinfo {pages} {097201} (\bibinfo {year}
  {2016})}\BibitemShut {NoStop}%
\bibitem [{\citenamefont {F\aa{}k}\ \emph {et~al.}(2012)\citenamefont
  {F\aa{}k}, \citenamefont {Kermarrec}, \citenamefont {Messio}, \citenamefont
  {Bernu}, \citenamefont {Lhuillier}, \citenamefont {Bert}, \citenamefont
  {Mendels}, \citenamefont {Koteswararao}, \citenamefont {Bouquet},
  \citenamefont {Ollivier}, \citenamefont {Hillier}, \citenamefont {Amato},
  \citenamefont {Colman},\ and\ \citenamefont
  {Wills}}]{PhysRevLett.109.037208}%
  \BibitemOpen
  \bibfield  {author} {\bibinfo {author} {\bibfnamefont {B.}~\bibnamefont
  {F\aa{}k}}, \bibinfo {author} {\bibfnamefont {E.}~\bibnamefont {Kermarrec}},
  \bibinfo {author} {\bibfnamefont {L.}~\bibnamefont {Messio}}, \bibinfo
  {author} {\bibfnamefont {B.}~\bibnamefont {Bernu}}, \bibinfo {author}
  {\bibfnamefont {C.}~\bibnamefont {Lhuillier}}, \bibinfo {author}
  {\bibfnamefont {F.}~\bibnamefont {Bert}}, \bibinfo {author} {\bibfnamefont
  {P.}~\bibnamefont {Mendels}}, \bibinfo {author} {\bibfnamefont
  {B.}~\bibnamefont {Koteswararao}}, \bibinfo {author} {\bibfnamefont
  {F.}~\bibnamefont {Bouquet}}, \bibinfo {author} {\bibfnamefont
  {J.}~\bibnamefont {Ollivier}}, \bibinfo {author} {\bibfnamefont {A.~D.}\
  \bibnamefont {Hillier}}, \bibinfo {author} {\bibfnamefont {A.}~\bibnamefont
  {Amato}}, \bibinfo {author} {\bibfnamefont {R.~H.}\ \bibnamefont {Colman}},\
  and\ \bibinfo {author} {\bibfnamefont {A.~S.}\ \bibnamefont {Wills}},\ }\href
  {https://doi.org/10.1103/PhysRevLett.109.037208} {\bibfield  {journal}
  {\bibinfo  {journal} {Phys. Rev. Lett.}\ }\textbf {\bibinfo {volume} {109}},\
  \bibinfo {pages} {037208} (\bibinfo {year} {2012})}\BibitemShut {NoStop}%
\bibitem [{\citenamefont {Kermarrec}\ \emph {et~al.}(2011)\citenamefont
  {Kermarrec}, \citenamefont {Mendels}, \citenamefont {Bert}, \citenamefont
  {Colman}, \citenamefont {Wills}, \citenamefont {Strobel}, \citenamefont
  {Bonville}, \citenamefont {Hillier},\ and\ \citenamefont
  {Amato}}]{PhysRevB.84.100401}%
  \BibitemOpen
  \bibfield  {author} {\bibinfo {author} {\bibfnamefont {E.}~\bibnamefont
  {Kermarrec}}, \bibinfo {author} {\bibfnamefont {P.}~\bibnamefont {Mendels}},
  \bibinfo {author} {\bibfnamefont {F.}~\bibnamefont {Bert}}, \bibinfo {author}
  {\bibfnamefont {R.~H.}\ \bibnamefont {Colman}}, \bibinfo {author}
  {\bibfnamefont {A.~S.}\ \bibnamefont {Wills}}, \bibinfo {author}
  {\bibfnamefont {P.}~\bibnamefont {Strobel}}, \bibinfo {author} {\bibfnamefont
  {P.}~\bibnamefont {Bonville}}, \bibinfo {author} {\bibfnamefont
  {A.}~\bibnamefont {Hillier}},\ and\ \bibinfo {author} {\bibfnamefont
  {A.}~\bibnamefont {Amato}},\ }\href
  {https://doi.org/10.1103/PhysRevB.84.100401} {\bibfield  {journal} {\bibinfo
  {journal} {Phys. Rev. B}\ }\textbf {\bibinfo {volume} {84}},\ \bibinfo
  {pages} {100401} (\bibinfo {year} {2011})}\BibitemShut {NoStop}%
\bibitem [{\citenamefont {Aronson}\ \emph {et~al.}(1995)\citenamefont
  {Aronson}, \citenamefont {Osborn}, \citenamefont {Robinson}, \citenamefont
  {Lynn}, \citenamefont {Chau}, \citenamefont {Seaman},\ and\ \citenamefont
  {Maple}}]{PhysRevLett.75.725}%
  \BibitemOpen
  \bibfield  {author} {\bibinfo {author} {\bibfnamefont {M.~C.}\ \bibnamefont
  {Aronson}}, \bibinfo {author} {\bibfnamefont {R.}~\bibnamefont {Osborn}},
  \bibinfo {author} {\bibfnamefont {R.~A.}\ \bibnamefont {Robinson}}, \bibinfo
  {author} {\bibfnamefont {J.~W.}\ \bibnamefont {Lynn}}, \bibinfo {author}
  {\bibfnamefont {R.}~\bibnamefont {Chau}}, \bibinfo {author} {\bibfnamefont
  {C.~L.}\ \bibnamefont {Seaman}},\ and\ \bibinfo {author} {\bibfnamefont
  {M.~B.}\ \bibnamefont {Maple}},\ }\href
  {https://doi.org/10.1103/PhysRevLett.75.725} {\bibfield  {journal} {\bibinfo
  {journal} {Phys. Rev. Lett.}\ }\textbf {\bibinfo {volume} {75}},\ \bibinfo
  {pages} {725} (\bibinfo {year} {1995})}\BibitemShut {NoStop}%
\bibitem [{\citenamefont {Arnold}\ \emph
  {et~al.}(2014{\natexlab{a}})\citenamefont {Arnold}, \citenamefont {Bilheux},
  \citenamefont {Borreguero}, \citenamefont {Buts}, \citenamefont {Campbell},
  \citenamefont {Chapon}, \citenamefont {Doucet}, \citenamefont {Draper},
  \citenamefont {{Ferraz Leal}}, \citenamefont {Gigg}, \citenamefont {Lynch},
  \citenamefont {Markvardsen}, \citenamefont {Mikkelson}, \citenamefont
  {Mikkelson}, \citenamefont {Miller}, \citenamefont {Palmen}, \citenamefont
  {Parker}, \citenamefont {Passos}, \citenamefont {Perring}, \citenamefont
  {Peterson}, \citenamefont {Ren}, \citenamefont {Reuter}, \citenamefont
  {Savici}, \citenamefont {Taylor}, \citenamefont {Taylor}, \citenamefont
  {Tolchenov}, \citenamefont {Zhou},\ and\ \citenamefont
  {Zikovsky}}]{ARNOLD2014156}%
  \BibitemOpen
  \bibfield  {author} {\bibinfo {author} {\bibfnamefont {O.}~\bibnamefont
  {Arnold}}, \bibinfo {author} {\bibfnamefont {J.}~\bibnamefont {Bilheux}},
  \bibinfo {author} {\bibfnamefont {J.}~\bibnamefont {Borreguero}}, \bibinfo
  {author} {\bibfnamefont {A.}~\bibnamefont {Buts}}, \bibinfo {author}
  {\bibfnamefont {S.}~\bibnamefont {Campbell}}, \bibinfo {author}
  {\bibfnamefont {L.}~\bibnamefont {Chapon}}, \bibinfo {author} {\bibfnamefont
  {M.}~\bibnamefont {Doucet}}, \bibinfo {author} {\bibfnamefont
  {N.}~\bibnamefont {Draper}}, \bibinfo {author} {\bibfnamefont
  {R.}~\bibnamefont {{Ferraz Leal}}}, \bibinfo {author} {\bibfnamefont
  {M.}~\bibnamefont {Gigg}}, \bibinfo {author} {\bibfnamefont {V.}~\bibnamefont
  {Lynch}}, \bibinfo {author} {\bibfnamefont {A.}~\bibnamefont {Markvardsen}},
  \bibinfo {author} {\bibfnamefont {D.}~\bibnamefont {Mikkelson}}, \bibinfo
  {author} {\bibfnamefont {R.}~\bibnamefont {Mikkelson}}, \bibinfo {author}
  {\bibfnamefont {R.}~\bibnamefont {Miller}}, \bibinfo {author} {\bibfnamefont
  {K.}~\bibnamefont {Palmen}}, \bibinfo {author} {\bibfnamefont
  {P.}~\bibnamefont {Parker}}, \bibinfo {author} {\bibfnamefont
  {G.}~\bibnamefont {Passos}}, \bibinfo {author} {\bibfnamefont
  {T.}~\bibnamefont {Perring}}, \bibinfo {author} {\bibfnamefont
  {P.}~\bibnamefont {Peterson}}, \bibinfo {author} {\bibfnamefont
  {S.}~\bibnamefont {Ren}}, \bibinfo {author} {\bibfnamefont {M.}~\bibnamefont
  {Reuter}}, \bibinfo {author} {\bibfnamefont {A.}~\bibnamefont {Savici}},
  \bibinfo {author} {\bibfnamefont {J.}~\bibnamefont {Taylor}}, \bibinfo
  {author} {\bibfnamefont {R.}~\bibnamefont {Taylor}}, \bibinfo {author}
  {\bibfnamefont {R.}~\bibnamefont {Tolchenov}}, \bibinfo {author}
  {\bibfnamefont {W.}~\bibnamefont {Zhou}},\ and\ \bibinfo {author}
  {\bibfnamefont {J.}~\bibnamefont {Zikovsky}},\ }\href
  {https://doi.org/https://doi.org/10.1016/j.nima.2014.07.029} {\bibfield
  {journal} {\bibinfo  {journal} {Nuclear Instruments and Methods in Physics
  Research Section A: Accelerators, Spectrometers, Detectors and Associated
  Equipment}\ }\textbf {\bibinfo {volume} {764}},\ \bibinfo {pages} {156}
  (\bibinfo {year} {2014}{\natexlab{a}})}\BibitemShut {NoStop}%
\bibitem [{\citenamefont {Pratt}(2000)}]{PRATT2000710}%
  \BibitemOpen
  \bibfield  {author} {\bibinfo {author} {\bibfnamefont {F.}~\bibnamefont
  {Pratt}},\ }\href
  {https://doi.org/https://doi.org/10.1016/S0921-4526(00)00328-8} {\bibfield
  {journal} {\bibinfo  {journal} {Physica B: Condensed Matter}\ }\textbf
  {\bibinfo {volume} {289-290}},\ \bibinfo {pages} {710} (\bibinfo {year}
  {2000})}\BibitemShut {NoStop}%
\bibitem [{\citenamefont {Bewley}\ \emph {et~al.}(2009)\citenamefont {Bewley},
  \citenamefont {Guidi},\ and\ \citenamefont {Bennington}}]{Bewley2009}%
  \BibitemOpen
  \bibfield  {author} {\bibinfo {author} {\bibfnamefont {R.~I.}\ \bibnamefont
  {Bewley}}, \bibinfo {author} {\bibfnamefont {T.}~\bibnamefont {Guidi}},\ and\
  \bibinfo {author} {\bibfnamefont {S.~M.}\ \bibnamefont {Bennington}},\
  }\href@noop {} {\bibfield  {journal} {\bibinfo  {journal} {Notiziario
  Neutroni e Luce di Sincrotrone}\ }\textbf {\bibinfo {volume} {14}},\ \bibinfo
  {pages} {22} (\bibinfo {year} {2009})}\BibitemShut {NoStop}%
\bibitem [{\citenamefont {Mustonen}\ \emph {et~al.}(2018)\citenamefont
  {Mustonen}, \citenamefont {Vasala}, \citenamefont {Sadrollahi}, \citenamefont
  {Schmidt}, \citenamefont {Baines}, \citenamefont {Walker}, \citenamefont
  {Terasaki}, \citenamefont {Litterst}, \citenamefont {Baggio-Saitovitch},\
  and\ \citenamefont {Karppinen}}]{mustonen2018spin}%
  \BibitemOpen
  \bibfield  {author} {\bibinfo {author} {\bibfnamefont {O.}~\bibnamefont
  {Mustonen}}, \bibinfo {author} {\bibfnamefont {S.}~\bibnamefont {Vasala}},
  \bibinfo {author} {\bibfnamefont {E.}~\bibnamefont {Sadrollahi}}, \bibinfo
  {author} {\bibfnamefont {K.}~\bibnamefont {Schmidt}}, \bibinfo {author}
  {\bibfnamefont {C.}~\bibnamefont {Baines}}, \bibinfo {author} {\bibfnamefont
  {H.}~\bibnamefont {Walker}}, \bibinfo {author} {\bibfnamefont
  {I.}~\bibnamefont {Terasaki}}, \bibinfo {author} {\bibfnamefont
  {F.}~\bibnamefont {Litterst}}, \bibinfo {author} {\bibfnamefont
  {E.}~\bibnamefont {Baggio-Saitovitch}},\ and\ \bibinfo {author}
  {\bibfnamefont {M.}~\bibnamefont {Karppinen}},\ }\href
  {https://doi.org/10.1038/s41467-018-03435-1} {\bibfield  {journal} {\bibinfo
  {journal} {Nature communications}\ }\textbf {\bibinfo {volume} {9}},\
  \bibinfo {pages} {1} (\bibinfo {year} {2018})}\BibitemShut {NoStop}%
\bibitem [{\citenamefont {Wang}\ \emph {et~al.}(2010)\citenamefont {Wang},
  \citenamefont {Lawrence}, \citenamefont {Christianson}, \citenamefont
  {Goremychkin}, \citenamefont {Fanelli}, \citenamefont {Gofryk}, \citenamefont
  {Bauer}, \citenamefont {Ronning}, \citenamefont {Thompson}, \citenamefont
  {de~Souza}, \citenamefont {Kolesnikov},\ and\ \citenamefont
  {Littrell}}]{PhysRevB.81.235132}%
  \BibitemOpen
  \bibfield  {author} {\bibinfo {author} {\bibfnamefont {C.~H.}\ \bibnamefont
  {Wang}}, \bibinfo {author} {\bibfnamefont {J.~M.}\ \bibnamefont {Lawrence}},
  \bibinfo {author} {\bibfnamefont {A.~D.}\ \bibnamefont {Christianson}},
  \bibinfo {author} {\bibfnamefont {E.~A.}\ \bibnamefont {Goremychkin}},
  \bibinfo {author} {\bibfnamefont {V.~R.}\ \bibnamefont {Fanelli}}, \bibinfo
  {author} {\bibfnamefont {K.}~\bibnamefont {Gofryk}}, \bibinfo {author}
  {\bibfnamefont {E.~D.}\ \bibnamefont {Bauer}}, \bibinfo {author}
  {\bibfnamefont {F.}~\bibnamefont {Ronning}}, \bibinfo {author} {\bibfnamefont
  {J.~D.}\ \bibnamefont {Thompson}}, \bibinfo {author} {\bibfnamefont {N.~R.}\
  \bibnamefont {de~Souza}}, \bibinfo {author} {\bibfnamefont {A.~I.}\
  \bibnamefont {Kolesnikov}},\ and\ \bibinfo {author} {\bibfnamefont {K.~C.}\
  \bibnamefont {Littrell}},\ }\href
  {https://doi.org/10.1103/PhysRevB.81.235132} {\bibfield  {journal} {\bibinfo
  {journal} {Phys. Rev. B}\ }\textbf {\bibinfo {volume} {81}},\ \bibinfo
  {pages} {235132} (\bibinfo {year} {2010})}\BibitemShut {NoStop}%
\bibitem [{\citenamefont {Uemura}\ \emph {et~al.}(1985)\citenamefont {Uemura},
  \citenamefont {Yamazaki}, \citenamefont {Harshman}, \citenamefont {Senba},\
  and\ \citenamefont {Ansaldo}}]{PhysRevB.31.546}%
  \BibitemOpen
  \bibfield  {author} {\bibinfo {author} {\bibfnamefont {Y.~J.}\ \bibnamefont
  {Uemura}}, \bibinfo {author} {\bibfnamefont {T.}~\bibnamefont {Yamazaki}},
  \bibinfo {author} {\bibfnamefont {D.~R.}\ \bibnamefont {Harshman}}, \bibinfo
  {author} {\bibfnamefont {M.}~\bibnamefont {Senba}},\ and\ \bibinfo {author}
  {\bibfnamefont {E.~J.}\ \bibnamefont {Ansaldo}},\ }\href
  {https://doi.org/10.1103/PhysRevB.31.546} {\bibfield  {journal} {\bibinfo
  {journal} {Phys. Rev. B}\ }\textbf {\bibinfo {volume} {31}},\ \bibinfo
  {pages} {546} (\bibinfo {year} {1985})}\BibitemShut {NoStop}%
\bibitem [{\citenamefont {Uemura}\ \emph {et~al.}(1994)\citenamefont {Uemura},
  \citenamefont {Keren}, \citenamefont {Kojima}, \citenamefont {Le},
  \citenamefont {Luke}, \citenamefont {Wu}, \citenamefont {Ajiro},
  \citenamefont {Asano}, \citenamefont {Kuriyama}, \citenamefont {Mekata},
  \citenamefont {Kikuchi},\ and\ \citenamefont
  {Kakurai}}]{PhysRevLett.73.3306}%
  \BibitemOpen
  \bibfield  {author} {\bibinfo {author} {\bibfnamefont {Y.~J.}\ \bibnamefont
  {Uemura}}, \bibinfo {author} {\bibfnamefont {A.}~\bibnamefont {Keren}},
  \bibinfo {author} {\bibfnamefont {K.}~\bibnamefont {Kojima}}, \bibinfo
  {author} {\bibfnamefont {L.~P.}\ \bibnamefont {Le}}, \bibinfo {author}
  {\bibfnamefont {G.~M.}\ \bibnamefont {Luke}}, \bibinfo {author}
  {\bibfnamefont {W.~D.}\ \bibnamefont {Wu}}, \bibinfo {author} {\bibfnamefont
  {Y.}~\bibnamefont {Ajiro}}, \bibinfo {author} {\bibfnamefont
  {T.}~\bibnamefont {Asano}}, \bibinfo {author} {\bibfnamefont
  {Y.}~\bibnamefont {Kuriyama}}, \bibinfo {author} {\bibfnamefont
  {M.}~\bibnamefont {Mekata}}, \bibinfo {author} {\bibfnamefont
  {H.}~\bibnamefont {Kikuchi}},\ and\ \bibinfo {author} {\bibfnamefont
  {K.}~\bibnamefont {Kakurai}},\ }\href
  {https://doi.org/10.1103/PhysRevLett.73.3306} {\bibfield  {journal} {\bibinfo
   {journal} {Phys. Rev. Lett.}\ }\textbf {\bibinfo {volume} {73}},\ \bibinfo
  {pages} {3306} (\bibinfo {year} {1994})}\BibitemShut {NoStop}%
\bibitem [{\citenamefont {Bono}\ \emph {et~al.}(2004)\citenamefont {Bono},
  \citenamefont {Mendels}, \citenamefont {Collin}, \citenamefont {Blanchard},
  \citenamefont {Bert}, \citenamefont {Amato}, \citenamefont {Baines},\ and\
  \citenamefont {Hillier}}]{PhysRevLett.93.187201}%
  \BibitemOpen
  \bibfield  {author} {\bibinfo {author} {\bibfnamefont {D.}~\bibnamefont
  {Bono}}, \bibinfo {author} {\bibfnamefont {P.}~\bibnamefont {Mendels}},
  \bibinfo {author} {\bibfnamefont {G.}~\bibnamefont {Collin}}, \bibinfo
  {author} {\bibfnamefont {N.}~\bibnamefont {Blanchard}}, \bibinfo {author}
  {\bibfnamefont {F.}~\bibnamefont {Bert}}, \bibinfo {author} {\bibfnamefont
  {A.}~\bibnamefont {Amato}}, \bibinfo {author} {\bibfnamefont
  {C.}~\bibnamefont {Baines}},\ and\ \bibinfo {author} {\bibfnamefont {A.~D.}\
  \bibnamefont {Hillier}},\ }\href
  {https://doi.org/10.1103/PhysRevLett.93.187201} {\bibfield  {journal}
  {\bibinfo  {journal} {Phys. Rev. Lett.}\ }\textbf {\bibinfo {volume} {93}},\
  \bibinfo {pages} {187201} (\bibinfo {year} {2004})}\BibitemShut {NoStop}%
\bibitem [{\citenamefont {Pratt}\ \emph {et~al.}(2011)\citenamefont {Pratt},
  \citenamefont {Baker}, \citenamefont {Blundell}, \citenamefont {Lancaster},
  \citenamefont {Ohira-Kawamura}, \citenamefont {Baines}, \citenamefont
  {Shimizu}, \citenamefont {Kanoda}, \citenamefont {Watanabe},\ and\
  \citenamefont {Saito}}]{Nature.471.612}%
  \BibitemOpen
  \bibfield  {author} {\bibinfo {author} {\bibfnamefont {F.~L.}\ \bibnamefont
  {Pratt}}, \bibinfo {author} {\bibfnamefont {P.~J.}\ \bibnamefont {Baker}},
  \bibinfo {author} {\bibfnamefont {S.~J.}\ \bibnamefont {Blundell}}, \bibinfo
  {author} {\bibfnamefont {T.}~\bibnamefont {Lancaster}}, \bibinfo {author}
  {\bibfnamefont {S.}~\bibnamefont {Ohira-Kawamura}}, \bibinfo {author}
  {\bibfnamefont {C.}~\bibnamefont {Baines}}, \bibinfo {author} {\bibfnamefont
  {Y.}~\bibnamefont {Shimizu}}, \bibinfo {author} {\bibfnamefont
  {K.}~\bibnamefont {Kanoda}}, \bibinfo {author} {\bibfnamefont
  {I.}~\bibnamefont {Watanabe}},\ and\ \bibinfo {author} {\bibfnamefont
  {G.}~\bibnamefont {Saito}},\ }\href {https://doi.org/10.1038/nature09910}
  {\bibfield  {journal} {\bibinfo  {journal} {Nature}\ }\textbf {\bibinfo
  {volume} {471}},\ \bibinfo {pages} {612–616} (\bibinfo {year}
  {2011})}\BibitemShut {NoStop}%
\bibitem [{\citenamefont {Szirmai}\ \emph {et~al.}(2020)\citenamefont
  {Szirmai}, \citenamefont {M{\'e}zi{\`e}re}, \citenamefont {Bastien},
  \citenamefont {Wzietek}, \citenamefont {Batail}, \citenamefont {Martino},
  \citenamefont {Mantulnikovs}, \citenamefont {Pisoni}, \citenamefont {Riedl},
  \citenamefont {Cottrell}, \citenamefont {Baines}, \citenamefont {Forr{\'o}},\
  and\ \citenamefont {N{\'a}fr{\'a}di}}]{Szirmai29555}%
  \BibitemOpen
  \bibfield  {author} {\bibinfo {author} {\bibfnamefont {P.}~\bibnamefont
  {Szirmai}}, \bibinfo {author} {\bibfnamefont {C.}~\bibnamefont
  {M{\'e}zi{\`e}re}}, \bibinfo {author} {\bibfnamefont {G.}~\bibnamefont
  {Bastien}}, \bibinfo {author} {\bibfnamefont {P.}~\bibnamefont {Wzietek}},
  \bibinfo {author} {\bibfnamefont {P.}~\bibnamefont {Batail}}, \bibinfo
  {author} {\bibfnamefont {E.}~\bibnamefont {Martino}}, \bibinfo {author}
  {\bibfnamefont {K.}~\bibnamefont {Mantulnikovs}}, \bibinfo {author}
  {\bibfnamefont {A.}~\bibnamefont {Pisoni}}, \bibinfo {author} {\bibfnamefont
  {K.}~\bibnamefont {Riedl}}, \bibinfo {author} {\bibfnamefont
  {S.}~\bibnamefont {Cottrell}}, \bibinfo {author} {\bibfnamefont
  {C.}~\bibnamefont {Baines}}, \bibinfo {author} {\bibfnamefont
  {L.}~\bibnamefont {Forr{\'o}}},\ and\ \bibinfo {author} {\bibfnamefont
  {B.}~\bibnamefont {N{\'a}fr{\'a}di}},\ }\href
  {https://doi.org/10.1073/pnas.2000188117} {\bibfield  {journal} {\bibinfo
  {journal} {Proc. Natl. Acad. Sci. USA}\ }\textbf {\bibinfo {volume} {117}},\
  \bibinfo {pages} {29555} (\bibinfo {year} {2020})}\BibitemShut {NoStop}%
\bibitem [{\citenamefont {Qi}\ \emph {et~al.}(2009)\citenamefont {Qi},
  \citenamefont {Xu},\ and\ \citenamefont {Sachdev}}]{PhysRevLett.102.176401}%
  \BibitemOpen
  \bibfield  {author} {\bibinfo {author} {\bibfnamefont {Y.}~\bibnamefont
  {Qi}}, \bibinfo {author} {\bibfnamefont {C.}~\bibnamefont {Xu}},\ and\
  \bibinfo {author} {\bibfnamefont {S.}~\bibnamefont {Sachdev}},\ }\href
  {https://doi.org/10.1103/PhysRevLett.102.176401} {\bibfield  {journal}
  {\bibinfo  {journal} {Phys. Rev. Lett.}\ }\textbf {\bibinfo {volume} {102}},\
  \bibinfo {pages} {176401} (\bibinfo {year} {2009})}\BibitemShut {NoStop}%
\bibitem [{\citenamefont {Sarkar}\ \emph {et~al.}(2019)\citenamefont {Sarkar},
  \citenamefont {Schlender}, \citenamefont {Grinenko}, \citenamefont
  {Haeussler}, \citenamefont {Baker}, \citenamefont {Doert},\ and\
  \citenamefont {Klauss}}]{PhysRevB.100.241116}%
  \BibitemOpen
  \bibfield  {author} {\bibinfo {author} {\bibfnamefont {R.}~\bibnamefont
  {Sarkar}}, \bibinfo {author} {\bibfnamefont {P.}~\bibnamefont {Schlender}},
  \bibinfo {author} {\bibfnamefont {V.}~\bibnamefont {Grinenko}}, \bibinfo
  {author} {\bibfnamefont {E.}~\bibnamefont {Haeussler}}, \bibinfo {author}
  {\bibfnamefont {P.~J.}\ \bibnamefont {Baker}}, \bibinfo {author}
  {\bibfnamefont {T.}~\bibnamefont {Doert}},\ and\ \bibinfo {author}
  {\bibfnamefont {H.-H.}\ \bibnamefont {Klauss}},\ }\href
  {https://doi.org/10.1103/PhysRevB.100.241116} {\bibfield  {journal} {\bibinfo
   {journal} {Phys. Rev. B}\ }\textbf {\bibinfo {volume} {100}},\ \bibinfo
  {pages} {241116} (\bibinfo {year} {2019})}\BibitemShut {NoStop}%
\bibitem [{\citenamefont {Clark}\ \emph {et~al.}(2013)\citenamefont {Clark},
  \citenamefont {Orain}, \citenamefont {Bert}, \citenamefont {De~Vries},
  \citenamefont {Aidoudi}, \citenamefont {Morris}, \citenamefont {Lightfoot},
  \citenamefont {Lord}, \citenamefont {Telling}, \citenamefont {Bonville},
  \citenamefont {Attfield}, \citenamefont {Mendels},\ and\ \citenamefont
  {Harrison}}]{PhysRevLett.110.207208}%
  \BibitemOpen
  \bibfield  {author} {\bibinfo {author} {\bibfnamefont {L.}~\bibnamefont
  {Clark}}, \bibinfo {author} {\bibfnamefont {J.~C.}\ \bibnamefont {Orain}},
  \bibinfo {author} {\bibfnamefont {F.}~\bibnamefont {Bert}}, \bibinfo {author}
  {\bibfnamefont {M.~A.}\ \bibnamefont {De~Vries}}, \bibinfo {author}
  {\bibfnamefont {F.~H.}\ \bibnamefont {Aidoudi}}, \bibinfo {author}
  {\bibfnamefont {R.~E.}\ \bibnamefont {Morris}}, \bibinfo {author}
  {\bibfnamefont {P.}~\bibnamefont {Lightfoot}}, \bibinfo {author}
  {\bibfnamefont {J.~S.}\ \bibnamefont {Lord}}, \bibinfo {author}
  {\bibfnamefont {M.~T.~F.}\ \bibnamefont {Telling}}, \bibinfo {author}
  {\bibfnamefont {P.}~\bibnamefont {Bonville}}, \bibinfo {author}
  {\bibfnamefont {J.~P.}\ \bibnamefont {Attfield}}, \bibinfo {author}
  {\bibfnamefont {P.}~\bibnamefont {Mendels}},\ and\ \bibinfo {author}
  {\bibfnamefont {A.}~\bibnamefont {Harrison}},\ }\href
  {https://doi.org/10.1103/PhysRevLett.110.207208} {\bibfield  {journal}
  {\bibinfo  {journal} {Phys. Rev. Lett.}\ }\textbf {\bibinfo {volume} {110}},\
  \bibinfo {pages} {207208} (\bibinfo {year} {2013})}\BibitemShut {NoStop}%
\bibitem [{\citenamefont {Mañas-Valero}\ \emph {et~al.}(2021)\citenamefont
  {Mañas-Valero}, \citenamefont {Huddart}, \citenamefont {Lancaster},
  \citenamefont {Coronado},\ and\ \citenamefont {Pratt}}]{npj.6.2397.4648}%
  \BibitemOpen
  \bibfield  {author} {\bibinfo {author} {\bibfnamefont {S.}~\bibnamefont
  {Mañas-Valero}}, \bibinfo {author} {\bibfnamefont {B.~M.}\ \bibnamefont
  {Huddart}}, \bibinfo {author} {\bibfnamefont {T.}~\bibnamefont {Lancaster}},
  \bibinfo {author} {\bibfnamefont {E.}~\bibnamefont {Coronado}},\ and\
  \bibinfo {author} {\bibfnamefont {F.~L.}\ \bibnamefont {Pratt}},\ }\href
  {https://doi.org/10.1038/s41535-021-00367-w} {\bibfield  {journal} {\bibinfo
  {journal} {npj Quantum Materials}\ }\textbf {\bibinfo {volume} {6}},\
  \bibinfo {pages} {2397} (\bibinfo {year} {2021})}\BibitemShut {NoStop}%
\bibitem [{\citenamefont {Wen}\ \emph {et~al.}(2019)\citenamefont {Wen},
  \citenamefont {Yu}, \citenamefont {Li}, \citenamefont {Yu},\ and\
  \citenamefont {Li}}]{npj.4.2019}%
  \BibitemOpen
  \bibfield  {author} {\bibinfo {author} {\bibfnamefont {J.}~\bibnamefont
  {Wen}}, \bibinfo {author} {\bibfnamefont {S.-L.}\ \bibnamefont {Yu}},
  \bibinfo {author} {\bibfnamefont {S.}~\bibnamefont {Li}}, \bibinfo {author}
  {\bibfnamefont {W.}~\bibnamefont {Yu}},\ and\ \bibinfo {author}
  {\bibfnamefont {J.-X.}\ \bibnamefont {Li}},\ }\href
  {https://doi.org/10.1038/s41535-019-0151-6} {\bibfield  {journal} {\bibinfo
  {journal} {npj Quantum Materials}\ }\textbf {\bibinfo {volume} {4}},\
  \bibinfo {pages} {2397} (\bibinfo {year} {2019})}\BibitemShut {NoStop}%
\bibitem [{\citenamefont {Gao}\ \emph {et~al.}(2019)\citenamefont {Gao},
  \citenamefont {Chen}, \citenamefont {Tam}, \citenamefont {Sasmal},
  \citenamefont {Adroja}, \citenamefont {Ye}, \citenamefont {Cao},
  \citenamefont {Sala}, \citenamefont {Stone}, \citenamefont {Verezhak},
  \citenamefont {Hu}, \citenamefont {Chung}, \citenamefont {Xu}, \citenamefont
  {Cheong}, \citenamefont {Nallaiyan}, \citenamefont {Spagna}, \citenamefont
  {Maple}, \citenamefont {Nevidomskyy},\ and\ \citenamefont
  {hen}}]{Nat.Phys.15.1052}%
  \BibitemOpen
  \bibfield  {author} {\bibinfo {author} {\bibfnamefont {B.}~\bibnamefont
  {Gao}}, \bibinfo {author} {\bibfnamefont {T.}~\bibnamefont {Chen}}, \bibinfo
  {author} {\bibfnamefont {C.-L.}\ \bibnamefont {Tam}, \bibfnamefont {David
  W.and~Huang}}, \bibinfo {author} {\bibfnamefont {K.}~\bibnamefont {Sasmal}},
  \bibinfo {author} {\bibfnamefont {D.~T.}\ \bibnamefont {Adroja}}, \bibinfo
  {author} {\bibfnamefont {F.}~\bibnamefont {Ye}}, \bibinfo {author}
  {\bibfnamefont {H.}~\bibnamefont {Cao}}, \bibinfo {author} {\bibfnamefont
  {G.}~\bibnamefont {Sala}}, \bibinfo {author} {\bibfnamefont {C.}~\bibnamefont
  {Stone}, \bibfnamefont {Matthew B.and~Baines}}, \bibinfo {author}
  {\bibfnamefont {J.~A.~T.}\ \bibnamefont {Verezhak}}, \bibinfo {author}
  {\bibfnamefont {H.}~\bibnamefont {Hu}}, \bibinfo {author} {\bibfnamefont
  {J.-H.}\ \bibnamefont {Chung}}, \bibinfo {author} {\bibfnamefont
  {X.}~\bibnamefont {Xu}}, \bibinfo {author} {\bibfnamefont {S.-W.}\
  \bibnamefont {Cheong}}, \bibinfo {author} {\bibfnamefont {M.}~\bibnamefont
  {Nallaiyan}}, \bibinfo {author} {\bibfnamefont {S.}~\bibnamefont {Spagna}},
  \bibinfo {author} {\bibfnamefont {M.~B.}\ \bibnamefont {Maple}}, \bibinfo
  {author} {\bibfnamefont {E.}~\bibnamefont {Nevidomskyy}, \bibfnamefont
  {Andriy H.and~Morosan}},\ and\ \bibinfo {author} {\bibfnamefont
  {P.}~\bibnamefont {hen}, \bibfnamefont {Gang~Dai}},\ }\href
  {https://doi.org/10.1038/s41567-019-0577-6} {\bibfield  {journal} {\bibinfo
  {journal} {Nat.Phys.}\ }\textbf {\bibinfo {volume} {15}},\ \bibinfo {pages}
  {1052–1057} (\bibinfo {year} {2019})}\BibitemShut {NoStop}%
\bibitem [{\citenamefont {Balz}\ \emph {et~al.}(2016)\citenamefont {Balz},
  \citenamefont {Lake}, \citenamefont {Reuther}, \citenamefont {Luetkens},
  \citenamefont {Schönemann}, \citenamefont {Herrmannsdörfer}, \citenamefont
  {Singh}, \citenamefont {Islam}, \citenamefont {Wheeler}, \citenamefont
  {Rodriguez-Rivera},\ and\ \citenamefont {Guidi}}]{Nat.Phys.12.942}%
  \BibitemOpen
  \bibfield  {author} {\bibinfo {author} {\bibfnamefont {C.}~\bibnamefont
  {Balz}}, \bibinfo {author} {\bibfnamefont {B.}~\bibnamefont {Lake}}, \bibinfo
  {author} {\bibfnamefont {J.}~\bibnamefont {Reuther}}, \bibinfo {author}
  {\bibfnamefont {H.}~\bibnamefont {Luetkens}}, \bibinfo {author}
  {\bibfnamefont {R.}~\bibnamefont {Schönemann}}, \bibinfo {author}
  {\bibfnamefont {T.}~\bibnamefont {Herrmannsdörfer}}, \bibinfo {author}
  {\bibfnamefont {Y.}~\bibnamefont {Singh}}, \bibinfo {author} {\bibfnamefont
  {A.}~\bibnamefont {Islam}}, \bibinfo {author} {\bibfnamefont
  {E.}~\bibnamefont {Wheeler}}, \bibinfo {author} {\bibfnamefont
  {J.}~\bibnamefont {Rodriguez-Rivera}},\ and\ \bibinfo {author} {\bibfnamefont
  {T.}~\bibnamefont {Guidi}},\ }\href {https://doi.org/10.1038/nphys3826}
  {\bibfield  {journal} {\bibinfo  {journal} {Nat. Phys.}\ }\textbf {\bibinfo
  {volume} {12}},\ \bibinfo {pages} {942} (\bibinfo {year} {2016})}\BibitemShut
  {NoStop}%
\bibitem [{SM()}]{SM}%
  \BibitemOpen
  \href@noop {} {}\bibinfo {note} {See Supplemental Material, "for a detailed
  analysis of muon and inelastic neutron results" at [URL will be inserted by
  publisher] for additional characterization.}\BibitemShut {Stop}%
\bibitem [{\citenamefont {Amato}(1997)}]{RevModPhys.69.1119}%
  \BibitemOpen
  \bibfield  {author} {\bibinfo {author} {\bibfnamefont {A.}~\bibnamefont
  {Amato}},\ }\href {https://doi.org/10.1103/RevModPhys.69.1119} {\bibfield
  {journal} {\bibinfo  {journal} {Rev. Mod. Phys.}\ }\textbf {\bibinfo {volume}
  {69}},\ \bibinfo {pages} {1119} (\bibinfo {year} {1997})}\BibitemShut
  {NoStop}%
\bibitem [{\citenamefont {Krishnamurthy}\ \emph {et~al.}(2002)\citenamefont
  {Krishnamurthy}, \citenamefont {Nagamine}, \citenamefont {Watanabe},
  \citenamefont {Nishiyama}, \citenamefont {Ohira}, \citenamefont {Ishikawa},
  \citenamefont {Eom}, \citenamefont {Ishikawa},\ and\ \citenamefont
  {Briere}}]{PhysRevLett.88.046402}%
  \BibitemOpen
  \bibfield  {author} {\bibinfo {author} {\bibfnamefont {V.~V.}\ \bibnamefont
  {Krishnamurthy}}, \bibinfo {author} {\bibfnamefont {K.}~\bibnamefont
  {Nagamine}}, \bibinfo {author} {\bibfnamefont {I.}~\bibnamefont {Watanabe}},
  \bibinfo {author} {\bibfnamefont {K.}~\bibnamefont {Nishiyama}}, \bibinfo
  {author} {\bibfnamefont {S.}~\bibnamefont {Ohira}}, \bibinfo {author}
  {\bibfnamefont {M.}~\bibnamefont {Ishikawa}}, \bibinfo {author}
  {\bibfnamefont {D.~H.}\ \bibnamefont {Eom}}, \bibinfo {author} {\bibfnamefont
  {T.}~\bibnamefont {Ishikawa}},\ and\ \bibinfo {author} {\bibfnamefont
  {T.~M.}\ \bibnamefont {Briere}},\ }\href
  {https://doi.org/10.1103/PhysRevLett.88.046402} {\bibfield  {journal}
  {\bibinfo  {journal} {Phys. Rev. Lett.}\ }\textbf {\bibinfo {volume} {88}},\
  \bibinfo {pages} {046402} (\bibinfo {year} {2002})}\BibitemShut {NoStop}%
\bibitem [{\citenamefont {Anand}\ \emph {et~al.}(2018)\citenamefont {Anand},
  \citenamefont {Adroja}, \citenamefont {Hillier}, \citenamefont {Shigetoh},
  \citenamefont {Takabatake}, \citenamefont {Park}, \citenamefont {McEwen},
  \citenamefont {Pixley},\ and\ \citenamefont
  {Si}}]{doi:10.7566/JPSJ.87.064708}%
  \BibitemOpen
  \bibfield  {author} {\bibinfo {author} {\bibfnamefont {V.~K.}\ \bibnamefont
  {Anand}}, \bibinfo {author} {\bibfnamefont {D.~T.}\ \bibnamefont {Adroja}},
  \bibinfo {author} {\bibfnamefont {A.~D.}\ \bibnamefont {Hillier}}, \bibinfo
  {author} {\bibfnamefont {K.}~\bibnamefont {Shigetoh}}, \bibinfo {author}
  {\bibfnamefont {T.}~\bibnamefont {Takabatake}}, \bibinfo {author}
  {\bibfnamefont {J.-G.}\ \bibnamefont {Park}}, \bibinfo {author}
  {\bibfnamefont {K.~A.}\ \bibnamefont {McEwen}}, \bibinfo {author}
  {\bibfnamefont {J.~H.}\ \bibnamefont {Pixley}},\ and\ \bibinfo {author}
  {\bibfnamefont {Q.}~\bibnamefont {Si}},\ }\href
  {https://doi.org/10.7566/JPSJ.87.064708} {\bibfield  {journal} {\bibinfo
  {journal} {J. Phys. Soc. Jpn.}\ }\textbf {\bibinfo {volume} {87}},\ \bibinfo
  {pages} {064708} (\bibinfo {year} {2018})}\BibitemShut {NoStop}%
\bibitem [{\citenamefont {Orain}\ \emph {et~al.}(2014)\citenamefont {Orain},
  \citenamefont {Clark}, \citenamefont {Bert}, \citenamefont {Mendels},
  \citenamefont {Attfield}, \citenamefont {Aidoudi}, \citenamefont {Morris},
  \citenamefont {Lightfoot}, \citenamefont {Amato},\ and\ \citenamefont
  {Baines}}]{Orain2014}%
  \BibitemOpen
  \bibfield  {author} {\bibinfo {author} {\bibfnamefont {J.~C.}\ \bibnamefont
  {Orain}}, \bibinfo {author} {\bibfnamefont {L.}~\bibnamefont {Clark}},
  \bibinfo {author} {\bibfnamefont {F.}~\bibnamefont {Bert}}, \bibinfo {author}
  {\bibfnamefont {P.}~\bibnamefont {Mendels}}, \bibinfo {author} {\bibfnamefont
  {P.}~\bibnamefont {Attfield}}, \bibinfo {author} {\bibfnamefont {F.~H.}\
  \bibnamefont {Aidoudi}}, \bibinfo {author} {\bibfnamefont {R.~E.}\
  \bibnamefont {Morris}}, \bibinfo {author} {\bibfnamefont {P.}~\bibnamefont
  {Lightfoot}}, \bibinfo {author} {\bibfnamefont {A.}~\bibnamefont {Amato}},\
  and\ \bibinfo {author} {\bibfnamefont {C.}~\bibnamefont {Baines}},\ }\href
  {https://doi.org/10.1088/1742-6596/551/1/012004} {\bibfield  {journal}
  {\bibinfo  {journal} {Journal of Physics: Conference Series}\ }\textbf
  {\bibinfo {volume} {551}},\ \bibinfo {pages} {012004} (\bibinfo {year}
  {2014})}\BibitemShut {NoStop}%
\bibitem [{\citenamefont {Fujihala}\ \emph {et~al.}(2020)\citenamefont
  {Fujihala}, \citenamefont {Morita}, \citenamefont {Mole}, \citenamefont
  {Mitsuda}, \citenamefont {Tohyama}, \citenamefont {Yano}, \citenamefont {Yu},
  \citenamefont {Sota}, \citenamefont {Kuwai}, \citenamefont {Koda},\ and\
  \citenamefont {Okabe}}]{Nature.com.11.1.7}%
  \BibitemOpen
  \bibfield  {author} {\bibinfo {author} {\bibfnamefont {M.}~\bibnamefont
  {Fujihala}}, \bibinfo {author} {\bibfnamefont {K.}~\bibnamefont {Morita}},
  \bibinfo {author} {\bibfnamefont {R.}~\bibnamefont {Mole}}, \bibinfo {author}
  {\bibfnamefont {S.}~\bibnamefont {Mitsuda}}, \bibinfo {author} {\bibfnamefont
  {T.}~\bibnamefont {Tohyama}}, \bibinfo {author} {\bibfnamefont
  {S.}~\bibnamefont {Yano}}, \bibinfo {author} {\bibfnamefont {D.}~\bibnamefont
  {Yu}}, \bibinfo {author} {\bibfnamefont {S.}~\bibnamefont {Sota}}, \bibinfo
  {author} {\bibfnamefont {T.}~\bibnamefont {Kuwai}}, \bibinfo {author}
  {\bibfnamefont {A.}~\bibnamefont {Koda}},\ and\ \bibinfo {author}
  {\bibfnamefont {H.}~\bibnamefont {Okabe}},\ }\href
  {https://doi.org/10.1038/s41467-020-17235-z} {\bibfield  {journal} {\bibinfo
  {journal} {Nature Communications}\ }\textbf {\bibinfo {volume} {11}},\
  \bibinfo {pages} {2041} (\bibinfo {year} {2020})}\BibitemShut {NoStop}%
\bibitem [{\citenamefont {Taku}\ \emph {et~al.}(2017)\citenamefont {Taku},
  \citenamefont {Franz}, \citenamefont {Toshiro},\ and\ \citenamefont
  {Mitsuru}}]{Sato}%
  \BibitemOpen
  \bibfield  {author} {\bibinfo {author} {\bibfnamefont {S.}~\bibnamefont
  {Taku}}, \bibinfo {author} {\bibfnamefont {D.}~\bibnamefont {Franz}},
  \bibinfo {author} {\bibfnamefont {T.}~\bibnamefont {Toshiro}},\ and\ \bibinfo
  {author} {\bibfnamefont {T.}~\bibnamefont {Mitsuru}},\ }\href
  {https://doi.org/10.5286/ISIS.E.RB1710315} {} (\bibinfo {year}
  {2017})\BibitemShut {NoStop}%
\bibitem [{\citenamefont {Xu}\ \emph {et~al.}(2013)\citenamefont {Xu},
  \citenamefont {Xu},\ and\ \citenamefont {Tranquada}}]{doi:10.1063/1.4818323}%
  \BibitemOpen
  \bibfield  {author} {\bibinfo {author} {\bibfnamefont {G.}~\bibnamefont
  {Xu}}, \bibinfo {author} {\bibfnamefont {Z.}~\bibnamefont {Xu}},\ and\
  \bibinfo {author} {\bibfnamefont {J.~M.}\ \bibnamefont {Tranquada}},\ }\href
  {https://doi.org/10.1063/1.4818323} {\bibfield  {journal} {\bibinfo
  {journal} {Review of Scientific Instruments}\ }\textbf {\bibinfo {volume}
  {84}},\ \bibinfo {pages} {083906} (\bibinfo {year} {2013})}\BibitemShut
  {NoStop}%
\bibitem [{\citenamefont {Park}\ \emph {et~al.}(2002)\citenamefont {Park},
  \citenamefont {Adroja}, \citenamefont {McEwen},\ and\ \citenamefont
  {Murani}}]{Park_2002}%
  \BibitemOpen
  \bibfield  {author} {\bibinfo {author} {\bibfnamefont {J.-G.}\ \bibnamefont
  {Park}}, \bibinfo {author} {\bibfnamefont {D.~T.}\ \bibnamefont {Adroja}},
  \bibinfo {author} {\bibfnamefont {K.~A.}\ \bibnamefont {McEwen}},\ and\
  \bibinfo {author} {\bibfnamefont {A.~P.}\ \bibnamefont {Murani}},\ }\href
  {https://doi.org/10.1088/0953-8984/14/15/302} {\bibfield  {journal} {\bibinfo
   {journal} {Journal of Physics: Condensed Matter}\ }\textbf {\bibinfo
  {volume} {14}},\ \bibinfo {pages} {3865} (\bibinfo {year}
  {2002})}\BibitemShut {NoStop}%
\bibitem [{\citenamefont {So}\ \emph {et~al.}(2002)\citenamefont {So},
  \citenamefont {Park}, \citenamefont {Adroja}, \citenamefont {McEwen},\ and\
  \citenamefont {Oh}}]{SO2002472}%
  \BibitemOpen
  \bibfield  {author} {\bibinfo {author} {\bibfnamefont {J.-Y.}\ \bibnamefont
  {So}}, \bibinfo {author} {\bibfnamefont {J.-G.}\ \bibnamefont {Park}},
  \bibinfo {author} {\bibfnamefont {D.}~\bibnamefont {Adroja}}, \bibinfo
  {author} {\bibfnamefont {K.}~\bibnamefont {McEwen}},\ and\ \bibinfo {author}
  {\bibfnamefont {S.-J.}\ \bibnamefont {Oh}},\ }\href
  {https://doi.org/https://doi.org/10.1016/S0921-4526(01)01326-6} {\bibfield
  {journal} {\bibinfo  {journal} {Physica B: Condensed Matter}\ }\textbf
  {\bibinfo {volume} {312-313}},\ \bibinfo {pages} {472} (\bibinfo {year}
  {2002})}\BibitemShut {NoStop}%
\bibitem [{\citenamefont {Adroja}\ \emph {et~al.}(2007)\citenamefont {Adroja},
  \citenamefont {Park}, \citenamefont {Jang}, \citenamefont {Walker},
  \citenamefont {McEwen},\ and\ \citenamefont {Takabatake}}]{ADROJA2007858}%
  \BibitemOpen
  \bibfield  {author} {\bibinfo {author} {\bibfnamefont {D.}~\bibnamefont
  {Adroja}}, \bibinfo {author} {\bibfnamefont {J.-G.}\ \bibnamefont {Park}},
  \bibinfo {author} {\bibfnamefont {K.-H.}\ \bibnamefont {Jang}}, \bibinfo
  {author} {\bibfnamefont {H.}~\bibnamefont {Walker}}, \bibinfo {author}
  {\bibfnamefont {K.}~\bibnamefont {McEwen}},\ and\ \bibinfo {author}
  {\bibfnamefont {T.}~\bibnamefont {Takabatake}},\ }\href
  {https://doi.org/https://doi.org/10.1016/j.jmmm.2006.10.1140} {\bibfield
  {journal} {\bibinfo  {journal} {J. Magn. Magn. Mater}\ }\textbf {\bibinfo
  {volume} {310}},\ \bibinfo {pages} {858} (\bibinfo {year}
  {2007})}\BibitemShut {NoStop}%
\bibitem [{\citenamefont {Helton}\ \emph {et~al.}(2010)\citenamefont {Helton},
  \citenamefont {Matan}, \citenamefont {Shores}, \citenamefont {Nytko},
  \citenamefont {Bartlett}, \citenamefont {Qiu}, \citenamefont {Nocera},\ and\
  \citenamefont {Lee}}]{PhysRevLett.104.147201}%
  \BibitemOpen
  \bibfield  {author} {\bibinfo {author} {\bibfnamefont {J.~S.}\ \bibnamefont
  {Helton}}, \bibinfo {author} {\bibfnamefont {K.}~\bibnamefont {Matan}},
  \bibinfo {author} {\bibfnamefont {M.~P.}\ \bibnamefont {Shores}}, \bibinfo
  {author} {\bibfnamefont {E.~A.}\ \bibnamefont {Nytko}}, \bibinfo {author}
  {\bibfnamefont {B.~M.}\ \bibnamefont {Bartlett}}, \bibinfo {author}
  {\bibfnamefont {Y.}~\bibnamefont {Qiu}}, \bibinfo {author} {\bibfnamefont
  {D.~G.}\ \bibnamefont {Nocera}},\ and\ \bibinfo {author} {\bibfnamefont
  {Y.~S.}\ \bibnamefont {Lee}},\ }\href
  {https://doi.org/10.1103/PhysRevLett.104.147201} {\bibfield  {journal}
  {\bibinfo  {journal} {Phys. Rev. Lett.}\ }\textbf {\bibinfo {volume} {104}},\
  \bibinfo {pages} {147201} (\bibinfo {year} {2010})}\BibitemShut {NoStop}%
\bibitem [{\citenamefont {Schr{\"o}der}\ \emph {et~al.}(2000)\citenamefont
  {Schr{\"o}der}, \citenamefont {Aeppli}, \citenamefont {Coldea}, \citenamefont
  {Adams}, \citenamefont {Stockert}, \citenamefont {L{\"o}hneysen},
  \citenamefont {Bucher}, \citenamefont {Ramazashvili},\ and\ \citenamefont
  {Coleman}}]{Schroder2000}%
  \BibitemOpen
  \bibfield  {author} {\bibinfo {author} {\bibfnamefont {A.}~\bibnamefont
  {Schr{\"o}der}}, \bibinfo {author} {\bibfnamefont {G.}~\bibnamefont
  {Aeppli}}, \bibinfo {author} {\bibfnamefont {R.}~\bibnamefont {Coldea}},
  \bibinfo {author} {\bibfnamefont {M.}~\bibnamefont {Adams}}, \bibinfo
  {author} {\bibfnamefont {O.}~\bibnamefont {Stockert}}, \bibinfo {author}
  {\bibfnamefont {H.}~\bibnamefont {L{\"o}hneysen}}, \bibinfo {author}
  {\bibfnamefont {E.}~\bibnamefont {Bucher}}, \bibinfo {author} {\bibfnamefont
  {R.}~\bibnamefont {Ramazashvili}},\ and\ \bibinfo {author} {\bibfnamefont
  {P.}~\bibnamefont {Coleman}},\ }\href {https://doi.org/10.1038/35030039}
  {\bibfield  {journal} {\bibinfo  {journal} {Nature}\ }\textbf {\bibinfo
  {volume} {407}},\ \bibinfo {pages} {351} (\bibinfo {year}
  {2000})}\BibitemShut {NoStop}%
\bibitem [{\citenamefont {Wilson}\ \emph {et~al.}(2005)\citenamefont {Wilson},
  \citenamefont {Dai}, \citenamefont {Adroja}, \citenamefont {Lee},
  \citenamefont {Chung}, \citenamefont {Lynn}, \citenamefont {Butch},\ and\
  \citenamefont {Maple}}]{PhysRevLett.94.056402}%
  \BibitemOpen
  \bibfield  {author} {\bibinfo {author} {\bibfnamefont {S.~D.}\ \bibnamefont
  {Wilson}}, \bibinfo {author} {\bibfnamefont {P.}~\bibnamefont {Dai}},
  \bibinfo {author} {\bibfnamefont {D.~T.}\ \bibnamefont {Adroja}}, \bibinfo
  {author} {\bibfnamefont {S.-H.}\ \bibnamefont {Lee}}, \bibinfo {author}
  {\bibfnamefont {J.-H.}\ \bibnamefont {Chung}}, \bibinfo {author}
  {\bibfnamefont {J.~W.}\ \bibnamefont {Lynn}}, \bibinfo {author}
  {\bibfnamefont {N.~P.}\ \bibnamefont {Butch}},\ and\ \bibinfo {author}
  {\bibfnamefont {M.~B.}\ \bibnamefont {Maple}},\ }\href
  {https://doi.org/10.1103/PhysRevLett.94.056402} {\bibfield  {journal}
  {\bibinfo  {journal} {Phys. Rev. Lett.}\ }\textbf {\bibinfo {volume} {94}},\
  \bibinfo {pages} {056402} (\bibinfo {year} {2005})}\BibitemShut {NoStop}%
\bibitem [{\citenamefont {Schr\"oder}\ \emph {et~al.}(1998)\citenamefont
  {Schr\"oder}, \citenamefont {Aeppli}, \citenamefont {Bucher}, \citenamefont
  {Ramazashvili},\ and\ \citenamefont {Coleman}}]{PhysRevLett.80.5623}%
  \BibitemOpen
  \bibfield  {author} {\bibinfo {author} {\bibfnamefont {A.}~\bibnamefont
  {Schr\"oder}}, \bibinfo {author} {\bibfnamefont {G.}~\bibnamefont {Aeppli}},
  \bibinfo {author} {\bibfnamefont {E.}~\bibnamefont {Bucher}}, \bibinfo
  {author} {\bibfnamefont {R.}~\bibnamefont {Ramazashvili}},\ and\ \bibinfo
  {author} {\bibfnamefont {P.}~\bibnamefont {Coleman}},\ }\href
  {https://doi.org/10.1103/PhysRevLett.80.5623} {\bibfield  {journal} {\bibinfo
   {journal} {Phys. Rev. Lett.}\ }\textbf {\bibinfo {volume} {80}},\ \bibinfo
  {pages} {5623} (\bibinfo {year} {1998})}\BibitemShut {NoStop}%
\bibitem [{\citenamefont {Tran}\ \emph {et~al.}(2012)\citenamefont {Tran},
  \citenamefont {Hillier}, \citenamefont {Adroja},\ and\ \citenamefont
  {Kaczorowski}}]{PhysRevB.86.094525}%
  \BibitemOpen
  \bibfield  {author} {\bibinfo {author} {\bibfnamefont {V.~H.}\ \bibnamefont
  {Tran}}, \bibinfo {author} {\bibfnamefont {A.~D.}\ \bibnamefont {Hillier}},
  \bibinfo {author} {\bibfnamefont {D.~T.}\ \bibnamefont {Adroja}},\ and\
  \bibinfo {author} {\bibfnamefont {D.}~\bibnamefont {Kaczorowski}},\ }\href
  {https://doi.org/10.1103/PhysRevB.86.094525} {\bibfield  {journal} {\bibinfo
  {journal} {Phys. Rev. B}\ }\textbf {\bibinfo {volume} {86}},\ \bibinfo
  {pages} {094525} (\bibinfo {year} {2012})}\BibitemShut {NoStop}%
\bibitem [{\citenamefont {Watanabe}\ and\ \citenamefont
  {Miyake}(2010)}]{PhysRevLett.105.186403}%
  \BibitemOpen
  \bibfield  {author} {\bibinfo {author} {\bibfnamefont {S.}~\bibnamefont
  {Watanabe}}\ and\ \bibinfo {author} {\bibfnamefont {K.}~\bibnamefont
  {Miyake}},\ }\href {https://doi.org/10.1103/PhysRevLett.105.186403}
  {\bibfield  {journal} {\bibinfo  {journal} {Phys. Rev. Lett.}\ }\textbf
  {\bibinfo {volume} {105}},\ \bibinfo {pages} {186403} (\bibinfo {year}
  {2010})}\BibitemShut {NoStop}%
\bibitem [{\citenamefont {Pixley}\ \emph {et~al.}(2012)\citenamefont {Pixley},
  \citenamefont {Kirchner}, \citenamefont {Ingersent},\ and\ \citenamefont
  {Si}}]{PhysRevLett.109.086403}%
  \BibitemOpen
  \bibfield  {author} {\bibinfo {author} {\bibfnamefont {J.~H.}\ \bibnamefont
  {Pixley}}, \bibinfo {author} {\bibfnamefont {S.}~\bibnamefont {Kirchner}},
  \bibinfo {author} {\bibfnamefont {K.}~\bibnamefont {Ingersent}},\ and\
  \bibinfo {author} {\bibfnamefont {Q.}~\bibnamefont {Si}},\ }\bibfield
  {title} {\bibinfo {title} {Kondo destruction and valence fluctuations in an
  anderson model},\ }\href {https://doi.org/10.1103/PhysRevLett.109.086403}
  {\bibfield  {journal} {\bibinfo  {journal} {Phys. Rev. Lett.}\ }\textbf
  {\bibinfo {volume} {109}},\ \bibinfo {pages} {086403} (\bibinfo {year}
  {2012})}\BibitemShut {NoStop}%
\bibitem [{\citenamefont {Cai}\ \emph {et~al.}(2020)\citenamefont {Cai},
  \citenamefont {Yu}, \citenamefont {Hu}, \citenamefont {Kirchner},\ and\
  \citenamefont {Si}}]{PhysRevLett.124.027205}%
  \BibitemOpen
  \bibfield  {author} {\bibinfo {author} {\bibfnamefont {A.}~\bibnamefont
  {Cai}}, \bibinfo {author} {\bibfnamefont {Z.}~\bibnamefont {Yu}}, \bibinfo
  {author} {\bibfnamefont {H.}~\bibnamefont {Hu}}, \bibinfo {author}
  {\bibfnamefont {S.}~\bibnamefont {Kirchner}},\ and\ \bibinfo {author}
  {\bibfnamefont {Q.}~\bibnamefont {Si}},\ }\href
  {https://doi.org/10.1103/PhysRevLett.124.027205} {\bibfield  {journal}
  {\bibinfo  {journal} {Phys. Rev. Lett.}\ }\textbf {\bibinfo {volume} {124}},\
  \bibinfo {pages} {027205} (\bibinfo {year} {2020})}\BibitemShut {NoStop}%
\bibitem [{\citenamefont {Zhu}\ \emph {et~al.}(2003)\citenamefont {Zhu},
  \citenamefont {Grempel},\ and\ \citenamefont {Si}}]{PhysRevLett.91.156404}%
  \BibitemOpen
  \bibfield  {author} {\bibinfo {author} {\bibfnamefont {J.-X.}\ \bibnamefont
  {Zhu}}, \bibinfo {author} {\bibfnamefont {D.~R.}\ \bibnamefont {Grempel}},\
  and\ \bibinfo {author} {\bibfnamefont {Q.}~\bibnamefont {Si}},\ }\href
  {https://doi.org/10.1103/PhysRevLett.91.156404} {\bibfield  {journal}
  {\bibinfo  {journal} {Phys. Rev. Lett.}\ }\textbf {\bibinfo {volume} {91}},\
  \bibinfo {pages} {156404} (\bibinfo {year} {2003})}\BibitemShut {NoStop}%
\bibitem [{\citenamefont {Grempel}\ and\ \citenamefont
  {Si}(2003)}]{PhysRevLett.91.026401}%
  \BibitemOpen
  \bibfield  {author} {\bibinfo {author} {\bibfnamefont {D.~R.}\ \bibnamefont
  {Grempel}}\ and\ \bibinfo {author} {\bibfnamefont {Q.}~\bibnamefont {Si}},\
  }\href {https://doi.org/10.1103/PhysRevLett.91.026401} {\bibfield  {journal}
  {\bibinfo  {journal} {Phys. Rev. Lett.}\ }\textbf {\bibinfo {volume} {91}},\
  \bibinfo {pages} {026401} (\bibinfo {year} {2003})}\BibitemShut {NoStop}%
\bibitem [{\citenamefont {Adroja}\ \emph
  {et~al.}(2003{\natexlab{a}})\citenamefont {Adroja}, \citenamefont {Park},
  \citenamefont {McEwen},\ and\ \citenamefont {Takabatake}}]{AdrojaMARI}%
  \BibitemOpen
  \bibfield  {author} {\bibinfo {author} {\bibfnamefont {D.~T.}\ \bibnamefont
  {Adroja}}, \bibinfo {author} {\bibfnamefont {J.-G.}\ \bibnamefont {Park}},
  \bibinfo {author} {\bibfnamefont {K.~A.}\ \bibnamefont {McEwen}},\ and\
  \bibinfo {author} {\bibfnamefont {T.}~\bibnamefont {Takabatake}},\ }\href
  {https://doi.org/doi.org/10.5286/ISIS.E.RB14460} {\bibfield  {journal}
  {\bibinfo  {journal} {ISIS experimental report}\ } (\bibinfo {year}
  {2003}{\natexlab{a}})}\BibitemShut {NoStop}%
\bibitem [{\citenamefont {Adroja}\ \emph
  {et~al.}(2003{\natexlab{b}})\citenamefont {Adroja}, \citenamefont {Park},
  \citenamefont {McEwen}, \citenamefont {Takeda}, \citenamefont {Ishikawa},\
  and\ \citenamefont {So}}]{PhysRevB.68.094425}%
  \BibitemOpen
  \bibfield  {author} {\bibinfo {author} {\bibfnamefont {D.~T.}\ \bibnamefont
  {Adroja}}, \bibinfo {author} {\bibfnamefont {J.-G.}\ \bibnamefont {Park}},
  \bibinfo {author} {\bibfnamefont {K.~A.}\ \bibnamefont {McEwen}}, \bibinfo
  {author} {\bibfnamefont {N.}~\bibnamefont {Takeda}}, \bibinfo {author}
  {\bibfnamefont {M.}~\bibnamefont {Ishikawa}},\ and\ \bibinfo {author}
  {\bibfnamefont {J.-Y.}\ \bibnamefont {So}},\ }\href
  {https://doi.org/10.1103/PhysRevB.68.094425} {\bibfield  {journal} {\bibinfo
  {journal} {Phys. Rev. B}\ }\textbf {\bibinfo {volume} {68}},\ \bibinfo
  {pages} {094425} (\bibinfo {year} {2003}{\natexlab{b}})}\BibitemShut
  {NoStop}%
\bibitem [{\citenamefont {Stevens}(1952)}]{Stevens_1952}%
  \BibitemOpen
  \bibfield  {author} {\bibinfo {author} {\bibfnamefont {K.~W.~H.}\
  \bibnamefont {Stevens}},\ }\href {https://doi.org/10.1088/0370-1298/65/3/308}
  {\bibfield  {journal} {\bibinfo  {journal} {Proceedings of the Physical
  Society. Section A}\ }\textbf {\bibinfo {volume} {65}},\ \bibinfo {pages}
  {209} (\bibinfo {year} {1952})}\BibitemShut {NoStop}%
\bibitem [{\citenamefont {Arnold}\ \emph
  {et~al.}(2014{\natexlab{b}})\citenamefont {Arnold}, \citenamefont {Bilheux},
  \citenamefont {Borreguero}, \citenamefont {Buts}, \citenamefont {Campbell},
  \citenamefont {Chapon}, \citenamefont {Doucet}, \citenamefont {Draper},
  \citenamefont {{Ferraz Leal}}, \citenamefont {Gigg}, \citenamefont {Lynch},
  \citenamefont {Markvardsen}, \citenamefont {Mikkelson}, \citenamefont
  {Mikkelson}, \citenamefont {Miller}, \citenamefont {Palmen}, \citenamefont
  {Parker}, \citenamefont {Passos}, \citenamefont {Perring}, \citenamefont
  {Peterson}, \citenamefont {Ren}, \citenamefont {Reuter}, \citenamefont
  {Savici}, \citenamefont {Taylor}, \citenamefont {Taylor}, \citenamefont
  {Tolchenov}, \citenamefont {Zhou},\ and\ \citenamefont
  {Zikovsky}}]{MantidCEF}%
  \BibitemOpen
  \bibfield  {author} {\bibinfo {author} {\bibfnamefont {O.}~\bibnamefont
  {Arnold}}, \bibinfo {author} {\bibfnamefont {J.~C.}\ \bibnamefont {Bilheux}},
  \bibinfo {author} {\bibfnamefont {J.~M.}\ \bibnamefont {Borreguero}},
  \bibinfo {author} {\bibfnamefont {A.}~\bibnamefont {Buts}}, \bibinfo {author}
  {\bibfnamefont {S.~I.}\ \bibnamefont {Campbell}}, \bibinfo {author}
  {\bibfnamefont {L.}~\bibnamefont {Chapon}}, \bibinfo {author} {\bibfnamefont
  {M.}~\bibnamefont {Doucet}}, \bibinfo {author} {\bibfnamefont
  {N.}~\bibnamefont {Draper}}, \bibinfo {author} {\bibfnamefont
  {R.}~\bibnamefont {{Ferraz Leal}}}, \bibinfo {author} {\bibfnamefont {M.~A.}\
  \bibnamefont {Gigg}}, \bibinfo {author} {\bibfnamefont {V.~E.}\ \bibnamefont
  {Lynch}}, \bibinfo {author} {\bibfnamefont {A.}~\bibnamefont {Markvardsen}},
  \bibinfo {author} {\bibfnamefont {D.~J.}\ \bibnamefont {Mikkelson}}, \bibinfo
  {author} {\bibfnamefont {R.~L.}\ \bibnamefont {Mikkelson}}, \bibinfo {author}
  {\bibfnamefont {R.}~\bibnamefont {Miller}}, \bibinfo {author} {\bibfnamefont
  {K.}~\bibnamefont {Palmen}}, \bibinfo {author} {\bibfnamefont
  {P.}~\bibnamefont {Parker}}, \bibinfo {author} {\bibfnamefont
  {G.}~\bibnamefont {Passos}}, \bibinfo {author} {\bibfnamefont {T.~G.}\
  \bibnamefont {Perring}}, \bibinfo {author} {\bibfnamefont {P.~F.}\
  \bibnamefont {Peterson}}, \bibinfo {author} {\bibfnamefont {S.}~\bibnamefont
  {Ren}}, \bibinfo {author} {\bibfnamefont {M.~A.}\ \bibnamefont {Reuter}},
  \bibinfo {author} {\bibfnamefont {A.~T.}\ \bibnamefont {Savici}}, \bibinfo
  {author} {\bibfnamefont {J.~W.}\ \bibnamefont {Taylor}}, \bibinfo {author}
  {\bibfnamefont {R.~J.}\ \bibnamefont {Taylor}}, \bibinfo {author}
  {\bibfnamefont {R.}~\bibnamefont {Tolchenov}}, \bibinfo {author}
  {\bibfnamefont {W.}~\bibnamefont {Zhou}},\ and\ \bibinfo {author}
  {\bibfnamefont {J.}~\bibnamefont {Zikovsky}},\ }\href
  {https://doi.org/https://doi.org/10.1016/j.nima.2014.07.029} {\bibfield
  {journal} {\bibinfo  {journal} {Nucl. Instrum. Meth. A}\ }\textbf {\bibinfo
  {volume} {764}},\ \bibinfo {pages} {156} (\bibinfo {year}
  {2014}{\natexlab{b}})}\BibitemShut {NoStop}%
\bibitem [{\citenamefont {Isikawa}\ \emph {et~al.}(1996)\citenamefont
  {Isikawa}, \citenamefont {Mizushima}, \citenamefont {Fukushima},
  \citenamefont {Kuwai}, \citenamefont {Sakurai},\ and\ \citenamefont
  {Kitzawa}}]{Isikawa}%
  \BibitemOpen
  \bibfield  {author} {\bibinfo {author} {\bibfnamefont {Y.}~\bibnamefont
  {Isikawa}}, \bibinfo {author} {\bibfnamefont {T.}~\bibnamefont {Mizushima}},
  \bibinfo {author} {\bibfnamefont {N.}~\bibnamefont {Fukushima}}, \bibinfo
  {author} {\bibfnamefont {T.}~\bibnamefont {Kuwai}}, \bibinfo {author}
  {\bibfnamefont {J.}~\bibnamefont {Sakurai}},\ and\ \bibinfo {author}
  {\bibfnamefont {H.}~\bibnamefont {Kitzawa}},\ }\href@noop {} {\bibfield
  {journal} {\bibinfo  {journal} {J. Phys. Soc. Jpn.}\ }\textbf {\bibinfo
  {volume} {65 Suppl. B}},\ \bibinfo {pages} {117} (\bibinfo {year}
  {1996})}\BibitemShut {NoStop}%
\bibitem [{\citenamefont
  {Rodríguez-Carvajal}(1993)}]{RODRIGUEZCARVAJAL199355}%
  \BibitemOpen
  \bibfield  {author} {\bibinfo {author} {\bibfnamefont {J.}~\bibnamefont
  {Rodríguez-Carvajal}},\ }\href
  {https://doi.org/https://doi.org/10.1016/0921-4526(93)90108-I} {\bibfield
  {journal} {\bibinfo  {journal} {Physica B: Condensed Matter}\ }\textbf
  {\bibinfo {volume} {192}},\ \bibinfo {pages} {55} (\bibinfo {year}
  {1993})}\BibitemShut {NoStop}%
\bibitem [{\citenamefont {Rodriguez-Carvajal}()}]{RodriguezFullProf}%
  \BibitemOpen
  \bibfield  {author} {\bibinfo {author} {\bibfnamefont {J.}~\bibnamefont
  {Rodriguez-Carvajal}},\ }\href {https://www.ill.eu/sites/fullprof/} {\emph
  {\bibinfo {title} {BASIREPS: a program for calculating irreducible
  representations of space groups and basis functions for axial and polar
  vector properties}}}\BibitemShut {NoStop}%
\bibitem [{\citenamefont {Ritter}(2011)}]{ritter2011}%
  \BibitemOpen
  \bibfield  {author} {\bibinfo {author} {\bibfnamefont {C.}~\bibnamefont
  {Ritter}},\ }\href {https://doi.org/10.4028/www.scientific.net/SSP.170.263}
  {\bibfield  {journal} {\bibinfo  {journal} {Sol. State. Phenom}\ }\textbf
  {\bibinfo {volume} {170}},\ \bibinfo {pages} {263} (\bibinfo {year}
  {2011})}\BibitemShut {NoStop}%
\bibitem [{\citenamefont {Dönni}\ \emph
  {et~al.}(1996{\natexlab{a}})\citenamefont {Dönni}, \citenamefont {Ehlers},
  \citenamefont {Maletta}, \citenamefont {Fischer}, \citenamefont {Kitazawa},\
  and\ \citenamefont {Zolliker}}]{Donni1996GeometricallyFM}%
  \BibitemOpen
  \bibfield  {author} {\bibinfo {author} {\bibfnamefont {A.}~\bibnamefont
  {Dönni}}, \bibinfo {author} {\bibfnamefont {G.}~\bibnamefont {Ehlers}},
  \bibinfo {author} {\bibfnamefont {H.}~\bibnamefont {Maletta}}, \bibinfo
  {author} {\bibfnamefont {P.}~\bibnamefont {Fischer}}, \bibinfo {author}
  {\bibfnamefont {H.}~\bibnamefont {Kitazawa}},\ and\ \bibinfo {author}
  {\bibfnamefont {M.}~\bibnamefont {Zolliker}},\ }\href
  {https://doi.org/10.1088/0953-8984/8/50/043} {\bibfield  {journal} {\bibinfo
  {journal} {Journal of Physics: Condensed Matter}\ }\textbf {\bibinfo {volume}
  {8}},\ \bibinfo {pages} {11213} (\bibinfo {year}
  {1996}{\natexlab{a}})}\BibitemShut {NoStop}%
\bibitem [{\citenamefont {Dönni}\ \emph
  {et~al.}(1996{\natexlab{b}})\citenamefont {Dönni}, \citenamefont {Ehlers},
  \citenamefont {Maletta}, \citenamefont {Fischer}, \citenamefont {Kitazawa},\
  and\ \citenamefont {Zolliker}}]{D_nni_1996}%
  \BibitemOpen
  \bibfield  {author} {\bibinfo {author} {\bibfnamefont {A.}~\bibnamefont
  {Dönni}}, \bibinfo {author} {\bibfnamefont {G.}~\bibnamefont {Ehlers}},
  \bibinfo {author} {\bibfnamefont {H.}~\bibnamefont {Maletta}}, \bibinfo
  {author} {\bibfnamefont {P.}~\bibnamefont {Fischer}}, \bibinfo {author}
  {\bibfnamefont {H.}~\bibnamefont {Kitazawa}},\ and\ \bibinfo {author}
  {\bibfnamefont {M.}~\bibnamefont {Zolliker}},\ }\href
  {https://doi.org/10.1088/0953-8984/8/50/043} {\bibfield  {journal} {\bibinfo
  {journal} {Journal of Physics: Condensed Matter}\ }\textbf {\bibinfo {volume}
  {8}},\ \bibinfo {pages} {11213} (\bibinfo {year}
  {1996}{\natexlab{b}})}\BibitemShut {NoStop}%
\bibitem [{\citenamefont {Fritsch}\ \emph {et~al.}(2014)\citenamefont
  {Fritsch}, \citenamefont {Bagrets}, \citenamefont {Goll}, \citenamefont
  {Kittler}, \citenamefont {Wolf}, \citenamefont {Grube}, \citenamefont
  {Huang},\ and\ \citenamefont {L\"ohneysen}}]{PhysRevB.89.054416}%
  \BibitemOpen
  \bibfield  {author} {\bibinfo {author} {\bibfnamefont {V.}~\bibnamefont
  {Fritsch}}, \bibinfo {author} {\bibfnamefont {N.}~\bibnamefont {Bagrets}},
  \bibinfo {author} {\bibfnamefont {G.}~\bibnamefont {Goll}}, \bibinfo {author}
  {\bibfnamefont {W.}~\bibnamefont {Kittler}}, \bibinfo {author} {\bibfnamefont
  {M.~J.}\ \bibnamefont {Wolf}}, \bibinfo {author} {\bibfnamefont
  {K.}~\bibnamefont {Grube}}, \bibinfo {author} {\bibfnamefont {C.-L.}\
  \bibnamefont {Huang}},\ and\ \bibinfo {author} {\bibfnamefont {H.~v.}\
  \bibnamefont {L\"ohneysen}},\ }\href
  {https://doi.org/10.1103/PhysRevB.89.054416} {\bibfield  {journal} {\bibinfo
  {journal} {Phys. Rev. B}\ }\textbf {\bibinfo {volume} {89}},\ \bibinfo
  {pages} {054416} (\bibinfo {year} {2014})}\BibitemShut {NoStop}%
\bibitem [{\citenamefont {Keller}\ \emph {et~al.}(1998)\citenamefont {Keller},
  \citenamefont {Dönni}, \citenamefont {Kitazawa}, \citenamefont {Tang},
  \citenamefont {Fauth},\ and\ \citenamefont {Zolliker}}]{KELLER1997660}%
  \BibitemOpen
  \bibfield  {author} {\bibinfo {author} {\bibfnamefont {L.}~\bibnamefont
  {Keller}}, \bibinfo {author} {\bibfnamefont {A.}~\bibnamefont {Dönni}},
  \bibinfo {author} {\bibfnamefont {H.}~\bibnamefont {Kitazawa}}, \bibinfo
  {author} {\bibfnamefont {J.}~\bibnamefont {Tang}}, \bibinfo {author}
  {\bibfnamefont {F.}~\bibnamefont {Fauth}},\ and\ \bibinfo {author}
  {\bibfnamefont {M.}~\bibnamefont {Zolliker}},\ }\href
  {https://doi.org/https://doi.org/10.1016/S0921-4526(97)00684-4} {\bibfield
  {journal} {\bibinfo  {journal} {Physica B: Condensed Matter}\ }\textbf
  {\bibinfo {volume} {241-243}},\ \bibinfo {pages} {660} (\bibinfo {year}
  {1998})}\BibitemShut {NoStop}%
\bibitem [{\citenamefont {Keller}\ \emph {et~al.}(2000)\citenamefont {Keller},
  \citenamefont {Dönni},\ and\ \citenamefont {Kitazawa}}]{KELLER2000672}%
  \BibitemOpen
  \bibfield  {author} {\bibinfo {author} {\bibfnamefont {L.}~\bibnamefont
  {Keller}}, \bibinfo {author} {\bibfnamefont {A.}~\bibnamefont {Dönni}},\
  and\ \bibinfo {author} {\bibfnamefont {H.}~\bibnamefont {Kitazawa}},\ }\href
  {https://doi.org/https://doi.org/10.1016/S0921-4526(99)01400-3} {\bibfield
  {journal} {\bibinfo  {journal} {Physica B: Condensed Matter}\ }\textbf
  {\bibinfo {volume} {276-278}},\ \bibinfo {pages} {672} (\bibinfo {year}
  {2000})}\BibitemShut {NoStop}%
\bibitem [{\citenamefont {Mihalik}\ \emph {et~al.}(2011)\citenamefont
  {Mihalik}, \citenamefont {Prokle\ifmmode~\check{s}\else \v{s}\fi{}ka},
  \citenamefont {Kamar\'ad}, \citenamefont {Proke\ifmmode~\check{s}\else
  \v{s}\fi{}}, \citenamefont {Isnard}, \citenamefont {McIntyre}, \citenamefont
  {D\"onni}, \citenamefont {Yoshii}, \citenamefont {Kitazawa}, \citenamefont
  {Sechovsk\'y},\ and\ \citenamefont {de~Boer}}]{PhysRevB.83.104403}%
  \BibitemOpen
  \bibfield  {author} {\bibinfo {author} {\bibfnamefont {M.}~\bibnamefont
  {Mihalik}}, \bibinfo {author} {\bibfnamefont {J.}~\bibnamefont
  {Prokle\ifmmode~\check{s}\else \v{s}\fi{}ka}}, \bibinfo {author}
  {\bibfnamefont {J.}~\bibnamefont {Kamar\'ad}}, \bibinfo {author}
  {\bibfnamefont {K.}~\bibnamefont {Proke\ifmmode~\check{s}\else \v{s}\fi{}}},
  \bibinfo {author} {\bibfnamefont {O.}~\bibnamefont {Isnard}}, \bibinfo
  {author} {\bibfnamefont {G.~J.}\ \bibnamefont {McIntyre}}, \bibinfo {author}
  {\bibfnamefont {A.}~\bibnamefont {D\"onni}}, \bibinfo {author} {\bibfnamefont
  {S.}~\bibnamefont {Yoshii}}, \bibinfo {author} {\bibfnamefont
  {H.}~\bibnamefont {Kitazawa}}, \bibinfo {author} {\bibfnamefont
  {V.}~\bibnamefont {Sechovsk\'y}},\ and\ \bibinfo {author} {\bibfnamefont
  {F.~R.}\ \bibnamefont {de~Boer}},\ }\href
  {https://doi.org/10.1103/PhysRevB.83.104403} {\bibfield  {journal} {\bibinfo
  {journal} {Phys. Rev. B}\ }\textbf {\bibinfo {volume} {83}},\ \bibinfo
  {pages} {104403} (\bibinfo {year} {2011})}\BibitemShut {NoStop}%
\bibitem [{\citenamefont {Helton}\ \emph {et~al.}(2007)\citenamefont {Helton},
  \citenamefont {Matan}, \citenamefont {Shores}, \citenamefont {Nytko},
  \citenamefont {Bartlett}, \citenamefont {Yoshida}, \citenamefont {Takano},
  \citenamefont {Suslov}, \citenamefont {Qiu}, \citenamefont {Chung},
  \citenamefont {Nocera},\ and\ \citenamefont {Lee}}]{PhysRevLett.98.107204}%
  \BibitemOpen
  \bibfield  {author} {\bibinfo {author} {\bibfnamefont {J.~S.}\ \bibnamefont
  {Helton}}, \bibinfo {author} {\bibfnamefont {K.}~\bibnamefont {Matan}},
  \bibinfo {author} {\bibfnamefont {M.~P.}\ \bibnamefont {Shores}}, \bibinfo
  {author} {\bibfnamefont {E.~A.}\ \bibnamefont {Nytko}}, \bibinfo {author}
  {\bibfnamefont {B.~M.}\ \bibnamefont {Bartlett}}, \bibinfo {author}
  {\bibfnamefont {Y.}~\bibnamefont {Yoshida}}, \bibinfo {author} {\bibfnamefont
  {Y.}~\bibnamefont {Takano}}, \bibinfo {author} {\bibfnamefont
  {A.}~\bibnamefont {Suslov}}, \bibinfo {author} {\bibfnamefont
  {Y.}~\bibnamefont {Qiu}}, \bibinfo {author} {\bibfnamefont {J.-H.}\
  \bibnamefont {Chung}}, \bibinfo {author} {\bibfnamefont {D.~G.}\ \bibnamefont
  {Nocera}},\ and\ \bibinfo {author} {\bibfnamefont {Y.~S.}\ \bibnamefont
  {Lee}},\ }\href {https://doi.org/10.1103/PhysRevLett.98.107204} {\bibfield
  {journal} {\bibinfo  {journal} {Phys. Rev. Lett.}\ }\textbf {\bibinfo
  {volume} {98}},\ \bibinfo {pages} {107204} (\bibinfo {year}
  {2007})}\BibitemShut {NoStop}%
\bibitem [{\citenamefont {Choi}\ \emph {et~al.}(2019)\citenamefont {Choi},
  \citenamefont {Lee}, \citenamefont {Lee}, \citenamefont {Yoon}, \citenamefont
  {Lee}, \citenamefont {Park}, \citenamefont {Ali}, \citenamefont {Singh},
  \citenamefont {Orain}, \citenamefont {Kim}, \citenamefont {Rhyee},
  \citenamefont {Chen}, \citenamefont {Chou},\ and\ \citenamefont
  {Choi}}]{PhysRevLett.122.167202}%
  \BibitemOpen
  \bibfield  {author} {\bibinfo {author} {\bibfnamefont {Y.~S.}\ \bibnamefont
  {Choi}}, \bibinfo {author} {\bibfnamefont {C.~H.}\ \bibnamefont {Lee}},
  \bibinfo {author} {\bibfnamefont {S.}~\bibnamefont {Lee}}, \bibinfo {author}
  {\bibfnamefont {S.}~\bibnamefont {Yoon}}, \bibinfo {author} {\bibfnamefont
  {W.-J.}\ \bibnamefont {Lee}}, \bibinfo {author} {\bibfnamefont
  {J.}~\bibnamefont {Park}}, \bibinfo {author} {\bibfnamefont {A.}~\bibnamefont
  {Ali}}, \bibinfo {author} {\bibfnamefont {Y.}~\bibnamefont {Singh}}, \bibinfo
  {author} {\bibfnamefont {J.-C.}\ \bibnamefont {Orain}}, \bibinfo {author}
  {\bibfnamefont {G.}~\bibnamefont {Kim}}, \bibinfo {author} {\bibfnamefont
  {J.-S.}\ \bibnamefont {Rhyee}}, \bibinfo {author} {\bibfnamefont {W.-T.}\
  \bibnamefont {Chen}}, \bibinfo {author} {\bibfnamefont {F.}~\bibnamefont
  {Chou}},\ and\ \bibinfo {author} {\bibfnamefont {K.-Y.}\ \bibnamefont
  {Choi}},\ }\href {https://doi.org/10.1103/PhysRevLett.122.167202} {\bibfield
  {journal} {\bibinfo  {journal} {Phys. Rev. Lett.}\ }\textbf {\bibinfo
  {volume} {122}},\ \bibinfo {pages} {167202} (\bibinfo {year}
  {2019})}\BibitemShut {NoStop}%
\bibitem [{\citenamefont {Lee}\ \emph {et~al.}(2017)\citenamefont {Lee},
  \citenamefont {Do}, \citenamefont {Yoon}, \citenamefont {Lee}, \citenamefont
  {Choi}, \citenamefont {Jang}, \citenamefont {Brando}, \citenamefont {Lee},
  \citenamefont {Choi}, \citenamefont {Ji}, \citenamefont {Jang}, \citenamefont
  {Suh},\ and\ \citenamefont {Choi}}]{PhysRevB.96.014432}%
  \BibitemOpen
  \bibfield  {author} {\bibinfo {author} {\bibfnamefont {W.-J.}\ \bibnamefont
  {Lee}}, \bibinfo {author} {\bibfnamefont {S.-H.}\ \bibnamefont {Do}},
  \bibinfo {author} {\bibfnamefont {S.}~\bibnamefont {Yoon}}, \bibinfo {author}
  {\bibfnamefont {S.}~\bibnamefont {Lee}}, \bibinfo {author} {\bibfnamefont
  {Y.~S.}\ \bibnamefont {Choi}}, \bibinfo {author} {\bibfnamefont {D.~J.}\
  \bibnamefont {Jang}}, \bibinfo {author} {\bibfnamefont {M.}~\bibnamefont
  {Brando}}, \bibinfo {author} {\bibfnamefont {M.}~\bibnamefont {Lee}},
  \bibinfo {author} {\bibfnamefont {E.~S.}\ \bibnamefont {Choi}}, \bibinfo
  {author} {\bibfnamefont {S.}~\bibnamefont {Ji}}, \bibinfo {author}
  {\bibfnamefont {Z.~H.}\ \bibnamefont {Jang}}, \bibinfo {author}
  {\bibfnamefont {B.~J.}\ \bibnamefont {Suh}},\ and\ \bibinfo {author}
  {\bibfnamefont {K.-Y.}\ \bibnamefont {Choi}},\ }\href
  {https://doi.org/10.1103/PhysRevB.96.014432} {\bibfield  {journal} {\bibinfo
  {journal} {Phys. Rev. B}\ }\textbf {\bibinfo {volume} {96}},\ \bibinfo
  {pages} {014432} (\bibinfo {year} {2017})}\BibitemShut {NoStop}%
\bibitem [{\citenamefont {Kim}\ \emph {et~al.}(2019)\citenamefont {Kim},
  \citenamefont {Wang}, \citenamefont {Huang}, \citenamefont {Wang},
  \citenamefont {Fang}, \citenamefont {Luo}, \citenamefont {Li}, \citenamefont
  {Wu}, \citenamefont {Mori}, \citenamefont {Kwok}, \citenamefont {Mun},
  \citenamefont {Zapf},\ and\ \citenamefont {Cheong}}]{PhysRevX.9.031005}%
  \BibitemOpen
  \bibfield  {author} {\bibinfo {author} {\bibfnamefont {J.}~\bibnamefont
  {Kim}}, \bibinfo {author} {\bibfnamefont {X.}~\bibnamefont {Wang}}, \bibinfo
  {author} {\bibfnamefont {F.-T.}\ \bibnamefont {Huang}}, \bibinfo {author}
  {\bibfnamefont {Y.}~\bibnamefont {Wang}}, \bibinfo {author} {\bibfnamefont
  {X.}~\bibnamefont {Fang}}, \bibinfo {author} {\bibfnamefont {X.}~\bibnamefont
  {Luo}}, \bibinfo {author} {\bibfnamefont {Y.}~\bibnamefont {Li}}, \bibinfo
  {author} {\bibfnamefont {M.}~\bibnamefont {Wu}}, \bibinfo {author}
  {\bibfnamefont {S.}~\bibnamefont {Mori}}, \bibinfo {author} {\bibfnamefont
  {D.}~\bibnamefont {Kwok}}, \bibinfo {author} {\bibfnamefont {E.~D.}\
  \bibnamefont {Mun}}, \bibinfo {author} {\bibfnamefont {V.~S.}\ \bibnamefont
  {Zapf}},\ and\ \bibinfo {author} {\bibfnamefont {S.-W.}\ \bibnamefont
  {Cheong}},\ }\href {https://doi.org/10.1103/PhysRevX.9.031005} {\bibfield
  {journal} {\bibinfo  {journal} {Phys. Rev. X}\ }\textbf {\bibinfo {volume}
  {9}},\ \bibinfo {pages} {031005} (\bibinfo {year} {2019})}\BibitemShut
  {NoStop}%
\bibitem [{\citenamefont {Shimura}\ \emph {et~al.}(2021)\citenamefont
  {Shimura}, \citenamefont {W\"orl}, \citenamefont {Sundermann}, \citenamefont
  {Tsuda}, \citenamefont {Adroja}, \citenamefont {Bhattacharyya}, \citenamefont
  {Strydom}, \citenamefont {Hillier}, \citenamefont {Pratt}, \citenamefont
  {Gloskovskii}, \citenamefont {Severing}, \citenamefont {Onimaru},
  \citenamefont {Gegenwart},\ and\ \citenamefont
  {Takabatake}}]{PhysRevLett.126.217202}%
  \BibitemOpen
  \bibfield  {author} {\bibinfo {author} {\bibfnamefont {Y.}~\bibnamefont
  {Shimura}}, \bibinfo {author} {\bibfnamefont {A.}~\bibnamefont {W\"orl}},
  \bibinfo {author} {\bibfnamefont {M.}~\bibnamefont {Sundermann}}, \bibinfo
  {author} {\bibfnamefont {S.}~\bibnamefont {Tsuda}}, \bibinfo {author}
  {\bibfnamefont {D.~T.}\ \bibnamefont {Adroja}}, \bibinfo {author}
  {\bibfnamefont {A.}~\bibnamefont {Bhattacharyya}}, \bibinfo {author}
  {\bibfnamefont {A.~M.}\ \bibnamefont {Strydom}}, \bibinfo {author}
  {\bibfnamefont {A.~D.}\ \bibnamefont {Hillier}}, \bibinfo {author}
  {\bibfnamefont {F.~L.}\ \bibnamefont {Pratt}}, \bibinfo {author}
  {\bibfnamefont {A.}~\bibnamefont {Gloskovskii}}, \bibinfo {author}
  {\bibfnamefont {A.}~\bibnamefont {Severing}}, \bibinfo {author}
  {\bibfnamefont {T.}~\bibnamefont {Onimaru}}, \bibinfo {author} {\bibfnamefont
  {P.}~\bibnamefont {Gegenwart}},\ and\ \bibinfo {author} {\bibfnamefont
  {T.}~\bibnamefont {Takabatake}},\ }\href
  {https://doi.org/10.1103/PhysRevLett.126.217202} {\bibfield  {journal}
  {\bibinfo  {journal} {Phys. Rev. Lett.}\ }\textbf {\bibinfo {volume} {126}},\
  \bibinfo {pages} {217202} (\bibinfo {year} {2021})}\BibitemShut {NoStop}%
\bibitem [{\citenamefont {Majumder}\ \emph {et~al.}(2022)\citenamefont
  {Majumder}, \citenamefont {Gupta}, \citenamefont {Luetkens}, \citenamefont
  {Khasanov}, \citenamefont {Stockert}, \citenamefont {Gegenwart},\ and\
  \citenamefont {Fritsch}}]{PhysRevB.105.L180402}%
  \BibitemOpen
  \bibfield  {author} {\bibinfo {author} {\bibfnamefont {M.}~\bibnamefont
  {Majumder}}, \bibinfo {author} {\bibfnamefont {R.}~\bibnamefont {Gupta}},
  \bibinfo {author} {\bibfnamefont {H.}~\bibnamefont {Luetkens}}, \bibinfo
  {author} {\bibfnamefont {R.}~\bibnamefont {Khasanov}}, \bibinfo {author}
  {\bibfnamefont {O.}~\bibnamefont {Stockert}}, \bibinfo {author}
  {\bibfnamefont {P.}~\bibnamefont {Gegenwart}},\ and\ \bibinfo {author}
  {\bibfnamefont {V.}~\bibnamefont {Fritsch}},\ }\href
  {https://doi.org/10.1103/PhysRevB.105.L180402} {\bibfield  {journal}
  {\bibinfo  {journal} {Phys. Rev. B}\ }\textbf {\bibinfo {volume} {105}},\
  \bibinfo {pages} {L180402} (\bibinfo {year} {2022})}\BibitemShut {NoStop}%
\bibitem [{\citenamefont {Sakai}\ \emph {et~al.}(2016)\citenamefont {Sakai},
  \citenamefont {Lucas}, \citenamefont {Gegenwart}, \citenamefont {Stockert},
  \citenamefont {v.~L\"ohneysen},\ and\ \citenamefont
  {Fritsch}}]{PhysRevB.94.220405}%
  \BibitemOpen
  \bibfield  {author} {\bibinfo {author} {\bibfnamefont {A.}~\bibnamefont
  {Sakai}}, \bibinfo {author} {\bibfnamefont {S.}~\bibnamefont {Lucas}},
  \bibinfo {author} {\bibfnamefont {P.}~\bibnamefont {Gegenwart}}, \bibinfo
  {author} {\bibfnamefont {O.}~\bibnamefont {Stockert}}, \bibinfo {author}
  {\bibfnamefont {H.}~\bibnamefont {v.~L\"ohneysen}},\ and\ \bibinfo {author}
  {\bibfnamefont {V.}~\bibnamefont {Fritsch}},\ }\href
  {https://doi.org/10.1103/PhysRevB.94.220405} {\bibfield  {journal} {\bibinfo
  {journal} {Phys. Rev. B}\ }\textbf {\bibinfo {volume} {94}},\ \bibinfo
  {pages} {220405} (\bibinfo {year} {2016})}\BibitemShut {NoStop}%
\bibitem [{\citenamefont {Zhao}\ \emph {et~al.}(2019)\citenamefont {Zhao},
  \citenamefont {Zhang}, \citenamefont {Lyu}, \citenamefont {Bachus},
  \citenamefont {Tokiwa}, \citenamefont {Gegenwart}, \citenamefont {Zhang},
  \citenamefont {Cheng}, \citenamefont {Yang}, \citenamefont {Chen} \emph
  {et~al.}}]{zhao2019quantum}%
  \BibitemOpen
  \bibfield  {author} {\bibinfo {author} {\bibfnamefont {H.}~\bibnamefont
  {Zhao}}, \bibinfo {author} {\bibfnamefont {J.}~\bibnamefont {Zhang}},
  \bibinfo {author} {\bibfnamefont {M.}~\bibnamefont {Lyu}}, \bibinfo {author}
  {\bibfnamefont {S.}~\bibnamefont {Bachus}}, \bibinfo {author} {\bibfnamefont
  {Y.}~\bibnamefont {Tokiwa}}, \bibinfo {author} {\bibfnamefont
  {P.}~\bibnamefont {Gegenwart}}, \bibinfo {author} {\bibfnamefont
  {S.}~\bibnamefont {Zhang}}, \bibinfo {author} {\bibfnamefont
  {J.}~\bibnamefont {Cheng}}, \bibinfo {author} {\bibfnamefont {Y.-f.}\
  \bibnamefont {Yang}}, \bibinfo {author} {\bibfnamefont {G.}~\bibnamefont
  {Chen}}, \emph {et~al.},\ }\bibfield  {title} {\bibinfo {title}
  {Quantum-critical phase from frustrated magnetism in a strongly correlated
  metal},\ }\href {https://doi.org/10.1038/s41567-019-0666-6} {\bibfield
  {journal} {\bibinfo  {journal} {Nature Physics}\ }\textbf {\bibinfo {volume}
  {15}},\ \bibinfo {pages} {1261} (\bibinfo {year} {2019})}\BibitemShut
  {NoStop}%
\bibitem [{\citenamefont {Binder}\ and\ \citenamefont
  {Young}(1986)}]{RevModPhys.58.801}%
  \BibitemOpen
  \bibfield  {author} {\bibinfo {author} {\bibfnamefont {K.}~\bibnamefont
  {Binder}}\ and\ \bibinfo {author} {\bibfnamefont {A.~P.}\ \bibnamefont
  {Young}},\ }\href {https://doi.org/10.1103/RevModPhys.58.801} {\bibfield
  {journal} {\bibinfo  {journal} {Rev. Mod. Phys.}\ }\textbf {\bibinfo {volume}
  {58}},\ \bibinfo {pages} {801} (\bibinfo {year} {1986})}\BibitemShut
  {NoStop}%
\bibitem [{\citenamefont {Mydosh}(1993)}]{Mydosh}%
  \BibitemOpen
  \bibfield  {author} {\bibinfo {author} {\bibfnamefont {J.}~\bibnamefont
  {Mydosh}},\ }\href {https://doi.org/10.1201/9781482295191} {\emph {\bibinfo
  {title} {Spin Glasses: An Experimental Introduction}}}\ (\bibinfo {year}
  {1993})\BibitemShut {NoStop}%
\bibitem [{\citenamefont {Mydosh}(1986)}]{Mydosh1}%
  \BibitemOpen
  \bibfield  {author} {\bibinfo {author} {\bibfnamefont {J.}~\bibnamefont
  {Mydosh}},\ }\href {https://doi.org/10.1007/BF02401580} {\bibfield  {journal}
  {\bibinfo  {journal} {Hyperfine Interactions}\ }\textbf {\bibinfo {volume}
  {31}},\ \bibinfo {pages} {347} (\bibinfo {year} {1986})}\BibitemShut
  {NoStop}%
\bibitem [{\citenamefont {Wu}\ \emph {et~al.}(2011)\citenamefont {Wu},
  \citenamefont {Wildeboer}, \citenamefont {Werner}, \citenamefont {Seidel},
  \citenamefont {Nussinov},\ and\ \citenamefont {Solin}}]{Wu_2011}%
  \BibitemOpen
  \bibfield  {author} {\bibinfo {author} {\bibfnamefont {J.}~\bibnamefont
  {Wu}}, \bibinfo {author} {\bibfnamefont {J.~S.}\ \bibnamefont {Wildeboer}},
  \bibinfo {author} {\bibfnamefont {F.}~\bibnamefont {Werner}}, \bibinfo
  {author} {\bibfnamefont {A.}~\bibnamefont {Seidel}}, \bibinfo {author}
  {\bibfnamefont {Z.}~\bibnamefont {Nussinov}},\ and\ \bibinfo {author}
  {\bibfnamefont {S.~A.}\ \bibnamefont {Solin}},\ }\href
  {https://doi.org/10.1209/0295-5075/93/67001} {\bibfield  {journal} {\bibinfo
  {journal} {{EPL} (Europhysics Letters)}\ }\textbf {\bibinfo {volume} {93}},\
  \bibinfo {pages} {67001} (\bibinfo {year} {2011})}\BibitemShut {NoStop}%
\bibitem [{\citenamefont {Ma}\ \emph {et~al.}(2018{\natexlab{a}})\citenamefont
  {Ma}, \citenamefont {Ran}, \citenamefont {Wang}, \citenamefont {Bao},
  \citenamefont {Cai}, \citenamefont {Li},\ and\ \citenamefont
  {Wen}}]{Ma_2018}%
  \BibitemOpen
  \bibfield  {author} {\bibinfo {author} {\bibfnamefont {Z.}~\bibnamefont
  {Ma}}, \bibinfo {author} {\bibfnamefont {K.}~\bibnamefont {Ran}}, \bibinfo
  {author} {\bibfnamefont {J.}~\bibnamefont {Wang}}, \bibinfo {author}
  {\bibfnamefont {S.}~\bibnamefont {Bao}}, \bibinfo {author} {\bibfnamefont
  {Z.}~\bibnamefont {Cai}}, \bibinfo {author} {\bibfnamefont {S.}~\bibnamefont
  {Li}},\ and\ \bibinfo {author} {\bibfnamefont {J.}~\bibnamefont {Wen}},\
  }\href {https://doi.org/10.1088/1674-1056/27/10/106101} {\bibfield  {journal}
  {\bibinfo  {journal} {Chinese Physics B}\ }\textbf {\bibinfo {volume} {27}},\
  \bibinfo {pages} {106101} (\bibinfo {year} {2018}{\natexlab{a}})}\BibitemShut
  {NoStop}%
\bibitem [{\citenamefont {Ma}\ \emph {et~al.}(2018{\natexlab{b}})\citenamefont
  {Ma}, \citenamefont {Wang}, \citenamefont {Dong}, \citenamefont {Zhang},
  \citenamefont {Li}, \citenamefont {Zheng}, \citenamefont {Yu}, \citenamefont
  {Wang}, \citenamefont {Che}, \citenamefont {Ran}, \citenamefont {Bao},
  \citenamefont {Cai}, \citenamefont {\ifmmode~\check{C}\else
  \v{C}\fi{}erm\'ak}, \citenamefont {Schneidewind}, \citenamefont {Yano},
  \citenamefont {Gardner}, \citenamefont {Lu}, \citenamefont {Yu},
  \citenamefont {Liu}, \citenamefont {Li}, \citenamefont {Li},\ and\
  \citenamefont {Wen}}]{PhysRevLett.120.087201}%
  \BibitemOpen
  \bibfield  {author} {\bibinfo {author} {\bibfnamefont {Z.}~\bibnamefont
  {Ma}}, \bibinfo {author} {\bibfnamefont {J.}~\bibnamefont {Wang}}, \bibinfo
  {author} {\bibfnamefont {Z.-Y.}\ \bibnamefont {Dong}}, \bibinfo {author}
  {\bibfnamefont {J.}~\bibnamefont {Zhang}}, \bibinfo {author} {\bibfnamefont
  {S.}~\bibnamefont {Li}}, \bibinfo {author} {\bibfnamefont {S.-H.}\
  \bibnamefont {Zheng}}, \bibinfo {author} {\bibfnamefont {Y.}~\bibnamefont
  {Yu}}, \bibinfo {author} {\bibfnamefont {W.}~\bibnamefont {Wang}}, \bibinfo
  {author} {\bibfnamefont {L.}~\bibnamefont {Che}}, \bibinfo {author}
  {\bibfnamefont {K.}~\bibnamefont {Ran}}, \bibinfo {author} {\bibfnamefont
  {S.}~\bibnamefont {Bao}}, \bibinfo {author} {\bibfnamefont {Z.}~\bibnamefont
  {Cai}}, \bibinfo {author} {\bibfnamefont {P.}~\bibnamefont
  {\ifmmode~\check{C}\else \v{C}\fi{}erm\'ak}}, \bibinfo {author}
  {\bibfnamefont {A.}~\bibnamefont {Schneidewind}}, \bibinfo {author}
  {\bibfnamefont {S.}~\bibnamefont {Yano}}, \bibinfo {author} {\bibfnamefont
  {J.~S.}\ \bibnamefont {Gardner}}, \bibinfo {author} {\bibfnamefont
  {X.}~\bibnamefont {Lu}}, \bibinfo {author} {\bibfnamefont {S.-L.}\
  \bibnamefont {Yu}}, \bibinfo {author} {\bibfnamefont {J.-M.}\ \bibnamefont
  {Liu}}, \bibinfo {author} {\bibfnamefont {S.}~\bibnamefont {Li}}, \bibinfo
  {author} {\bibfnamefont {J.-X.}\ \bibnamefont {Li}},\ and\ \bibinfo {author}
  {\bibfnamefont {J.}~\bibnamefont {Wen}},\ }\href
  {https://doi.org/10.1103/PhysRevLett.120.087201} {\bibfield  {journal}
  {\bibinfo  {journal} {Phys. Rev. Lett.}\ }\textbf {\bibinfo {volume} {120}},\
  \bibinfo {pages} {087201} (\bibinfo {year} {2018}{\natexlab{b}})}\BibitemShut
  {NoStop}%
\bibitem [{\citenamefont {Shen}\ \emph {et~al.}(2016)\citenamefont {Shen},
  \citenamefont {Li}, \citenamefont {Wo}, \citenamefont {Li}, \citenamefont
  {Shen}, \citenamefont {Pan}, \citenamefont {Wang}, \citenamefont {Walker},
  \citenamefont {Steffens}, \citenamefont {Boehm} \emph
  {et~al.}}]{shen2016evidence}%
  \BibitemOpen
  \bibfield  {author} {\bibinfo {author} {\bibfnamefont {Y.}~\bibnamefont
  {Shen}}, \bibinfo {author} {\bibfnamefont {Y.-D.}\ \bibnamefont {Li}},
  \bibinfo {author} {\bibfnamefont {H.}~\bibnamefont {Wo}}, \bibinfo {author}
  {\bibfnamefont {Y.}~\bibnamefont {Li}}, \bibinfo {author} {\bibfnamefont
  {S.}~\bibnamefont {Shen}}, \bibinfo {author} {\bibfnamefont {B.}~\bibnamefont
  {Pan}}, \bibinfo {author} {\bibfnamefont {Q.}~\bibnamefont {Wang}}, \bibinfo
  {author} {\bibfnamefont {H.}~\bibnamefont {Walker}}, \bibinfo {author}
  {\bibfnamefont {P.}~\bibnamefont {Steffens}}, \bibinfo {author}
  {\bibfnamefont {M.}~\bibnamefont {Boehm}}, \emph {et~al.},\ }\href
  {https://doi.org/10.1038/nature20614} {\bibfield  {journal} {\bibinfo
  {journal} {Nature}\ }\textbf {\bibinfo {volume} {540}},\ \bibinfo {pages}
  {559} (\bibinfo {year} {2016})}\BibitemShut {NoStop}%
\bibitem [{\citenamefont {Paddison}\ \emph {et~al.}(2017)\citenamefont
  {Paddison}, \citenamefont {Daum}, \citenamefont {Dun}, \citenamefont
  {Ehlers}, \citenamefont {Liu}, \citenamefont {Stone}, \citenamefont {Zhou},\
  and\ \citenamefont {Mourigal}}]{paddison2017continuous}%
  \BibitemOpen
  \bibfield  {author} {\bibinfo {author} {\bibfnamefont {J.~A.}\ \bibnamefont
  {Paddison}}, \bibinfo {author} {\bibfnamefont {M.}~\bibnamefont {Daum}},
  \bibinfo {author} {\bibfnamefont {Z.}~\bibnamefont {Dun}}, \bibinfo {author}
  {\bibfnamefont {G.}~\bibnamefont {Ehlers}}, \bibinfo {author} {\bibfnamefont
  {Y.}~\bibnamefont {Liu}}, \bibinfo {author} {\bibfnamefont {M.~B.}\
  \bibnamefont {Stone}}, \bibinfo {author} {\bibfnamefont {H.}~\bibnamefont
  {Zhou}},\ and\ \bibinfo {author} {\bibfnamefont {M.}~\bibnamefont
  {Mourigal}},\ }\href {https://doi.org/10.1038/nphys3971} {\bibfield
  {journal} {\bibinfo  {journal} {Nature Physics}\ }\textbf {\bibinfo {volume}
  {13}},\ \bibinfo {pages} {117} (\bibinfo {year} {2017})}\BibitemShut
  {NoStop}%
\bibitem [{\citenamefont {Xu}\ \emph {et~al.}(2016)\citenamefont {Xu},
  \citenamefont {Zhang}, \citenamefont {Li}, \citenamefont {Yu}, \citenamefont
  {Hong}, \citenamefont {Zhang},\ and\ \citenamefont
  {Li}}]{PhysRevLett.117.267202}%
  \BibitemOpen
  \bibfield  {author} {\bibinfo {author} {\bibfnamefont {Y.}~\bibnamefont
  {Xu}}, \bibinfo {author} {\bibfnamefont {J.}~\bibnamefont {Zhang}}, \bibinfo
  {author} {\bibfnamefont {Y.~S.}\ \bibnamefont {Li}}, \bibinfo {author}
  {\bibfnamefont {Y.~J.}\ \bibnamefont {Yu}}, \bibinfo {author} {\bibfnamefont
  {X.~C.}\ \bibnamefont {Hong}}, \bibinfo {author} {\bibfnamefont {Q.~M.}\
  \bibnamefont {Zhang}},\ and\ \bibinfo {author} {\bibfnamefont {S.~Y.}\
  \bibnamefont {Li}},\ }\href {https://doi.org/10.1103/PhysRevLett.117.267202}
  {\bibfield  {journal} {\bibinfo  {journal} {Phys. Rev. Lett.}\ }\textbf
  {\bibinfo {volume} {117}},\ \bibinfo {pages} {267202} (\bibinfo {year}
  {2016})}\BibitemShut {NoStop}%
\bibitem [{\citenamefont {Han}\ \emph {et~al.}(2012)\citenamefont {Han},
  \citenamefont {Helton}, \citenamefont {Chu}, \citenamefont {Nocera},
  \citenamefont {Rodriguez-Rivera}, \citenamefont {Broholm},\ and\
  \citenamefont {Lee}}]{han2012fractionalized}%
  \BibitemOpen
  \bibfield  {author} {\bibinfo {author} {\bibfnamefont {T.-H.}\ \bibnamefont
  {Han}}, \bibinfo {author} {\bibfnamefont {J.~S.}\ \bibnamefont {Helton}},
  \bibinfo {author} {\bibfnamefont {S.}~\bibnamefont {Chu}}, \bibinfo {author}
  {\bibfnamefont {D.~G.}\ \bibnamefont {Nocera}}, \bibinfo {author}
  {\bibfnamefont {J.~A.}\ \bibnamefont {Rodriguez-Rivera}}, \bibinfo {author}
  {\bibfnamefont {C.}~\bibnamefont {Broholm}},\ and\ \bibinfo {author}
  {\bibfnamefont {Y.~S.}\ \bibnamefont {Lee}},\ }\href
  {https://doi.org/10.1038/nature11659} {\bibfield  {journal} {\bibinfo
  {journal} {Nature}\ }\textbf {\bibinfo {volume} {492}},\ \bibinfo {pages}
  {406} (\bibinfo {year} {2012})}\BibitemShut {NoStop}%
\bibitem [{\citenamefont {Norman}(2016)}]{RevModPhys.88.041002}%
  \BibitemOpen
  \bibfield  {author} {\bibinfo {author} {\bibfnamefont {M.~R.}\ \bibnamefont
  {Norman}},\ }\href {https://doi.org/10.1103/RevModPhys.88.041002} {\bibfield
  {journal} {\bibinfo  {journal} {Rev. Mod. Phys.}\ }\textbf {\bibinfo {volume}
  {88}},\ \bibinfo {pages} {041002} (\bibinfo {year} {2016})}\BibitemShut
  {NoStop}%
\bibitem [{\citenamefont {Adroja}\ \emph {et~al.}(2019)\citenamefont {Adroja},
  \citenamefont {Takabatake}, \citenamefont {Walker}, \citenamefont
  {Bhattacharyya},\ and\ \citenamefont {Yang}}]{MERLIN1}%
  \BibitemOpen
  \bibfield  {author} {\bibinfo {author} {\bibfnamefont {D.}~\bibnamefont
  {Adroja}}, \bibinfo {author} {\bibfnamefont {T.}~\bibnamefont {Takabatake}},
  \bibinfo {author} {\bibfnamefont {H.}~\bibnamefont {Walker}}, \bibinfo
  {author} {\bibfnamefont {A.}~\bibnamefont {Bhattacharyya}},\ and\ \bibinfo
  {author} {\bibfnamefont {C.}~\bibnamefont {Yang}},\ }\href
  {https://doi.org/10.5286/ISIS.E.RB1620476} {} (\bibinfo {year}
  {2019})\BibitemShut {NoStop}%
\bibitem [{\citenamefont {Adroja}(2006)}]{MARI1}%
  \BibitemOpen
  \bibfield  {author} {\bibinfo {author} {\bibfnamefont {D.}~\bibnamefont
  {Adroja}},\ }\href {https://doi.org/10.5286/ISIS.E.RB14460} {} (\bibinfo
  {year} {2006})\BibitemShut {NoStop}%
\bibitem [{\citenamefont {Adroja}\ \emph {et~al.}(2020)\citenamefont {Adroja},
  \citenamefont {Takabatake}, \citenamefont {Yang},\ and\ \citenamefont
  {Hillier}}]{MUSR1}%
  \BibitemOpen
  \bibfield  {author} {\bibinfo {author} {\bibfnamefont {D.}~\bibnamefont
  {Adroja}}, \bibinfo {author} {\bibfnamefont {T.}~\bibnamefont {Takabatake}},
  \bibinfo {author} {\bibfnamefont {C.}~\bibnamefont {Yang}},\ and\ \bibinfo
  {author} {\bibfnamefont {A.~D.}\ \bibnamefont {Hillier}},\ }\href
  {https://doi.org/10.5286/ISIS.E.RB1710457} {} (\bibinfo {year}
  {2020})\BibitemShut {NoStop}%
\bibitem [{\citenamefont {Adroja.}\ \emph {et~al.}(2017)\citenamefont
  {Adroja.}, \citenamefont {Koza},\ and\ \citenamefont {Takabatake}}]{IN6}%
  \BibitemOpen
  \bibfield  {author} {\bibinfo {author} {\bibfnamefont {D.}~\bibnamefont
  {Adroja.}}, \bibinfo {author} {\bibfnamefont {M.}~\bibnamefont {Koza}},\ and\
  \bibinfo {author} {\bibfnamefont {T.}~\bibnamefont {Takabatake}},\ }\href
  {https://doi.org/10.5291/ILL-DATA.4-01-1552} {} (\bibinfo {year}
  {2017})\BibitemShut {NoStop}%
\bibitem [{\citenamefont {Ding}\ \emph {et~al.}(2019)\citenamefont {Ding},
  \citenamefont {Manuel}, \citenamefont {Bachus}, \citenamefont {Gru\ss{}ler},
  \citenamefont {Gegenwart}, \citenamefont {Singleton}, \citenamefont
  {Johnson}, \citenamefont {Walker}, \citenamefont {Adroja}, \citenamefont
  {Hillier},\ and\ \citenamefont {Tsirlin}}]{PhysRevB.100.144432}%
  \BibitemOpen
  \bibfield  {author} {\bibinfo {author} {\bibfnamefont {L.}~\bibnamefont
  {Ding}}, \bibinfo {author} {\bibfnamefont {P.}~\bibnamefont {Manuel}},
  \bibinfo {author} {\bibfnamefont {S.}~\bibnamefont {Bachus}}, \bibinfo
  {author} {\bibfnamefont {F.}~\bibnamefont {Gru\ss{}ler}}, \bibinfo {author}
  {\bibfnamefont {P.}~\bibnamefont {Gegenwart}}, \bibinfo {author}
  {\bibfnamefont {J.}~\bibnamefont {Singleton}}, \bibinfo {author}
  {\bibfnamefont {R.~D.}\ \bibnamefont {Johnson}}, \bibinfo {author}
  {\bibfnamefont {H.~C.}\ \bibnamefont {Walker}}, \bibinfo {author}
  {\bibfnamefont {D.~T.}\ \bibnamefont {Adroja}}, \bibinfo {author}
  {\bibfnamefont {A.~D.}\ \bibnamefont {Hillier}},\ and\ \bibinfo {author}
  {\bibfnamefont {A.~A.}\ \bibnamefont {Tsirlin}},\ }\href
  {https://doi.org/10.1103/PhysRevB.100.144432} {\bibfield  {journal} {\bibinfo
   {journal} {Phys. Rev. B}\ }\textbf {\bibinfo {volume} {100}},\ \bibinfo
  {pages} {144432} (\bibinfo {year} {2019})}\BibitemShut {NoStop}%
\bibitem [{\citenamefont {Mendels}\ \emph {et~al.}(2007)\citenamefont
  {Mendels}, \citenamefont {Bert}, \citenamefont {de~Vries}, \citenamefont
  {Olariu}, \citenamefont {Harrison}, \citenamefont {Duc}, \citenamefont
  {Trombe}, \citenamefont {Lord}, \citenamefont {Amato},\ and\ \citenamefont
  {Baines}}]{PhysRevLett.98.077204}%
  \BibitemOpen
  \bibfield  {author} {\bibinfo {author} {\bibfnamefont {P.}~\bibnamefont
  {Mendels}}, \bibinfo {author} {\bibfnamefont {F.}~\bibnamefont {Bert}},
  \bibinfo {author} {\bibfnamefont {M.~A.}\ \bibnamefont {de~Vries}}, \bibinfo
  {author} {\bibfnamefont {A.}~\bibnamefont {Olariu}}, \bibinfo {author}
  {\bibfnamefont {A.}~\bibnamefont {Harrison}}, \bibinfo {author}
  {\bibfnamefont {F.}~\bibnamefont {Duc}}, \bibinfo {author} {\bibfnamefont
  {J.~C.}\ \bibnamefont {Trombe}}, \bibinfo {author} {\bibfnamefont {J.~S.}\
  \bibnamefont {Lord}}, \bibinfo {author} {\bibfnamefont {A.}~\bibnamefont
  {Amato}},\ and\ \bibinfo {author} {\bibfnamefont {C.}~\bibnamefont
  {Baines}},\ }\href {https://doi.org/10.1103/PhysRevLett.98.077204} {\bibfield
   {journal} {\bibinfo  {journal} {Phys. Rev. Lett.}\ }\textbf {\bibinfo
  {volume} {98}},\ \bibinfo {pages} {077204} (\bibinfo {year}
  {2007})}\BibitemShut {NoStop}%
\bibitem [{\citenamefont {Arh}\ \emph {et~al.}(2022)\citenamefont {Arh},
  \citenamefont {Sana}, \citenamefont {Pregelj}, \citenamefont {Khuntia},
  \citenamefont {Jagli{\v{c}}i{\'c}}, \citenamefont {Le}, \citenamefont
  {Biswas}, \citenamefont {Manuel}, \citenamefont {Mangin-Thro}, \citenamefont
  {Ozarowski} \emph {et~al.}}]{arh2022ising}%
  \BibitemOpen
  \bibfield  {author} {\bibinfo {author} {\bibfnamefont {T.}~\bibnamefont
  {Arh}}, \bibinfo {author} {\bibfnamefont {B.}~\bibnamefont {Sana}}, \bibinfo
  {author} {\bibfnamefont {M.}~\bibnamefont {Pregelj}}, \bibinfo {author}
  {\bibfnamefont {P.}~\bibnamefont {Khuntia}}, \bibinfo {author} {\bibfnamefont
  {Z.}~\bibnamefont {Jagli{\v{c}}i{\'c}}}, \bibinfo {author} {\bibfnamefont
  {M.}~\bibnamefont {Le}}, \bibinfo {author} {\bibfnamefont {P.}~\bibnamefont
  {Biswas}}, \bibinfo {author} {\bibfnamefont {P.}~\bibnamefont {Manuel}},
  \bibinfo {author} {\bibfnamefont {L.}~\bibnamefont {Mangin-Thro}}, \bibinfo
  {author} {\bibfnamefont {A.}~\bibnamefont {Ozarowski}}, \emph {et~al.},\
  }\href {https://doi.org/10.1038/s41563-021-01169-y} {\bibfield  {journal}
  {\bibinfo  {journal} {Nature Materials}\ }\textbf {\bibinfo {volume} {21}},\
  \bibinfo {pages} {416} (\bibinfo {year} {2022})}\BibitemShut {NoStop}%
\bibitem [{\citenamefont {Chen}\ \emph {et~al.}(2022)\citenamefont {Chen},
  \citenamefont {Sinclair}, \citenamefont {Akbari-Sharbaf}, \citenamefont
  {Huang}, \citenamefont {Dun}, \citenamefont {Choi}, \citenamefont {Mourigal},
  \citenamefont {Verrier}, \citenamefont {Rouane}, \citenamefont
  {Bazier-Matte}, \citenamefont {Quilliam}, \citenamefont {Aczel},\ and\
  \citenamefont {Zhou}}]{PhysRevMaterials.6.044414}%
  \BibitemOpen
  \bibfield  {author} {\bibinfo {author} {\bibfnamefont {Q.}~\bibnamefont
  {Chen}}, \bibinfo {author} {\bibfnamefont {R.}~\bibnamefont {Sinclair}},
  \bibinfo {author} {\bibfnamefont {A.}~\bibnamefont {Akbari-Sharbaf}},
  \bibinfo {author} {\bibfnamefont {Q.}~\bibnamefont {Huang}}, \bibinfo
  {author} {\bibfnamefont {Z.}~\bibnamefont {Dun}}, \bibinfo {author}
  {\bibfnamefont {E.~S.}\ \bibnamefont {Choi}}, \bibinfo {author}
  {\bibfnamefont {M.}~\bibnamefont {Mourigal}}, \bibinfo {author}
  {\bibfnamefont {A.}~\bibnamefont {Verrier}}, \bibinfo {author} {\bibfnamefont
  {R.}~\bibnamefont {Rouane}}, \bibinfo {author} {\bibfnamefont
  {X.}~\bibnamefont {Bazier-Matte}}, \bibinfo {author} {\bibfnamefont {J.~A.}\
  \bibnamefont {Quilliam}}, \bibinfo {author} {\bibfnamefont {A.~A.}\
  \bibnamefont {Aczel}},\ and\ \bibinfo {author} {\bibfnamefont {H.~D.}\
  \bibnamefont {Zhou}},\ }\href
  {https://doi.org/10.1103/PhysRevMaterials.6.044414} {\bibfield  {journal}
  {\bibinfo  {journal} {Phys. Rev. Materials}\ }\textbf {\bibinfo {volume}
  {6}},\ \bibinfo {pages} {044414} (\bibinfo {year} {2022})}\BibitemShut
  {NoStop}%
\bibitem [{\citenamefont {\ifmmode \check{Z}\else
  \v{Z}\fi{}ivkovi\ifmmode~\acute{c}\else \'{c}\fi{}}\ \emph
  {et~al.}(2021)\citenamefont {\ifmmode \check{Z}\else
  \v{Z}\fi{}ivkovi\ifmmode~\acute{c}\else \'{c}\fi{}}, \citenamefont {Favre},
  \citenamefont {Salazar~Mejia}, \citenamefont {Jeschke}, \citenamefont
  {Magrez}, \citenamefont {Dabholkar}, \citenamefont {Noculak}, \citenamefont
  {Freitas}, \citenamefont {Jeong}, \citenamefont {Hegde}, \citenamefont
  {Testa}, \citenamefont {Babkevich}, \citenamefont {Su}, \citenamefont
  {Manuel}, \citenamefont {Luetkens}, \citenamefont {Baines}, \citenamefont
  {Baker}, \citenamefont {Wosnitza}, \citenamefont {Zaharko}, \citenamefont
  {Iqbal}, \citenamefont {Reuther},\ and\ \citenamefont
  {R\o{}nnow}}]{PhysRevLett.127.157204}%
  \BibitemOpen
  \bibfield  {author} {\bibinfo {author} {\bibfnamefont {I.}~\bibnamefont
  {\ifmmode \check{Z}\else \v{Z}\fi{}ivkovi\ifmmode~\acute{c}\else
  \'{c}\fi{}}}, \bibinfo {author} {\bibfnamefont {V.}~\bibnamefont {Favre}},
  \bibinfo {author} {\bibfnamefont {C.}~\bibnamefont {Salazar~Mejia}}, \bibinfo
  {author} {\bibfnamefont {H.~O.}\ \bibnamefont {Jeschke}}, \bibinfo {author}
  {\bibfnamefont {A.}~\bibnamefont {Magrez}}, \bibinfo {author} {\bibfnamefont
  {B.}~\bibnamefont {Dabholkar}}, \bibinfo {author} {\bibfnamefont
  {V.}~\bibnamefont {Noculak}}, \bibinfo {author} {\bibfnamefont {R.~S.}\
  \bibnamefont {Freitas}}, \bibinfo {author} {\bibfnamefont {M.}~\bibnamefont
  {Jeong}}, \bibinfo {author} {\bibfnamefont {N.~G.}\ \bibnamefont {Hegde}},
  \bibinfo {author} {\bibfnamefont {L.}~\bibnamefont {Testa}}, \bibinfo
  {author} {\bibfnamefont {P.}~\bibnamefont {Babkevich}}, \bibinfo {author}
  {\bibfnamefont {Y.}~\bibnamefont {Su}}, \bibinfo {author} {\bibfnamefont
  {P.}~\bibnamefont {Manuel}}, \bibinfo {author} {\bibfnamefont
  {H.}~\bibnamefont {Luetkens}}, \bibinfo {author} {\bibfnamefont
  {C.}~\bibnamefont {Baines}}, \bibinfo {author} {\bibfnamefont {P.~J.}\
  \bibnamefont {Baker}}, \bibinfo {author} {\bibfnamefont {J.}~\bibnamefont
  {Wosnitza}}, \bibinfo {author} {\bibfnamefont {O.}~\bibnamefont {Zaharko}},
  \bibinfo {author} {\bibfnamefont {Y.}~\bibnamefont {Iqbal}}, \bibinfo
  {author} {\bibfnamefont {J.}~\bibnamefont {Reuther}},\ and\ \bibinfo {author}
  {\bibfnamefont {H.~M.}\ \bibnamefont {R\o{}nnow}},\ }\href
  {https://doi.org/10.1103/PhysRevLett.127.157204} {\bibfield  {journal}
  {\bibinfo  {journal} {Phys. Rev. Lett.}\ }\textbf {\bibinfo {volume} {127}},\
  \bibinfo {pages} {157204} (\bibinfo {year} {2021})}\BibitemShut {NoStop}%
\bibitem [{\citenamefont {Lee}\ \emph {et~al.}(2021)\citenamefont {Lee},
  \citenamefont {Lee}, \citenamefont {Berlie}, \citenamefont {Hillier},
  \citenamefont {Adroja}, \citenamefont {Zhong}, \citenamefont {Cava},
  \citenamefont {Jang},\ and\ \citenamefont {Choi}}]{PhysRevB.103.024413}%
  \BibitemOpen
  \bibfield  {author} {\bibinfo {author} {\bibfnamefont {S.}~\bibnamefont
  {Lee}}, \bibinfo {author} {\bibfnamefont {C.~H.}\ \bibnamefont {Lee}},
  \bibinfo {author} {\bibfnamefont {A.}~\bibnamefont {Berlie}}, \bibinfo
  {author} {\bibfnamefont {A.~D.}\ \bibnamefont {Hillier}}, \bibinfo {author}
  {\bibfnamefont {D.~T.}\ \bibnamefont {Adroja}}, \bibinfo {author}
  {\bibfnamefont {R.}~\bibnamefont {Zhong}}, \bibinfo {author} {\bibfnamefont
  {R.~J.}\ \bibnamefont {Cava}}, \bibinfo {author} {\bibfnamefont {Z.~H.}\
  \bibnamefont {Jang}},\ and\ \bibinfo {author} {\bibfnamefont {K.-Y.}\
  \bibnamefont {Choi}},\ }\href {https://doi.org/10.1103/PhysRevB.103.024413}
  {\bibfield  {journal} {\bibinfo  {journal} {Phys. Rev. B}\ }\textbf {\bibinfo
  {volume} {103}},\ \bibinfo {pages} {024413} (\bibinfo {year}
  {2021})}\BibitemShut {NoStop}%
\end{thebibliography}%



	\section*{Supplemental Material:}
	
	\subsection*{Abstract}
	
		In this supplementary materials, we have presented an additional analysis of the zero-field $\mu$SR data of the CeRh$_{1-x}$Pd$_x$Sn with $x=0.1$ single crystal and $x=0.2$ polycrystal using a two-component fit and the neutron diffraction (ND) data of the polycrystalline sample with $x=0.2$ down to 0.1~K. The ND data do not reveal any clear sign of long-range magnetic ordering, which is consistent with elastic neutron data from the inelastic measurements of the polycrystalline sample with $x=0.1$. Furthermore, we have compared quasi-elastic scattering as a function of $x$. High energy INS data are shown as color coded 2D intensity maps, momentum transfer ($Q$) vs energy transfer along with the estimated magnetic scattering. We have also provided $Q$-integrated 1D energy cuts from the raw data of the Ce and La compounds, from low-$Q$ and from high-$Q$, for the comparison.

		\section{Muon spin relaxation}
		
		\begin{figure}
			\includegraphics[width=5.5cm, keepaspectratio]{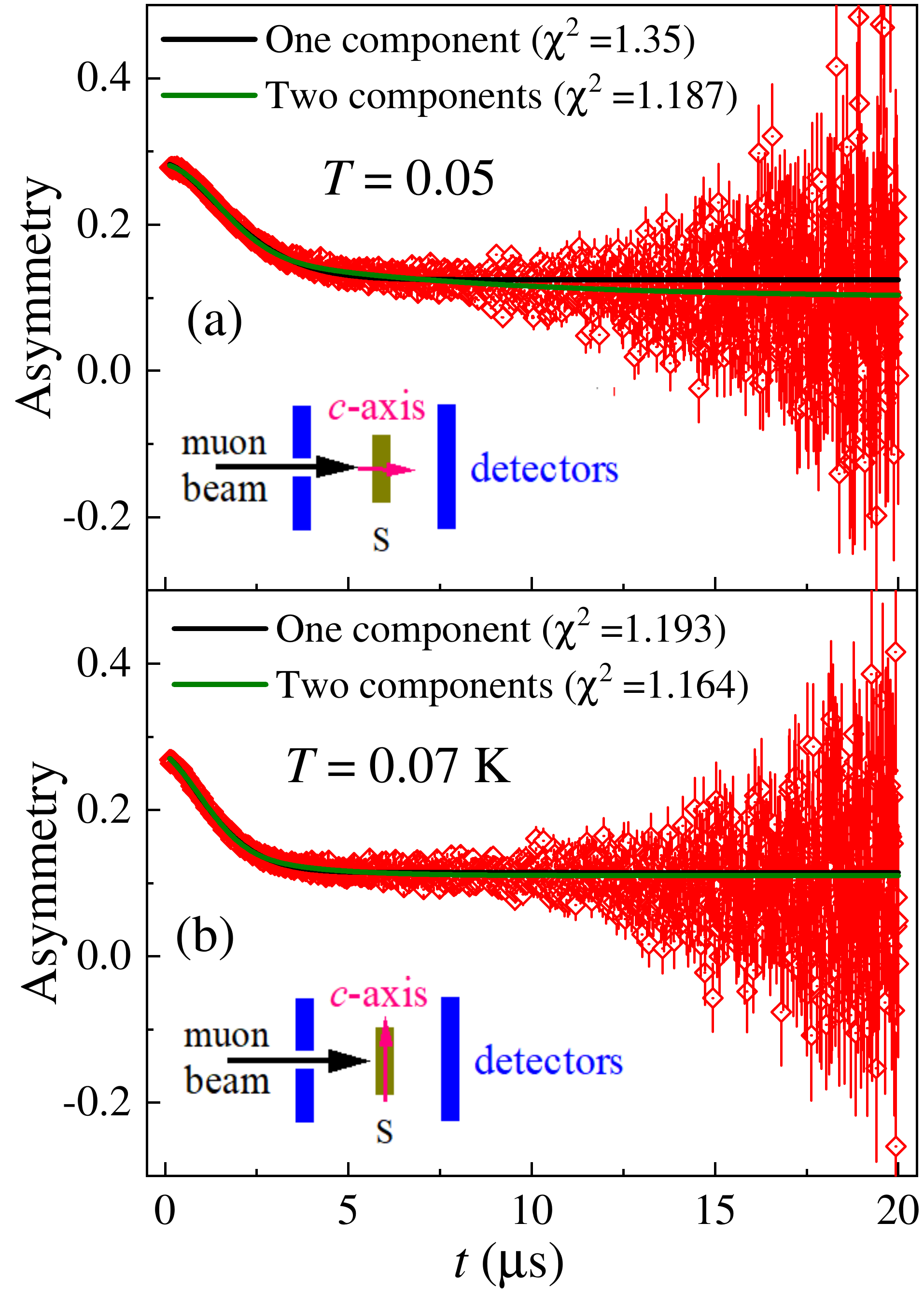}
			\caption{ZF-$\mu$SR spectra of CeRh$_{1-x}$Pd$_{x}$Sn with $x=0.1$ for the muon beam: (a) along the $c$-axis at 0.05~K and (b) perpendicular to the $c$-axis at 0.07~K. The solid lines represent fits to the relaxation functions (see the text for details).}
			\label{OneVsTwoComponent}
		\end{figure}

		\begin{figure}
			\includegraphics[width=7.5cm, keepaspectratio]{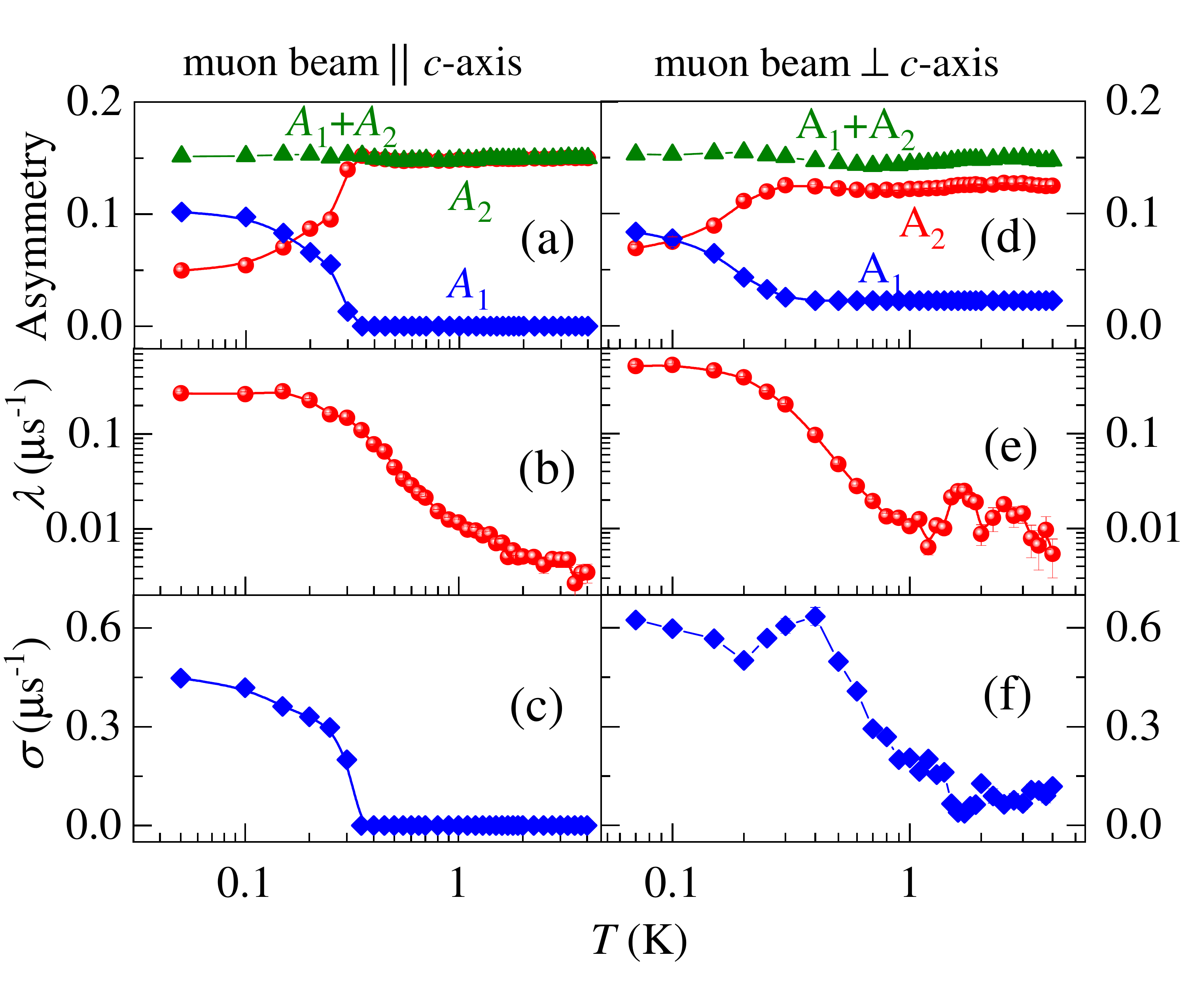}
			\caption{Temperature dependence of the initial asymmetry, the Lorentzian depolarization ($\lambda$) rate, and the Gaussian depolarization rate ($\sigma$), respectively, for $x=0.1$ single crystal sample with the muons beam parallel (a) -- (c) and perpendicular (d) -- (f) to the $c$-axis.}
			\label{MUSR_x_0pt1}
		\end{figure}
		
		\begin{figure}
			\includegraphics[width=5.5cm, keepaspectratio]{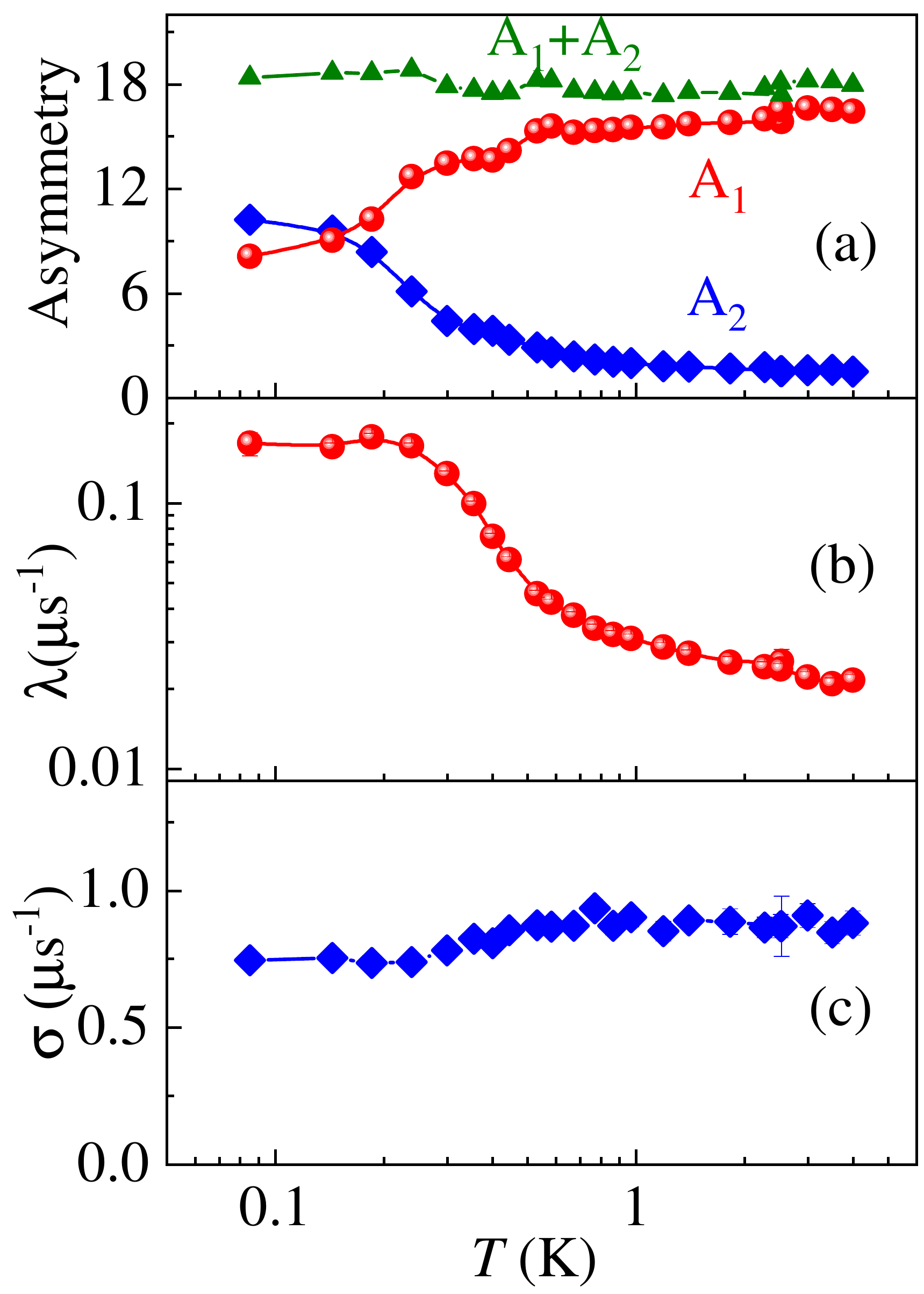}
			\caption{Temperature dependence of (a) the initial asymmetry, (b) the Lorentzian depolarization rate, and (c) the Gaussian depolarization rate for $x=0.2$ polycrystalline sample.}
			\label{MUSR_x_0pt2}
		\end{figure}
		
		\begin{figure}
			
			\includegraphics[width=7.5cm, keepaspectratio]{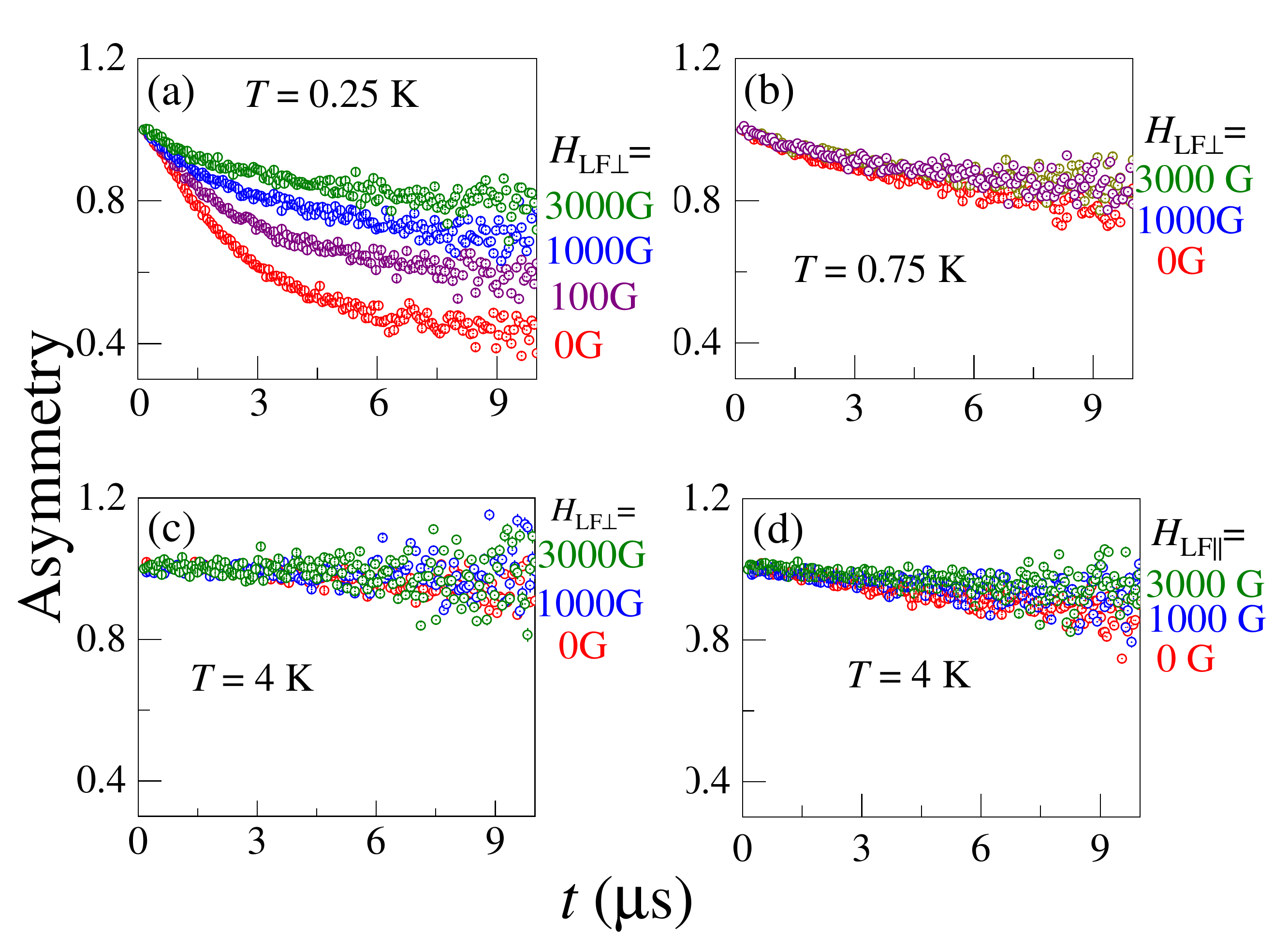}
			\caption{LF-$\mu$SR time spectra of CeRh$_{1-x}$Pd$_x$Sn for $x$ = 0.1 at different temperatures $T$ = 0.25, 0.75, and 4~K and at different applied magnetic field $H_{LF\perp}$ and $H_{LF\parallel}$ up to 3000~G.}
			\label{LF4K_x_0pt1}
		\end{figure}
		
		\begin{figure}
			
			\includegraphics[width=7.0cm, keepaspectratio]{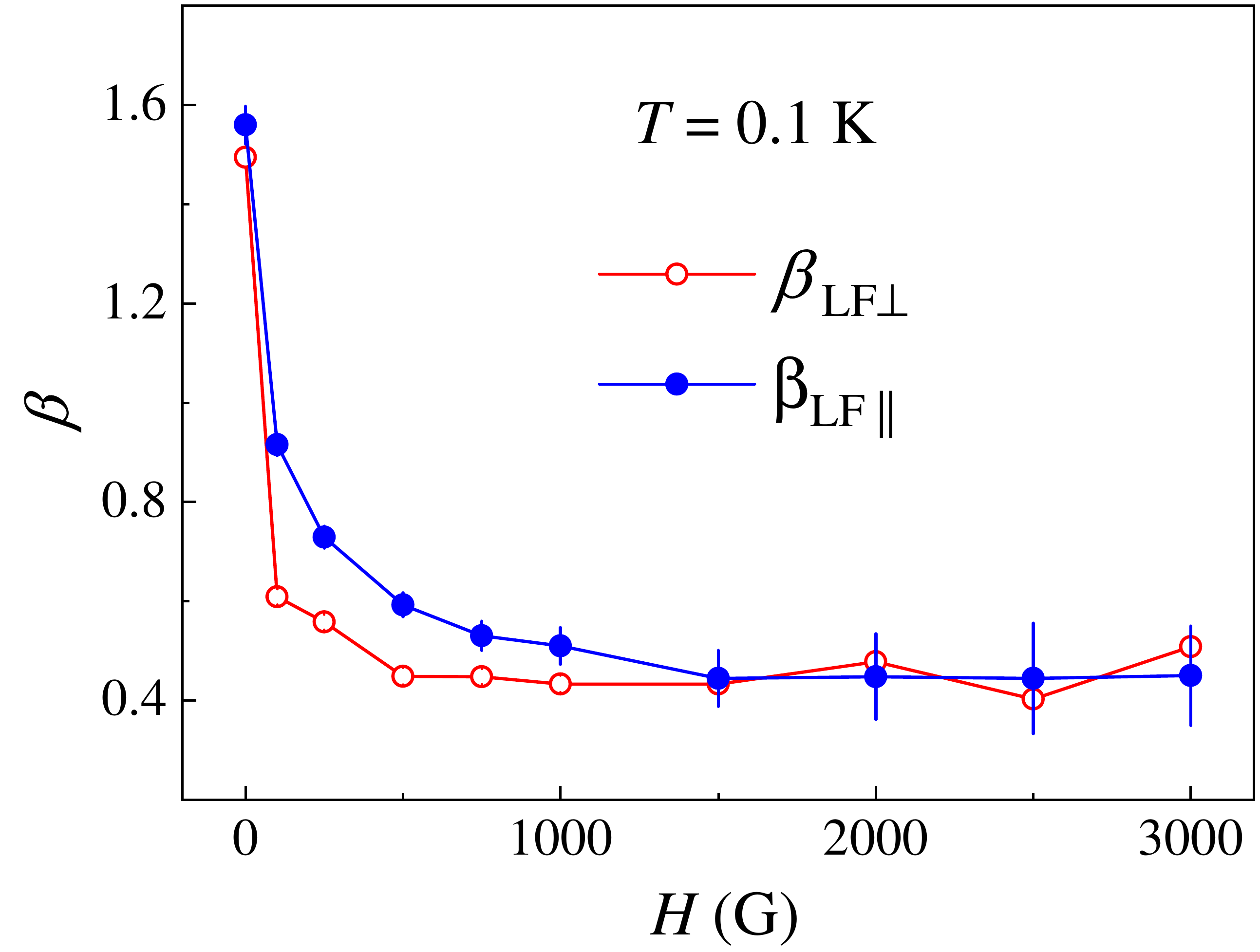}
			\caption{LF-dependence of the stretched exponent $\beta$ obtained from the fit of the LF-spectra [main text: Figs.8(c)-8(d)] with Eq. (5) in the main text with the initial polarisation of muons along (inset) and perpendicular to the $c$-axis.}
			\label{LFBeta_x_0pt1}
		\end{figure}

		In the main body of the paper, we presented the analysis of ZF-$\mu$SR spectra of CeRh$_{1-x}$Pd$_x$Sn with $x=0.1$ single crystal and $x=0.2$ polycrystalline samples using fits with one component for the stretched exponent. Here we present a comparison between one component and two-component fits to the ZF-$\mu$SR  spectra of $x=0.1$ and 0.2. Figures~\ref{OneVsTwoComponent}(a) and~\ref{OneVsTwoComponent}(b) show that the ZF-$\mu$SR spectra for a single crystal sample of $x =$ 0.1 with the initial polarisation of muons along and perpendicular to the $c$-axis, respectively, described by Eq.~(1) (solid black line) given in the main paper, can also be described with the two-component function (solid olive line) given by the equation below. 
		
		\begin{equation}
			A(t) = A_1 \exp(-\lambda t) + A_2 \exp(\frac{-\sigma^2t^2}{2}) + A_{\rm BG}
		\end{equation}
		
		where $A$ = $A_1$ + $A_2$ represents the effective initial asymmetry of the signal, $\lambda$ is the Lorentzian depolarization rate which accounts for the dynamic magnetic fluctuations, and $\sigma$ is the Gaussian depolarization rate accounts for isotropic Gaussian distribution of static fields. $A_{\text{BG}}$ is a constant background arising from muons stopping on the silver sample holder. The temperature dependence of the fitting parameters for $x=0.1$ with initial polarisation of muons along and perpendicular to the $c$-axis and for $x=0.2$ are presented in Figs.~\ref{MUSR_x_0pt1}(a)-~\ref{MUSR_x_0pt1}(f) and Fig.~\ref{MUSR_x_0pt2}, respectively. The analysis shows that the ZF-$\mu$SR spectra of $x=0.1$ (and also $x$ = 0.2) can be fitted equally well to either the one or two-component function. In both the methods, the Lorentzian depolarization rate initially increases below 1~K and then saturates at low temperature, indicating the presence of dynamical spin fluctuations as observed in many quantum spin liquid materials ~\cite{PhysRevLett.117.097201, PhysRevB.100.241116, PhysRevB.100.144432}. On the other hand, the temperature dependence of the Gaussian depolarization rate is almost zero down to 0.3~K and sharply rises upon further cooling, indicating the slowing down of spin fluctuations.
		
		Figures~\ref{LF4K_x_0pt1}(a) -~\ref{LF4K_x_0pt1}(c) show $\mu$SR asymmetry data of the $x = 0.1$ single crystal at 0.25, 0.75, and 4~K in longitudinal external field with initial polarisation of muons perpendicular to the $c$-axis ($H_{LF\perp}$ $\sim$ 0~G to 3000~G) and Fig.~\ref{LF4K_x_0pt1}(d) shows $\mu$SR asymmetry data at 4~K in longitudinal external field with initial polarisation of muons parallel to the $c$-axis ($H_{LF\parallel}$ $\sim$ 0~G to 3000~G). It is clear from the LF spectra in Fig.~\ref{LF4K_x_0pt1} that 100~G LF is enough for almost complete suppression of the depolarization at 0.75 and 4~K (PM region) [Figs.~\ref{LF4K_x_0pt1}(b) -~\ref{LF4K_x_0pt1}(d)]. However, at 0.25~K (in the cross-over region), the muon depolarization is gradually reduced by much higher fields [Figure~\ref{LF4K_x_0pt1}(a)] and is not suppressed completely even with an applied LF of 3000~G. This indicates that the observed relaxation is due to a dynamic field generated by the Ce magnetic moments.
		
		Figure~\ref{LFBeta_x_0pt1} shows LF dependence of the stretched exponent $\beta$ obtained from the fit of the LF-spectra with the initial polarisation of muons along (inset) and perpendicular to the $c$-axis with Eq.~(1) in the main paper for the $x = 0.1$ single crystal. At the low field, $\beta$ decreases and then remains constant up to 3000~G for both directions.

		\section{Neutron Diffraction (ND)}

\begin{figure}
			\includegraphics[width=7.5cm, keepaspectratio]{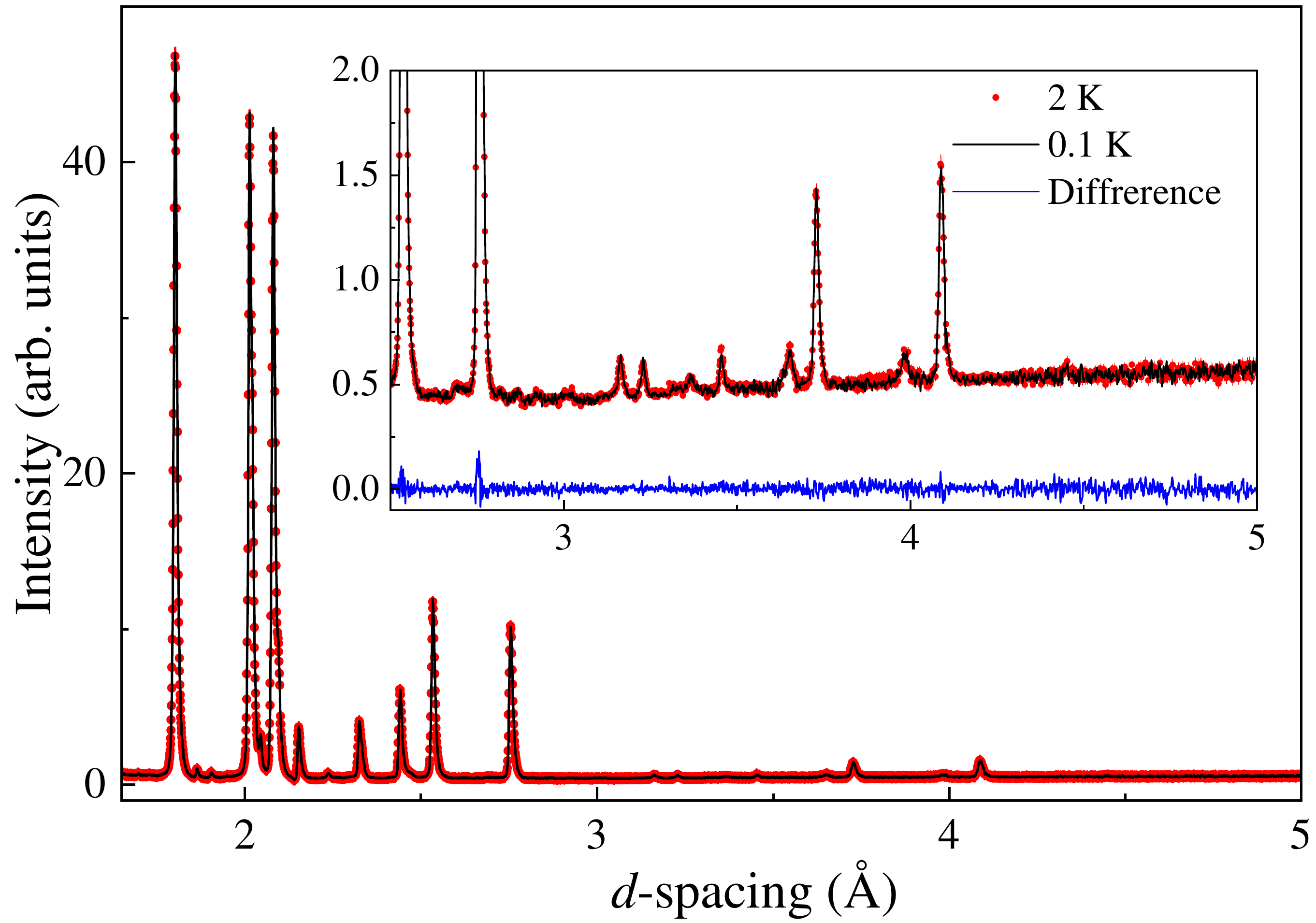}
			\caption{ND data collected on a polycrystalline sample of CeRh$_{1-x}$Pd$_x$Sn with $x = 0.2$ using the OSIRIS diffractometer at $T$ = 0.1~K (solid black line) and 2~K (red circle). Inset shows the enlarged view.}
			\label{0p2_Diffraction}
		\end{figure}

\begin{figure}
			\includegraphics[width=7.5cm, keepaspectratio]{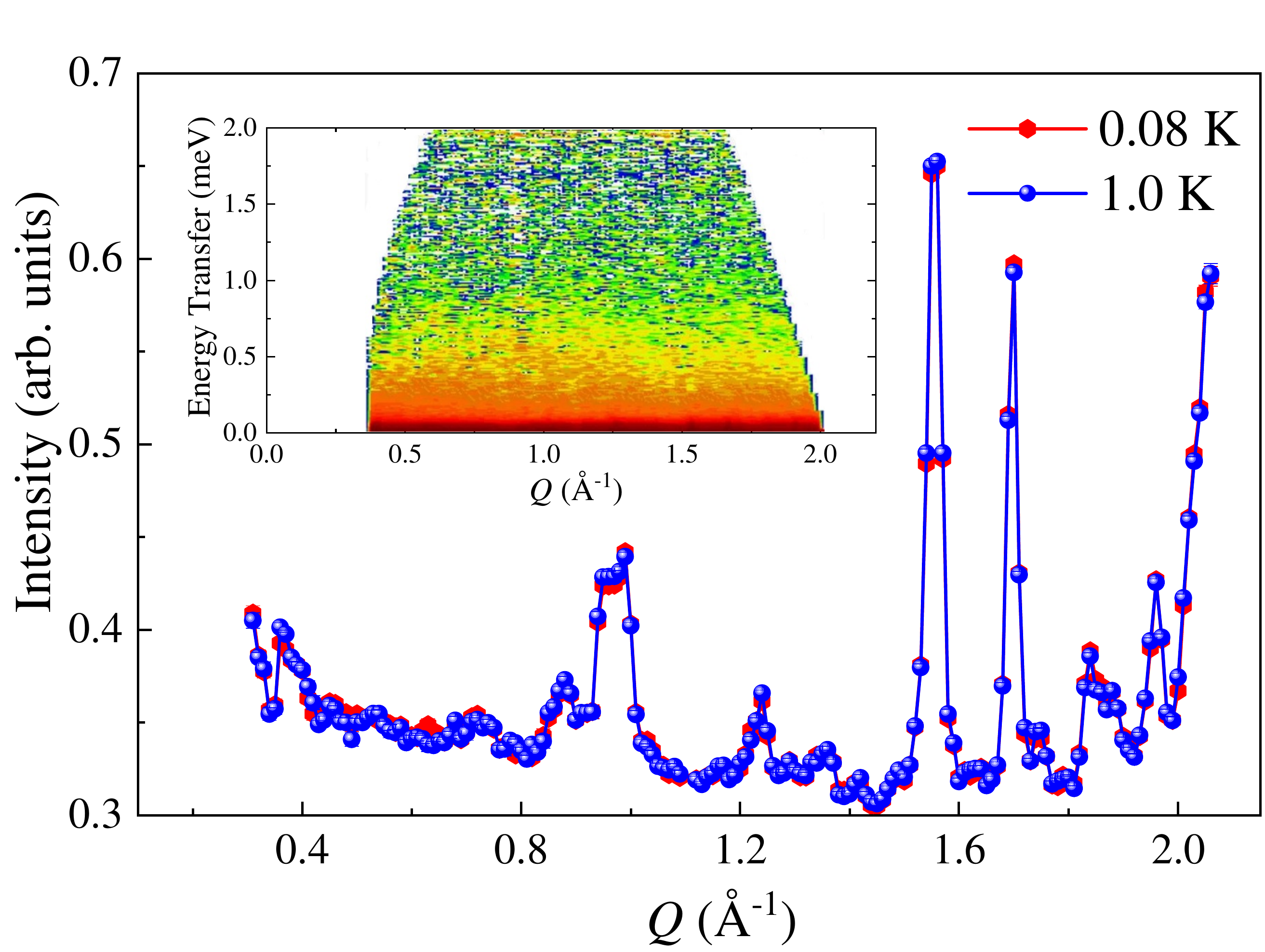}
			\caption{The energy integrated ($\Delta E$ = $\pm$ 0.5~meV) elastic cuts from the IN6 data plotted as intensity vs $Q$ at 0.08~K and 1~K for CeRh$_{1-x}$Pd$_{x}$Sn with $x=0.1$ polycrystalline sample. The data show no extra Bragg peaks present at 0.08~K compared to 1~K data. The inset shows the color coded scattering intensity plotted as energy transfer vs $Q$ at 0.08~K from IN6.}
			\label{0p1_Diffraction}
		\end{figure}

\begin{figure}
			\includegraphics[width=7.5cm, keepaspectratio]{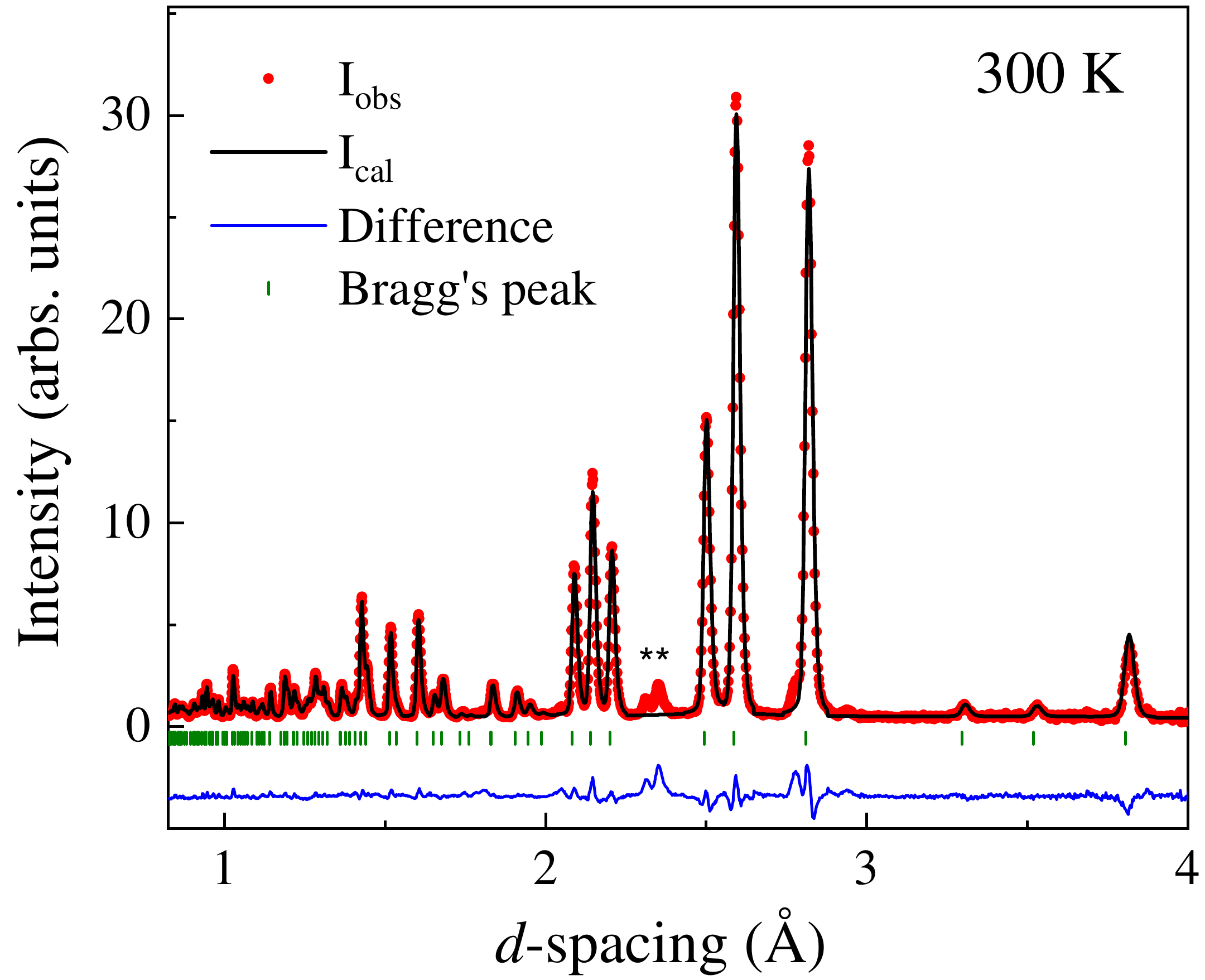}
			\caption{Rietveld refinement of the ND pattern for the sample with $x=0.75$ at room temperature The data were collected using the GEM TOF diffractometer. The two peaks marked with a star symbol show a small unknown impurity phase.}
			\label{0p75_Diffraction}
		\end{figure}

		Figure~\ref{0p2_Diffraction} shows the ND data collected on a polycrystalline sample of CeRh$_{1-x}$Pd$_x$Sn with $x = 0.2$ using the time of flight OSIRIS diffractometer at 0.1~K and 2~K. No clear sign of magnetic Bragg peaks could be observed in the diffraction data collected down to 0.1~K. The absence of magnetic ordering in $x=$ 0.1 down to 0.08~K [Fig.~\ref{0p1_Diffraction}] is also supported through the INS data in the elastic region (-0.5$\leq E \leq$ 0.5). As we noted an anomaly in $\chi_{ac}$($T$) and $C/T$ at around $T$ = 0.13~K for $x$ = 0.1, and at around $T$ = 0.2~K [in $C/T$ vs $T$] for $x$ = 0.2, we expect a difference in the ND data for temperature above and below the temperature of this anomaly, if its origin is due to a long-rang magnetic ordering with a reasonable size of the ordered state Ce moment. However, we did not see any additional Bragg reflections for both $x=0.1$ (Fig.~\ref{0p1_Diffraction}) and 0.2 (Fig.~\ref{0p2_Diffraction}) samples, indicating that either the ground state is non-magnetic or the ordered state moment is too small to be detected. As we have not seen any clear sign of the magnetic ordering in our ZF-$\mu$SR, we concluded that the ground state of $x$ = 0.1 and 0.2 is a non-magnetic metallic spin liquid. Figure~\ref{0p75_Diffraction} shows Rietveld refinement of the ND data for CeRh$_{1-x}$Pd$_x$Sn with $x = 0.75$, collected from one of the GEM detector banks at 300~K. The Rietveld refinement confirms that the $x=0.75$ compound crystallizes in the hexagonal ZrNiAl-type structure with space group $P\bar{6}2m$. 
		
		\begin{figure}
			
			\includegraphics[width=6.0cm, keepaspectratio]{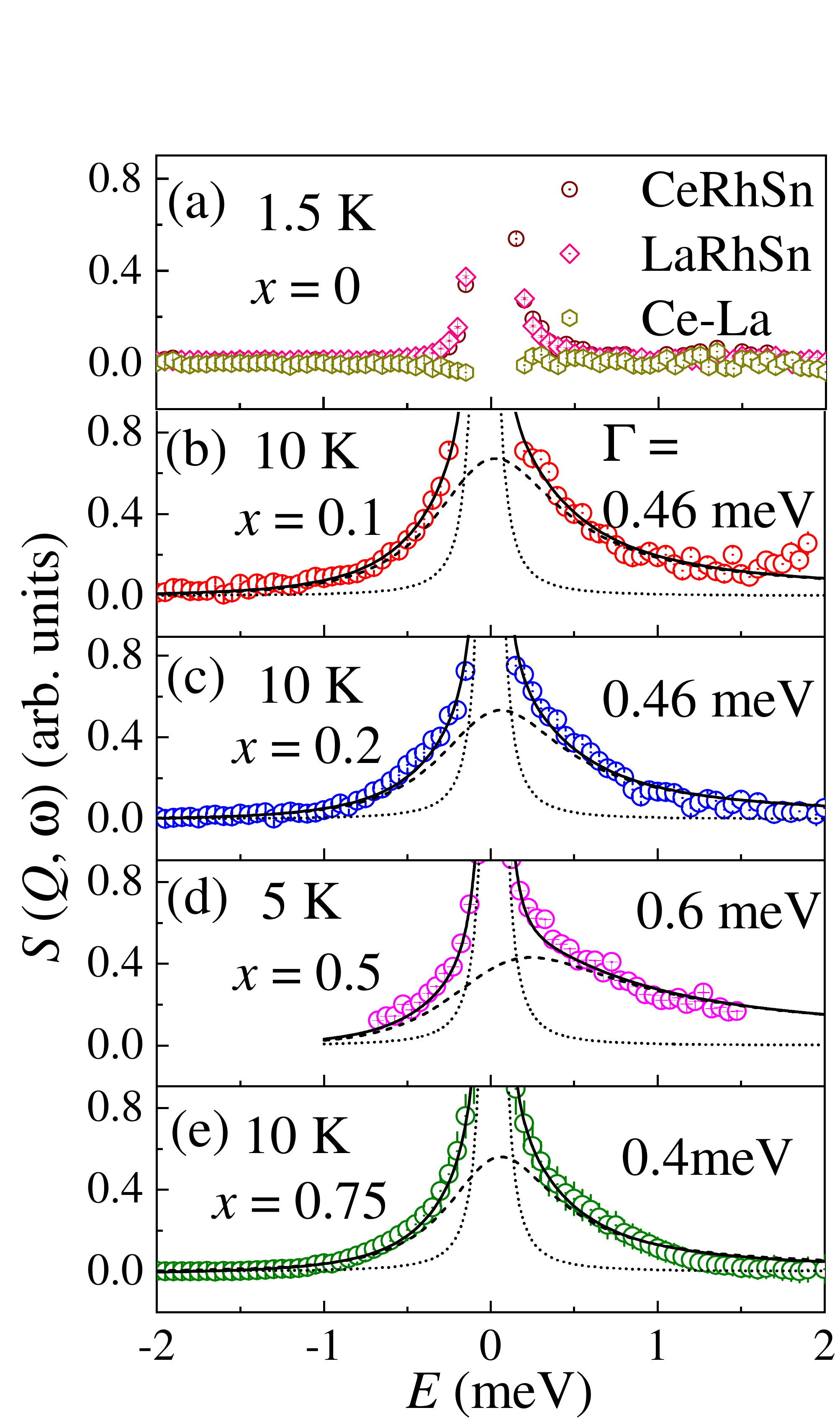}
			\caption{ $Q$-integrated ($0 < Q < 2\mathrm{\AA}^{-1}$) scattering intensity $S$($Q$, $\omega$) versus $E$ for $x=0$, 0.1, 0.2, 0.5, and 0.75 at low temperatures. The $y$-scale for $x$ = 0.75 is divided by 10 to make the similar $y$-scale. The solid lines show the fitting using an elastic line and a quasi-elastic Lorentzian function multiplied by the Bose factor. The dotted lines show the elastic peak and the dashed line shows the quasi-elastic peak. It is to be noted that the magnetic scattering in $x = 0$ was too weak to be analyzed.}
			\label{INSLow_x}
		\end{figure}
		
		\begin{figure}
			\includegraphics[width=6.0cm, keepaspectratio]{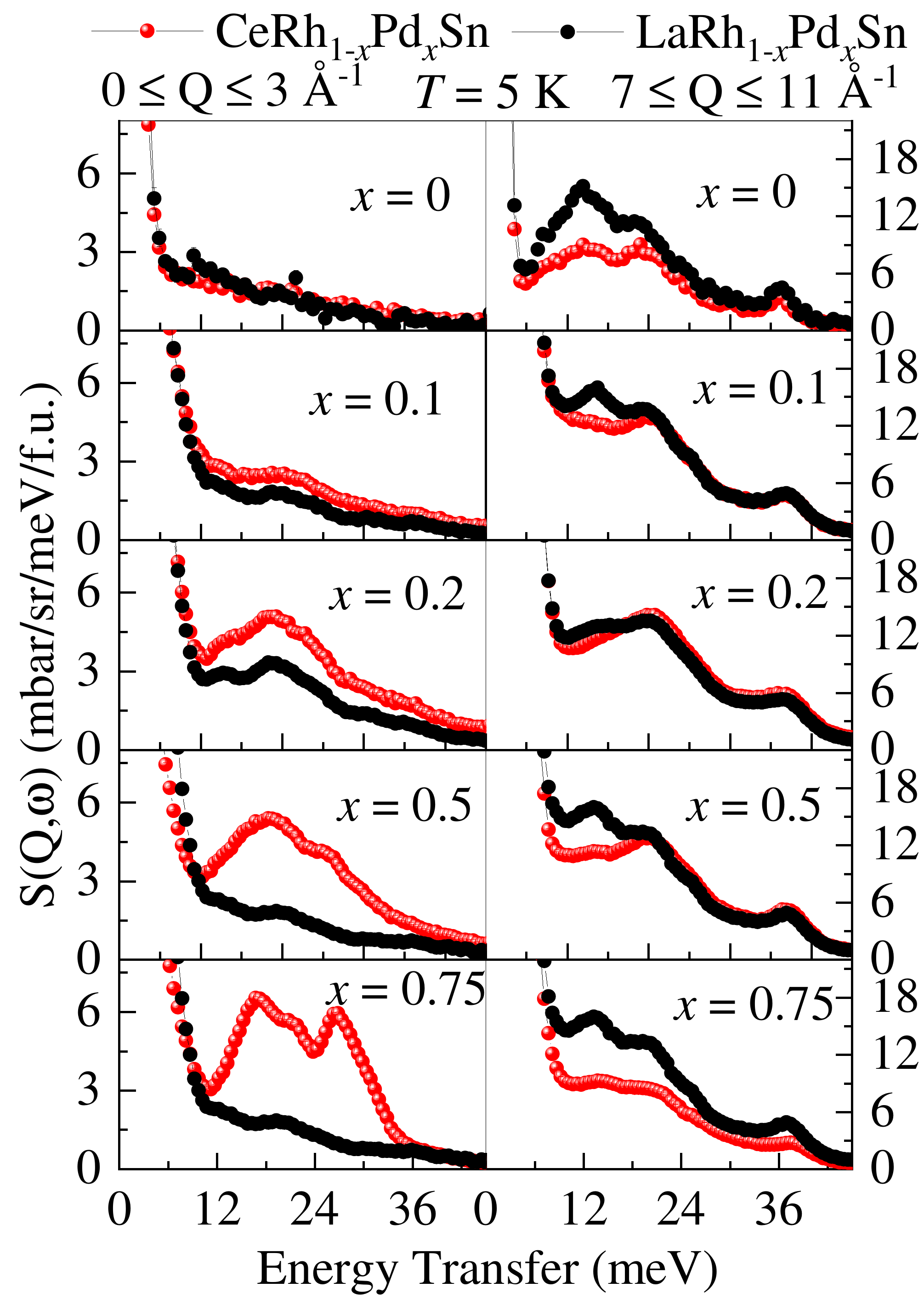}
			\caption{One dimensional (1D) cut from the raw INS data of CeRh$_{1-x}$Pd$_x$Sn (red symbols) and LaRh$_{1-x}$Pd$_x$Sn (black symbols) at 5~K with neutron incident energy of 60.0~meV using the MERLIN TOF spectrometer. Data for $x=0$ are obtained using the MARI TOF spectrometer.}
			\label{INS_La_Ce}
		\end{figure}
		
		\begin{figure}
			\begin{center}
				\includegraphics[width=8.5cm, keepaspectratio]{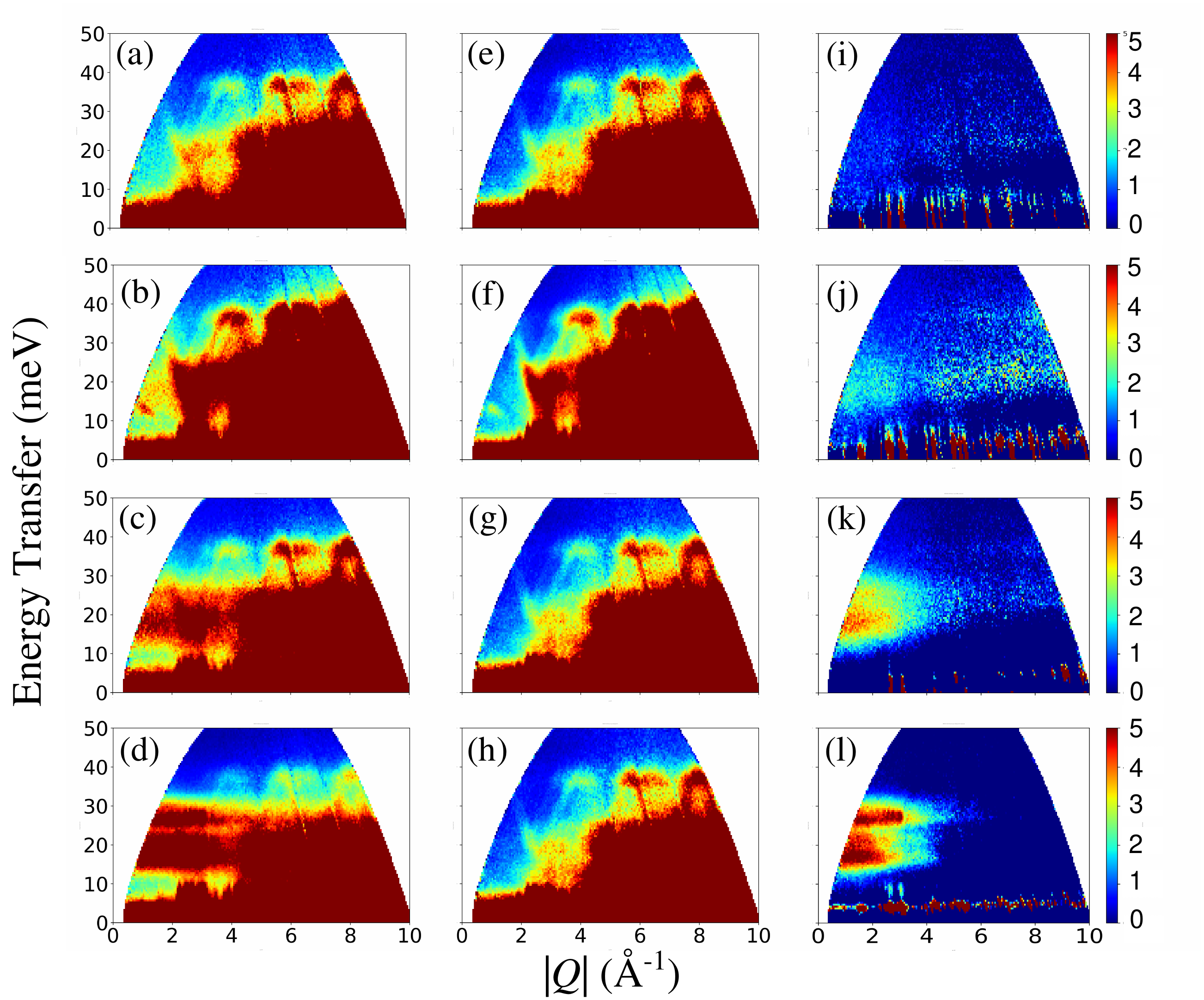}
				\caption{ INS spectra with an incident neutron energy of 60.00~meV of CeRh$_{1-x}$Pd$_x$Sn with (a) $x$ = 0.1 at 5~K, (b) $x$ = 0.2 at 1.8~K, (c) $x$ = 0.5 at 5~K, (d) $x$ = 0.75 at 5~K, and of LaRh$_{1-y}$Pd$_y$Sn with (e) $y$ = 0.1 at 5~K, (f) $y$ = 0.1 at 1.8~K, (g) $y$ = 0.5 at 5~K, and (h) $y$ = 0.5 at 5~K. The magnetic scattering (Ce-La) for (i) $x$ = 0.1 at 5~K, (j) $x$ = 0.2 at 1.8~K, (k) $x$ = 0.5 at 5~K, and (l) $x$ = 0.75 at 5~K.}
				\label{INS_spectra}
			\end{center}
		\end{figure}
		
		\section{Inelastic Neutron scattering (INS)}

		Fig.~\ref{INSLow_x} shows the $Q$-integrated (0 \text{$< Q <2\mathrm{\AA}^{-1}$}) scattering intensity $S$($Q$, \text{$\omega$}) versus $E$ for $x= 0$, 0.1, 0.2, 0.5, and 0.75 at low temperatures. The data of $x=0$, 0.1, 0.2, and 0.75 are obtained using the direct geometry spectrometer IN6 at ILL, while the data of $x = 0.5$ are obtained using the inverted geometry spectrometer OSIRIS at the ISIS Facility. The fit to the quasi-elastic linewidth using a Lorentzian function multiplied by the Bose factor reveals that the quasi-elastic linewidth does not change much with doping $x$ and remains between 0.4 (1) to 0.6 (1)meV for $x = 0.1$ to 0.75. The low temperature value of the quasi-elastic linewidth can give an estimation of the Kondo temperature. Hence the present results suggest that low energy Kondo temperature associated with the ground state remains nearly constant for $x=0.1$ to 0.75.

		Fig.~\ref{INS_La_Ce} shows the $Q$-integrated energy vs intensity 1D cuts from low-$Q$ and high-$Q$ regions for neutron incident energy of $E_i= 60$. The estimated magnetic scattering after subtracting the phonon scattering are shown in Fig.~7 of the main text. We have fitted the INS peaks for $x = 0.1$, 0.2, and 0.5 with a double Lorentzian function multiplied by the population factor (i.e., Bose factor) and for $x = 0.75$, with a four Lorentzian function multiplied by the population factors, shown by the solid black curves in Figs.~8(a) - 8(d) of the main text. The fit parameters, i.e., the peak positions (c), widths (w), and the amplitudes (h), are given in Table~\ref{INSTable}.
		
		\begin{table}
			\caption{Lorentzian fit parameters [center, c (meV), width, w (meV), and amplitude, h(arb. units)] for the high energy INS data of CeRh$_{1-x}$Pd$_{x}$Sn for $x = 0.1$, 0.2, 0.5, and 0.75.}
			\vskip .2cm
			\addtolength{\tabcolsep}{+1pt}
			\begin{ruledtabular}
				\begin  {tabular}{ccccc }
				LFP &  $x=0.1$  & $x=0.2$ & $x=0.5$ & $x=0.75$ \\ \hline 
				
				c1  &14(2)& 15.6(2)& 16.83(6)& 16.42(8) \\ 
				w1  &23(4)& 19.7(7)& 10.3(3)& 5.45(3) \\
				h1  &1.7(7)& 3.2(2)& 3.3(1)& 2.3(1) \\ \hline 
				
				c2  &-& -& 25.8(1)& 21.2(3) \\
				w2  &-& -& 7.4(3)& 4.3(7) \\ 
				h2  &-& -& 0.74(5)& 0.65(1) \\ \hline 
				
				c3  &-& -& -& 27.1(6)\\
				w3  &-& -& -& 5.3 \\ 
				h3  &-& -& -& 1.4(7) \\  \hline 
				
				c4  &-& -& -& 30.5\\
				w4  &-& -& -& 1.13 \\ 
				h4  &-& -& -& 0.04(7) \\  	
			\end{tabular}
		\end{ruledtabular}
		\label{INSTable}
	\end{table}

	\begin{figure}
		\includegraphics[width=8.0 cm, keepaspectratio]{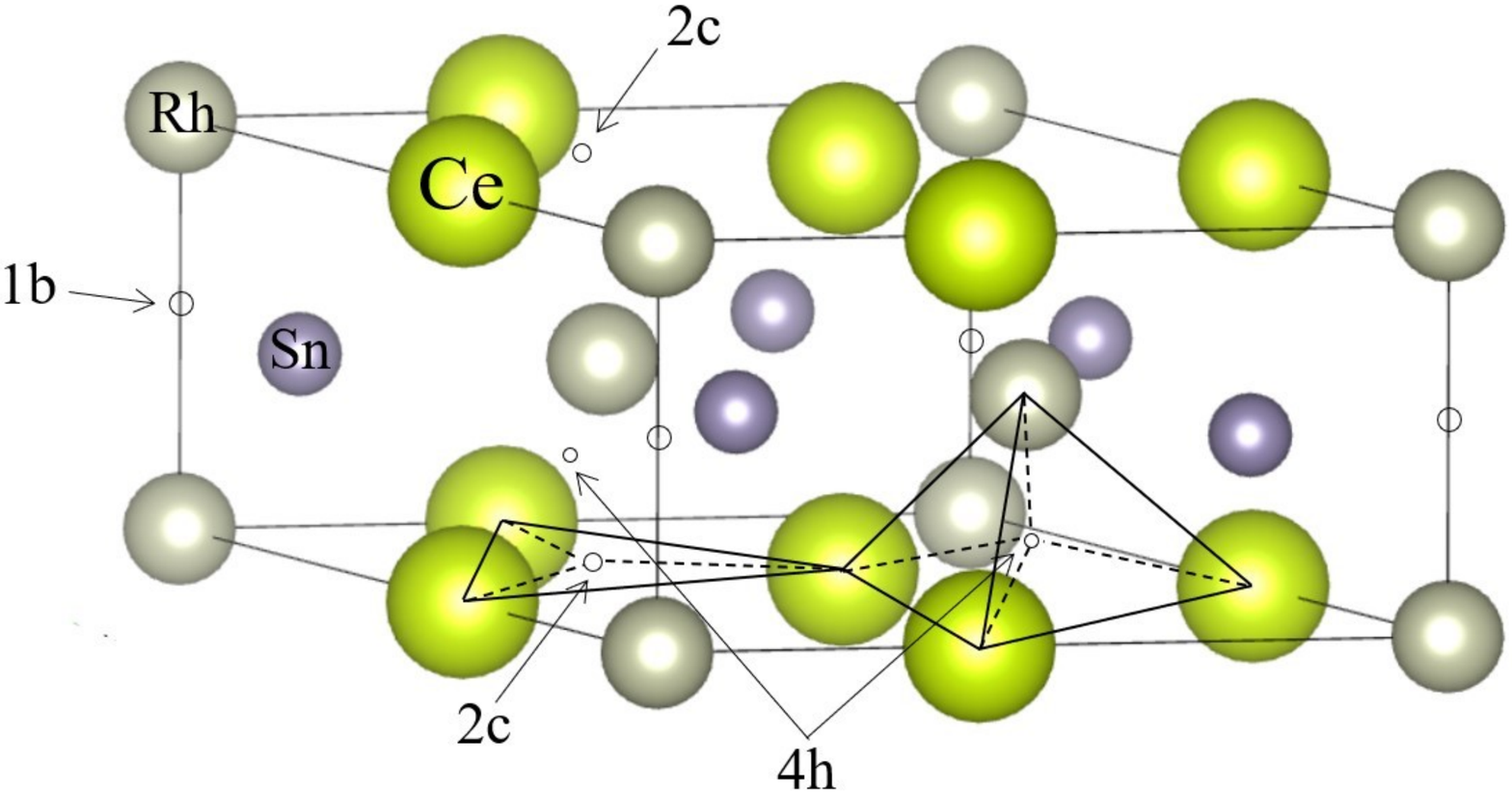}
		\caption{Crystal structure of CeRhSn and interstitial muon topping sites 1b, 2c, and 4h (z = 0.18) proposed from the TF-$\mu$SR measurements from ref~\cite{doi:10.1143/JPSJ.73.3099}.}
		\label{muonsite}
	\end{figure}

	Figs.~\ref{INS_spectra}(a) -~\ref{INS_spectra}(h) show the inelastic neutron-scattering spectra with an incident neutron energy of 60.0~meV of CeRh$_{1-x}$Pd$_x$Sn and LaRh$_{1-y}$Pd$_y$Sn for (a) $x$ = 0.1 at 5~K, (b) $x$ = 0.2 at 1.8~K, (c) $x$ = 0.5 at 5~K, (d) $x$ = 0.75 at 5~K, (e) $y$ = 0.1 at 5~K, (f) $y$ = 0.1 at 1.8~K, (g) $y$ = 0.5 at 5~K, and (h) $y$ = 0.5 at 5~K. Figs.~\ref{INS_spectra}(i) -~\ref{INS_spectra}(l) show and magnetic scattering (Ce-La) for (i) $x$ = 0.1 at 5~K, (j) $x$ = 0.2 at 1.8~K, (k) $x$ = 0.5 at 5~K, and (l) $x$ = 0.75 at 5~K.
	
	The muon site in CeRhSn single crystal has been estimated from the Knight Shift measurements by A. Schenck et al.~\cite{doi:10.1143/JPSJ.73.3099}. Interstitial sites available to the implanted muons are the axially symmetric sites 1b at position (0 0 1/2) and 4h at the generic position (1/3 2/3 z), as well as various non-axially symmetric sites 2c (see Fig.~\ref {muonsite}). From TF-measurements and comparing the estimated dipolar field with minimizing the electrostatic energy, the proposed possible muon site is 2c in CeRhSn (see Fig.~\ref {muonsite}). As our ZF musr relaxation rate exhibits very similar behavior in both the directions (see Fig. 4c in main paper), hence we ruled out the effect of a geometric factor on our LF -relaxation rate. 
	
	\begin{table*}
		
		\caption{ZF-$\mu$SR relaxation rate ($\lambda_{\text{ZF}}$) at the base temperature, the stretched exponents at high ($\beta_{\text{high T}}$) and low ($\beta_{\text{low T}}$) temperatures, the fluctuation rate ($\nu = 1/\tau$), $H_{\text{loc}}$, and the ratio $\gamma_\mu\times H_{\text{loc}}/ \nu$ derived from the Redfield equation at the base temperature, for geometrically frustrated quantum-spin-liquid candidates.}
		\begin{ruledtabular}
			\begin  {tabular}{cccccccc }
			Materials  & Base T (K) & $\lambda_{\text{ZF}}$($\mu s^{-1}$) & $\beta_{\text{high T}}$/$\beta_{\text{low T}}$  & $\nu$ (MHz)& $H_{\text{loc}}$(G) &$\gamma_\mu H_{\text{loc}}/\nu$ & Ref.  \\ \hline
			CeRh$_{0.9}$Pd$_{0.1}$Sn   &0.05& 0.6& 1/1.6 &33.3& 25 & 0.1& --\\ 
			CePdAl    &0.3& 0.3& 1/0.7& -- &--&--& \cite{PhysRevB.105.L180402}\\ 
			CeRhBi    &0.05& 0.22& --& 23 &15&0.054 & \cite{doi:10.7566/JPSJ.87.064708}\\ 
			Ce$_2$Zr$_2$O$_{9}$    &0.02& 0.3& 0.65/0.9&  -- &--&--& \cite{Nat.Phys.15.1052}\\ 
			YbMgGaO$_4$   &0.04& 0.3& 0.9/0.65&9.5&--& --& \cite{PhysRevLett.117.097201}\\ 
			NaYbS$_2$   &0.05& 0.3& -- &  1.7 &--&-- & \cite{PhysRevB.100.241116}\\
			Ca$_{10}$Cr$_7$O$_{28}$   &0.02& 0.6& 0.9/1.3& -- &--&--& \cite{Nat.Phys.12.942}\\
			Cu$_2$IrO$_3$   &0.03& 0.6& 1/0.5& 72  &25&0.03& \cite{PhysRevLett.122.167202}\\
			SrCr$_8$Ga$_4$O$_{19}$  &0.1 & 10$^*$& 1/2& -- &--&--& \cite{PhysRevLett.73.3306}\\ 
			Zn$_x$Cu$_{4-x}$(OH)$_6$Cl$_2$    &0.05& 0.05& --&  150 &18&0.01 & \cite{PhysRevLett.98.077204}\\ 
			NdTa$_7$O$_{19}$    &0.05& 2& 1/0.7& -- &--&--& \cite{arh2022ising}\\ 
			Na$_3$Sc$_2$Mo$_5$O$_{16}$    &0.02& 1.8& --&  480 &24&0.04& \cite{PhysRevMaterials.6.044414}\\ 
			NaYbO$_{2}$    &0.1& 0.9& --&  -- &--&--& \cite{PhysRevB.100.144432}\\ 		
			K$_{2}$Ni$_{2}$(SO$_{4}$)$_{3}$    &0.02& 1.2& 1/2&  -- &--&--& \cite{PhysRevLett.127.157204}\\ 
			Na$_{2}$BaCo(PO$_{4}$)$_{2}$ &0.06& 0.6& 1.6/0.8&  7.8 &--&--& \cite{PhysRevB.103.024413}\\ 
			%					H$_{3}$LiIr$_{2}$O$_{6}$ &0.06& 0.6& 1.6/0.8&  7.8 &--&--& \cite{arxiv.2201.12978}\\ 
			%						Ag$_{3}$LiIr$_{2}$O$_{6}$ &0.06& 0.6& 1.6/0.8&  7.8 &--&--& \cite{PhysRevLett.123.237203}\\ 
		\end{tabular}
		
	\end{ruledtabular}
	
	\label{edxTable}
	*LF = 100G	
\end{table*}

\end{document}